\documentclass[lettersize,journal]{IEEEtran}
\usepackage{amsmath,amsfonts,amsthm,amssymb}
\usepackage{algorithmic}
\usepackage{algorithm}
\usepackage{array}
\usepackage{textcomp}
\usepackage{subfig}
\usepackage{float}
\usepackage{enumerate}
\usepackage{tabularx}
\usepackage{makecell}
\usepackage{stfloats}
\usepackage{url}
\usepackage{hyperref}
\usepackage{url}
\usepackage{xcolor,soul,framed} 
\usepackage{verbatim}
\usepackage{graphicx}
\usepackage{epstopdf}
\usepackage{cite}
\usepackage{booktabs}
\usepackage{multirow}
\begin{document}
\renewcommand{\algorithmicrequire}{\textbf{Input:}} 
\renewcommand{\algorithmicensure}{\textbf{Output:}}

\newcommand {\zhengyu}[1]{{\color{BLUE}\sf{[Zhengyu: #1]}}}

\title{Design of Reconfigurable Intelligent Surfaces for Wireless Communication: A Review}

\author{Rujing~Xiong,~Jianan~Zhang, Fuhai~Wang, Zhengyu~Wang,~Xiang~Ren, Junshuo~Liu, ~Jialong~Lu,~\IEEEmembership{Student Member,~IEEE,}
		Kai,~Wan,~Tiebin~Mi,~\IEEEmembership{Member,~IEEE,}
 		Robert~Caiming~Qiu,~\IEEEmembership{Fellow,~IEEE,}
\thanks{R.~Xiong, J.~Zhang, F.~Wang, Z.~Wang, X.~Ren, J.~Lu, J.~Liu, K.~Wan, T.~Mi and R.~Qiu are with the School of Electronic Information and Communications, Huazhong University of Science and Technology, Wuhan 430074, China ( e-mail: \{rujing,~zhangjn,~wangfuhai,~wangzhengyu,~m202272470,~m202272434,~junshuo\_liu, ~kai\_wan,~mitiebin,~caiming\}@hust.edu.cn).}
\thanks{National Foundation, NO.12141107, supports this work.}
\thanks{This manuscript is to appear in Journal of Huazhong University of Science \&~Technology (Natural Science Edition),~2023,~51(9): 1-32  (DOI:10.13245/j.hust.230901).}}



\maketitle

\begin{abstract}
This paper addresses the hardware structure of Reconfigurable Intelligent Surfaces (RIS) and presents a comprehensive overview of RIS design, considering both unit design and prototype systems. It commences by tracing the evolutionary trajectory of RIS, originating from static cell-structured hypersurfaces. The article conducts a meticulous examination from the standpoint of adaptability, elucidating the diverse array of unit structures and design philosophies that underlie existing RIS frameworks. Following this, the study systematically categorizes and synthesizes channel modeling research for RIS-facilitated wireless communication, leveraging both physical insights and statistical data. Additionally, the article provides a detailed exposition of current RIS experimental setups and their corresponding empirical findings, delving into the attributes of prototype design and system functionalities. Moreover, this work introduces an in-house developed RIS prototype. The prototype undergoes rigorous empirical evaluation, encompassing multi-hop RIS signal amplification, image reconstruction, and real-world indoor signal coverage experiments. The empirical results robustly affirm the efficacy of RIS in effectively mitigating signal coverage blind spots and enabling radio wave imaging. With RIS-enhanced augmentation, the average indoor signal gain surpasses 8 dB.
\end{abstract}


\section{Introduction}
\IEEEPARstart{W}{ith} the development of Internet of Things (IoT) systems, the demand for wireless capacity is increasing in the context of the sixth-generation (6G) mobile communication system. According to a report by Cisco~\cite{website2019cisco}, by 2030, global mobile traffic will grow to 5,016 EB/month (1 EB = 1,024 PB, 1 PB = 1,024 TB), a 65-fold increase from 2023. Undoubtedly, in the current large-scale commercial deployment of 5G, techniques like massive multiple-input multiple-output (MIMO) and millimeter-wave communication have significantly enhanced system capacity. Massive MIMO employs exceptionally large antenna arrays and employs simple signal processing techniques to mitigate interference within a unit~\cite{ngo2013energy}. Millimeter-wave communication offers higher spectrum efficiency. The improvements from these technologies complement novel unit architecture topologies, ultimately improving user experience. However, it's important to note that these enhancements come at the cost of complex hardware requirements and exceedingly high energy consumption~\cite{hashmi2018efficiency,hashmi2020enhancing}. Additionally, millimeter-wave and even terahertz communication hold the promise of leveraging vast bandwidths in higher spectra to enhance data rates. Nevertheless, increased frequency leads to significant electromagnetic wave attenuation. Moreover, environmental factors like water vapor and other mediums can absorb electromagnetic waves, resulting in substantial path loss and impacting coverage range. It can be stated that energy loss and the wireless transmission environment have become critical bottlenecks limiting the application of new communication key technologies.

The concept of smart radio environments (SREs) has emerged from fundamental research on 6G~\cite{di2020smart}. SREs refer to a perspective where the wireless environment is not regarded as a random and uncontrollable entity but is considered a part of network design parameters, allowing people to tailor the wireless environment according to their preferences. To materialize SREs, the concept of RIS, as an entirely new technology, has captivated the attention of the wireless communications research domain.

RIS, initially stemming from engineered composite materials~\cite{mendis2014artificial,pendry1996extremely}, primarily refers to metamaterials (or metasurfaces)~\cite{della2014digital}. Metasurfaces are artificially crafted to possess special properties that natural media cannot provide, such as controlling electromagnetic waves. Initially, metasurfaces are single-layered structures comprised of various patterned units. These could induce abrupt phase changes in incident wavefronts. Through ingeniously designed discontinuous interface profiles, such as mushroom-shaped, annular, or cross-shaped patterns~\cite{sievenpiper1999high,yen2004terahertz,tao2008highly,tao2009reconfigurable}, arbitrary manipulation of the wavefront could be achieved. However, due to the inflexibility and high tuning costs associated with these metasurfaces, efforts are redirected toward designs that could yield variable phase shifts. This led to the birth of programmable metasurfaces, where phase variation could be achieved through physical factors like the adjustable height and length of components~\cite{gianvittorio2006reconfigurable,cabria2009active}.

The idea of programmable design of RISs can be traced back to 2011 when Capasso et al. exploited the spatial phase variations and subwavelength separation properties of a two-dimensional array of optical resonators~\cite{yu2011light}. This results in discontinuous phase changes during light reflection and transmission, enabling the design of different beams. They introduce the generalized Snell's law and provide insights into the microwave and millimeter-wave domains. Subsequently, gradient phase, geometric phase metasurfaces, and others are proposed~\cite{sun2012gradient,sun2012high,huang2012dispersionless}. The team led by Cui achieved a groundbreaking milestone by employing PIN diodes (positive-intrinsic-negative diode) for the first time to enable digital programmable control of metasurfaces~\cite{cui2014coding}. This digitally encoded metasurface, also known as RIS, enables real-time programmability in 1-bit or multi-bit configurations, marking a revolutionary leap compared to traditional metasurfaces. With this achievement, RIS has established a bridge between physical space and information space, offering new dimensions for communication, radar, imaging, and other information systems.

The primary feature of RIS lies in its intelligent capability to reflect impinging signals towards desired directions (beamforming), hence, appropriately designing the phase shifts of RIS can fully exploit its potential. Currently, typical objectives for RIS phase design primarily focus on maximizing Signal-to-Noise Ratio (SNR) or transmission rate~\cite{yang2019irs,guo2019weighted}, minimizing transmit power~\cite{wu2019intelligent,wu2019beamforming}, and maximizing energy efficiency~\cite{huang2019reconfigurable,yang2020energy}. By setting such optimization objectives and solving the associated problems to obtain suitable RIS phase values, it's important to note that due to inherent constraints within RIS, these problems are typically non-convex. The primary methods for solving these problems encompass continuous convex approximation~\cite{song2022intelligent}, alternating direction method of multipliers~\cite{liang2016unimodular}, semi-definite relaxation~\cite{wu2019beamforming}, manifold optimization~\cite{xiong2023ris}, and other techniques.

As an artificial electromagnetic surface structure with programmable electromagnetic properties, RISs typically consist of a large number of carefully designed electromagnetic units. By applying control signals to controllable elements on these units, control over the state of each unit on the surface can be achieved. This enables the creation of an electromagnetic field with controllable parameters such as amplitude, phase, polarization, and resonance frequency, actively manipulating spatial electromagnetic waves. Deploying RIS on the surfaces of various objects in the wireless transmission environment can break through the uncontrollability of traditional wireless channels and establish SREs. RIS exhibits several prominent characteristics:
(i) They are nearly passive. Ideally, they require no dedicated power source. (ii) They are regarded as a continuous surface. Ideally, any array unit can reshape the waveform of incident electromagnetic waves through soft programming. (iii) They are immune to receiver noise. RISs do not require Analog-to-Digital Converters (ADCs) and Digital-to-Analog Converters (DACs), as well as power amplifiers. This means that they neither amplify nor introduce noise during signal detection, they providing inherent full-duplex transmission. (iv) They have a full-band response and can operate at any frequency. (v) They can be easily deployed on building facades, factory ceilings, indoor spaces, human clothing, etc. It is worth noting that different RIS unit structures can be designed for the same physical property requirements.

The existing reviews on RIS predominantly center on its historical evolution~\cite{basar2019wireless}, signal processing~\cite{bjornson2022reconfigurable,pan2022overview,mei2022intelligent}, theoretical modeling~\cite{di2022communication}, among other related areas. A comprehensive review exclusively dedicated to RIS unit design and prototypes has yet to emerge. To bridge this gap, this work provides an overview of the electromagnetic unit design of RIS, the system of RIS-assisted communication prototypes, and associated experimental tests. The objective is to furnish readers with a profound comprehension of the operational principles and empirical efficacy of RIS. Furthermore, it introduces a novel RIS prototype, along with its experimental trials aimed at augmenting signal coverage and refining target imaging within wireless communication systems.

The remaining sections of the paper are organized as follows: Section~\ref{Section2} delves into the fundamental model of RIS, encompassing the categorization of various RIS unit structures and detailed explanations of their electromagnetic structural models. Additionally, it covers the classification and examination of the channel model pertaining to RIS-assisted communication. Section~\ref{Section3} categorizes RIS prototypes according to their mission requirements and presents associated experimental evaluations. This section highlights the proposed RIS prototype system operating at 5.8 GHz, along with its corresponding experimental tests. Finally, Section~\ref{Section4} concludes the paper and offers insights into future directions for the hardware design of RIS.

\section{Basic Models}\label{Section2}
\subsection{Electromagnetic Structure Model}

This chapter mainly explains the development process of RIS from the unit design of the metasurface. The unit design of metasurfaces can be divided into fixed unit models and adjustable unit models according to their functional characteristics. First, the development of fixed-structure metasurfaces is reviewed, and then the development of RIS among metasurfaces is further classified and elaborated.

\subsection*{(1) Unit with fixed structure}

\begin{figure}[htbp]
\centerline{\includegraphics[width=.8\columnwidth]{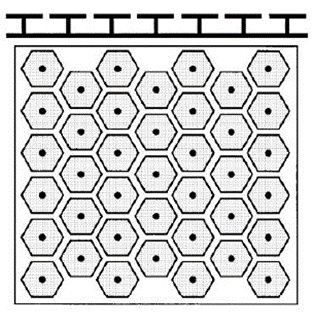}}
\caption{Mushroom-shaped metallic electromagnetic structure\cite{sievenpiper1999high}.}
\label{F1-1}
\end{figure}

The prototype of metasurface can be traced back to 1999. Sievenpiper et al. of the University of California, USA designed a `mushroom cloud' type two-dimensional electromagnetic band gap electromagnetic structure  (Fig.~\ref{F1-1}). They periodically arranged regular hexagonal units on a two-dimensional plane, which is used to control the transmission of different modes of surface waves~\cite{sievenpiper1999high}. However, whether it is the manipulation of surface waves or space waves, due to the influence of the inherent research system of metasurfaces in the early stage, people still design and study metasurfaces from the perspective of equivalent medium parameters (such as surface polarizability, surface impedance, etc.), resulting in most metasurfaces being uniform and all electromagnetic units having the same electromagnetic response.

\begin{figure}[htbp]
\centerline{\includegraphics[width=.9\columnwidth]{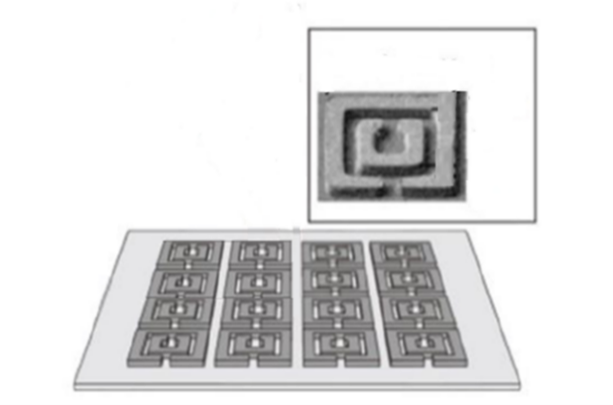}}
\caption{Metasurface based on SRR-resonance structure~\cite{yen2004terahertz}.}
\label{F1-2}
\end{figure}

 Ye et al. achieved an artificial magnetic resonant response at 1 THz using a split ring resonator (SRR) resonant structure fabricated on a silicon wafer (shown in Fig.~\ref{F1-2})~\cite{yen2004terahertz}. Chen et al.x designed a ring electrically tuned metasurface in the terahertz band by introducing a DC bias signal using a GaAs process~\cite{chen2006active}. Gianvittorio et al. used air as a medium to change the reflection phase by controlling the height of the patch from the substrate through electrostatic forces in the same year~\cite{gianvittorio2006reconfigurable}. Tao et al. experimentally verified the absorption performance of an ELC resonant structure with a metal backplane in the terahertz band for the first time~\cite{tao2008highly}. The physical mechanism of electromagnetic wave absorption is based on the strong resonance of the metal structure in the super-surface structure, resulting in the loss of most of the energy in the incident electromagnetic waves in the form of heat in the medium, thus achieving the effect of wave absorption. Tao et al. designed an RIS that adjusts the angle of the SRR resonant ring by temperature using the different thermal expansion coefficients between silicon nitride and gold ~\cite{tao2009reconfigurable}. They experimentally showed that its transmission coefficient changes with the orientation (temperature control) of the SRR ring, which can be used for the design of devices such as temperature detection and reconfigurable filters.

\begin{figure}[htbp]
\centerline{\includegraphics[width=.8\columnwidth]{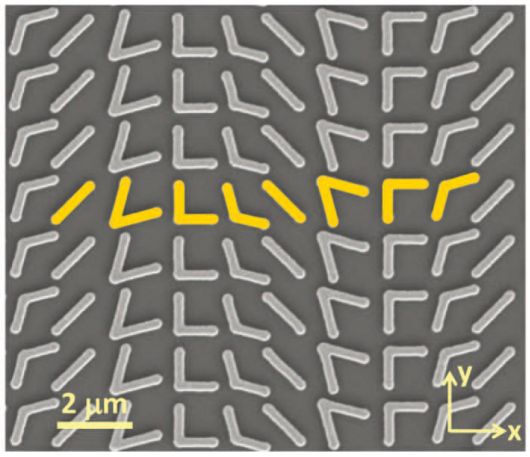}}
\caption{V-shaped metasurface~\cite{yu2011light}.}
\label{F1-3}
\end{figure}

 Capasso et al. proposed a generalized Snell's law, which broke the inherent design method of metasurfaces. This method allowed for the design of non-uniformly distributed metasurfaces, such as gradient metasurfaces, to demonstrate the non-uniform regulation of electromagnetic waves~\cite{yu2011light}. The same concept was reflected in the time-reversal technique~\cite{qiu2007time}. The researchers introduced a new idea of phase mutation, achieving phase tuning of cross-polarized electromagnetic waves with full 360° coverage on an ultra-thin dividing surface by controlling the tensor angle and orientation of the designed `V' unit structure, as shown in Fig.~\ref{F1-3}. Sun et al. designed an H-shaped unit structure and realized a gradient-phase reflective super-surface ~\cite{sun2012gradient}. In this design, researchers controlled the gradient period of the metasurface within one wavelength, and for the first time, they used the surface to efficiently convert space waves into surface waves. Subsequently, he achieved high-efficiency anomalous reflection of incident waves within broadband using a gradient-phase metasurface~\cite{sun2012high}. In addition, Geometric phase-based metasurfaces, also known as Pancharatnam-Berry (PB) phase metasurfaces, have been widely investigated for modulating circularly polarized waves~\cite{huang2012dispersionless}. Using the same structure of the unit to rotate a certain Angle to produce a phase mutation, used to regulate the electromagnetic wave, can realize the vortex beam, holographic imaging, plane lens and other applications.

In order to further study the metasurface of fixed structure, Zhang et al. designed a low-scattering Fabry-Perot antenna to address this issue~\cite{zhang2017realization}. The peak measured gain of this antenna was 19.8 dB in the frequency range of 8 to 12 GHz, and the scattering was significantly reduced. Yuan et al. proposed an effective method to reduce the broadband radar cross section~\cite{yuan2017broadband}. By encoding eight linear phase gradients in a spiral pattern, the reflected wave is uniformly spread, thus achieving a backscatter RCS reduction in the ultra-wideband range.

 Li and Yao constructed a multi-bit digitally encoded metasurface using PB phases based on metal particles of the same size but with different orientations~\cite{li2018manipulation}. Wang et al. proposed an ultrathin phase metasurface structure based on complementary square split ring (SSR) nanoantennas, which enables the realization of terahertz optical hypermorphic lenses with multi-dimensionality and multi-focusing. They also introduced the concept of split-phase mode to realize a longitudinal dual-focused super configuration lens, providing a flexible and convenient method to achieve the desired focusing performance~\cite{wang2019terahertz}. Shao L et al. proposed a scalable design strategy for a bifunctional anisotropic encoded metasurface based on a fully dielectric resonant cavity. The strategy exploits two orthogonal polarizations of electromagnetic waves propagating on the metasurface to encode different sequences for polarization-dependent beam manipulation~\cite{shao2019dual}. Al-Nuaim et al. proposed non-uniform, single-layer, fully dielectric, non-absorbing 1- and 2-bit encoded diffuse reflected-ray metasurfaces that are designed for nearly uniform low-level at large incidence angle ranges (up to $60^{\circ}$) around 12 GHz electromagnetic wave diffusion. The proposed metasurface is implemented using subwavelength periodic all-dielectric units to achieve the desired reflection phase correction (and effective dielectric constant) on each unit within the small-size dielectric reflection rays and to achieve 1-bit and 2-bit encoded sequences of encoded particles~\cite{al2019design}. Wang et al. proposed a method to realize a single-fed multibeam metasurface antenna by combining phase gradients and coding sequences. This technique can produce multibeam antennas with different far-field patterns by superimposing different kinds of phase coding sequences. The fabricated antenna can produce four perfect pencil-shaped beams over a wide frequency range of 11-13 GHz, achieving a gain of about 19.9 dB at the center frequency~\cite{wang2019multi}.
Lin et al. proposed a circularly polarized (CP)-sustained metasurface that enables ultra-broadband CP-maintained reflection, resulting in a co-polarized reflection coefficient close to 1.0 at CP incidence in the 6.2-26.4 GHz band. Furthermore, by rotating its unit-unit structure, a PB phase was achieved~\cite{lin2021ultra}. Wan et al. proposed a metasurface combined absorption frequency selective structure (AFSS) scheme that provides low-frequency scattering, broadband transparent window, and high-frequency absorption~\cite{wan2021composite}. Jing et al. proposed a novel principle of additive coded grating that can add and subtract two or more conventional coded element grating sequences to obtain a new encoded meta-grating sequence. The encoded meta-grating can be flexibly adjusted to the scattering angle, providing a new degree of freedom for flexible modulation of terahertz wavefronts~\cite{jing2022enhancement}.

 Abdullah and Koziel investigated the design of coded metasurface units with broadband RCS reduction properties based on supervised learning. They employed a two-stage optimization process to maximize the RCS-abbreviated bandwidth and designed lattices with four unique geometries, corresponding to `00', `01', `10', and `11' binary codes, representing four phase reflection states of $0, \pi/2, \pi$, and $3\pi/2$, respectively~\cite{abdullah2021supervised}. 
 Patel SK investigated three different graphene-based refractive index sensors (split-ring resonator (SRR), fine-wire split-ring resonator (SRRTW), and fine line refractive index sensor) with 1-bit encoding of `0' and `1' by changing the graphene chemical potential~\cite{patel2022encoding}.
Zhu et al. designed, fabricated, and measured a monolayer meta-hologram that provides two different information channels in the electromagnetic wave emission and reflection regions. The proposed encoded meta-atom consists of a cross-shaped patch structure and a slot formed by two merged rectangles on both sides of the dielectric substrate. By adjusting the geometric parameters of the patch and slot structures, 1-bit phase and amplitude modulation can be achieved for different incident wave polarizations~\cite{zhu2021full}. Subsequently, they also proposed a flexible dual-broadband polarization-insensitive coded metasurface, capable of handling electromagnetic scattering in the microwave band. The 1-bit coded units `0' and `1' consist of meta-atoms with different orientation PB phases. The encoded surfaces conform to metal column surfaces of different curvature radii while maintaining good diffuse scattering properties in the double broadband band. The scattering properties of the conformal metasurface become progressively better as the radius of curvature decreases for a certain size of the flexible surface~\cite{fu2021combining}.
 Yuan et al. explored the correlation between the geometric entropy of the electromagnetic metasurface map and the scattering entropy by incorporating the power spectrum into the coding sequence generation. They found that the two-dimensional broadband power spectrum with a central depression has the highest entropy value in the scattering mode. To verify that, the authors designed, fabricated, and experimentally measured a metasurface sample~\cite{yuan2021impact}.

Sawant et al. designed and tested a perforated planar dielectric plate that can convert a Gaussian beam into a 2nd-order orbital angular momentum (OAM) beam with better amplitude uniformity in the millimeter-wave frequency range. They also fabricated and tested a transmissive metasurface design to obtain the best amplitude uniformity in OAM beam generation~\cite{sawant2022amplitude}. Al-Nuaimi et al. proposed a 1-bit metasurface design for wideband radar cross-section reduction from $60$ GHz to $120$ GHz. Their design does not require complex optimization algorithms or time-consuming simulations to achieve an optimized phase distribution map~\cite{al2022design}.

Chen et al. proposed a symmetrical circular quasi-Minkowski closed-loop shape for a transparent broadband scattering meatasurface~\cite{chen2022transparent}. They utilized an ultrathin linewidth metal ring for the top and bottom layers of the unit to enhance its optical transmittance. The measurement results demonstrated that the designed surface achieved two phase variations of `0' and `1' in the broadband range of 10.5 to 19.5 GHz with a phase difference of $180^{\circ}\pm 30^{\circ}$.

Ghosh et al. investigated a graphene-based metasurface for broadband linear to circular polarization conversion in the terahertz range (2.25-6 THz). The proposed structure consists of a three-layer configuration, with the meta-atomic layer comprising a patterned graphene layer on a gold-base silica substrate. The polarization converter operates in reflection and produces both right and left-hand circular polarizations in the operating spectrum, as confirmed by a circuit model~\cite{ghosh2022terahertz}.

Guirado et al. proposed an effective anisotropic uniaxial model for polymer network liquid crystal (PNLC) mixtures, with the aim of enhancing response time. This model operates within the supercell unit of a metasurface at 100 GHz, achieving a 50-fold improvement in response time at the cost of a threefold reduction in tunability~\cite {guirado2022mm}.

This section describes the history of the development of fixed-structure metasurfaces, which during this period had limited ability to change the electromagnetic field due to the fixed electromagnetic structure. However, it has also been proposed to change the electromagnetic characteristics by modifying the geometric parameters of the reflection unit, such as shape, size, spacing, etc.  Although this kind of metasurface is not enough to improve the wireless communication environment compared with RIS, it also shows the development prospect of metasurface and promotes the development of RIS.

\subsection*{(2) Unit with adjustable structure}
General `non-reconfigurable metamaterials and metasurfaces' have fixed topological geometry. Once the physical structure of metamaterials and metasurfaces is processed and formed, it can only achieve a certain function and cannot be changed in real time according to the demand. However, in some practical application scenarios, metasurface is usually required to switch between different functions in real time.
RIS is a kind of metasurface that utilizes the reconfigurable characteristics of metasurface units to realize the regulation function of electromagnetic wave signals in a real-time programmable/controllable way.
At present, according to the design structure classification, the RIS unit structure mainly includes microstrip patch type and slot-coupling type, in addition to other metasurfaces with refraction and other functions.

\subsection*{a. Microstrip patch type unit}
 The microstrip patch antenna is formed by etching the metal radiation plate on the dielectric substrate with a thin metal layer as the ground on the back. It has two main feeding modes: microstrip line and coaxial line. The microstrip antenna excites the radiation field between the metal patch and the metal ground and radiates outward through the gap between the periphery of the patch and the ground. At present, most RIS units are based on microstrip patch antennae.

\begin{figure}[htbp]
\centering
 \subfloat{\includegraphics[width=.8\columnwidth]{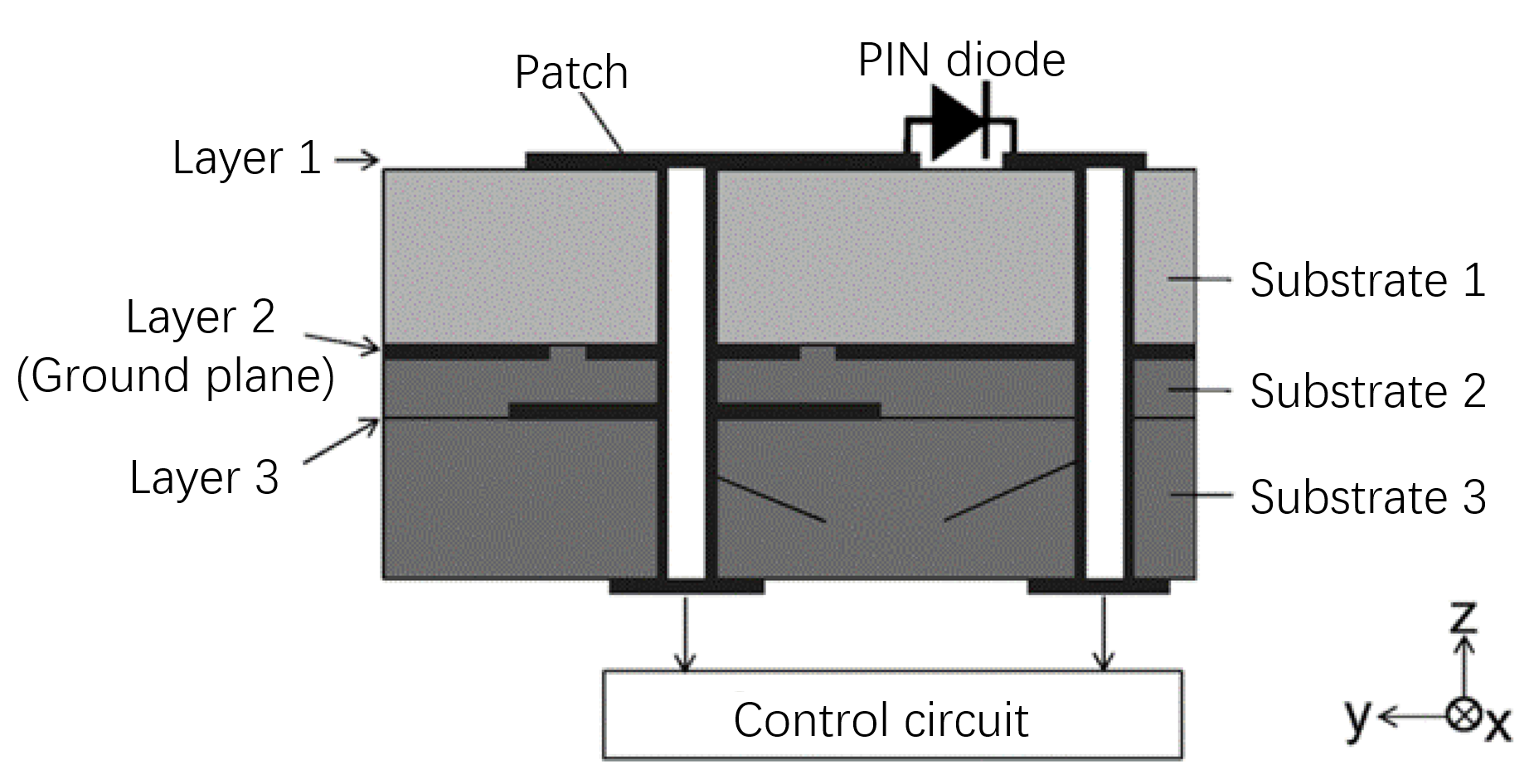}
 \label{F1-4-1}
 }
 \\
 \subfloat{ 
 \includegraphics[width=.9\columnwidth]{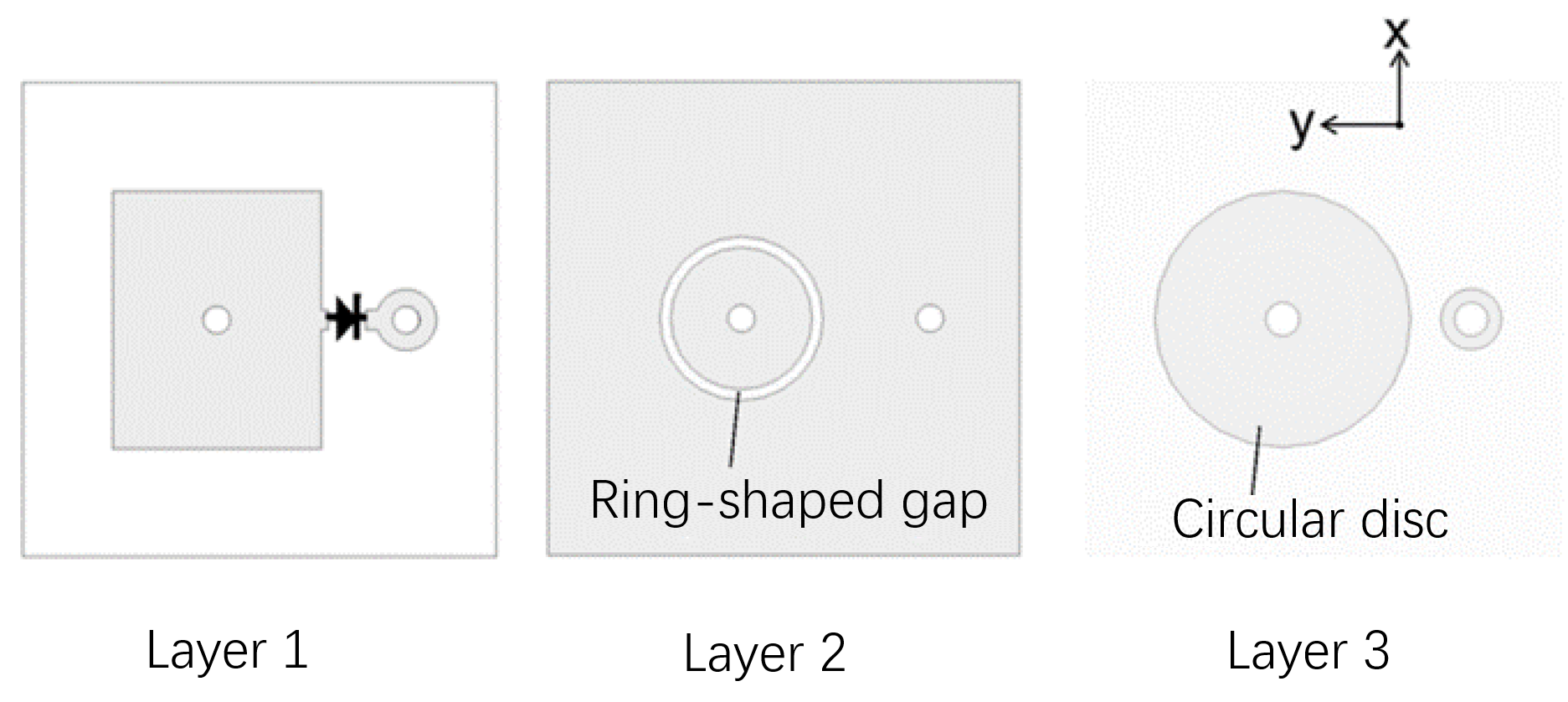}
 \label{F1-4-2} 
 }
 \caption{Digital phase-shifting structure based on PIN diodes~\cite{kamoda201160}}
 \label{F1-4}
\end{figure}

\begin{figure}[htbp]
\centerline{\includegraphics[width=.9\columnwidth]{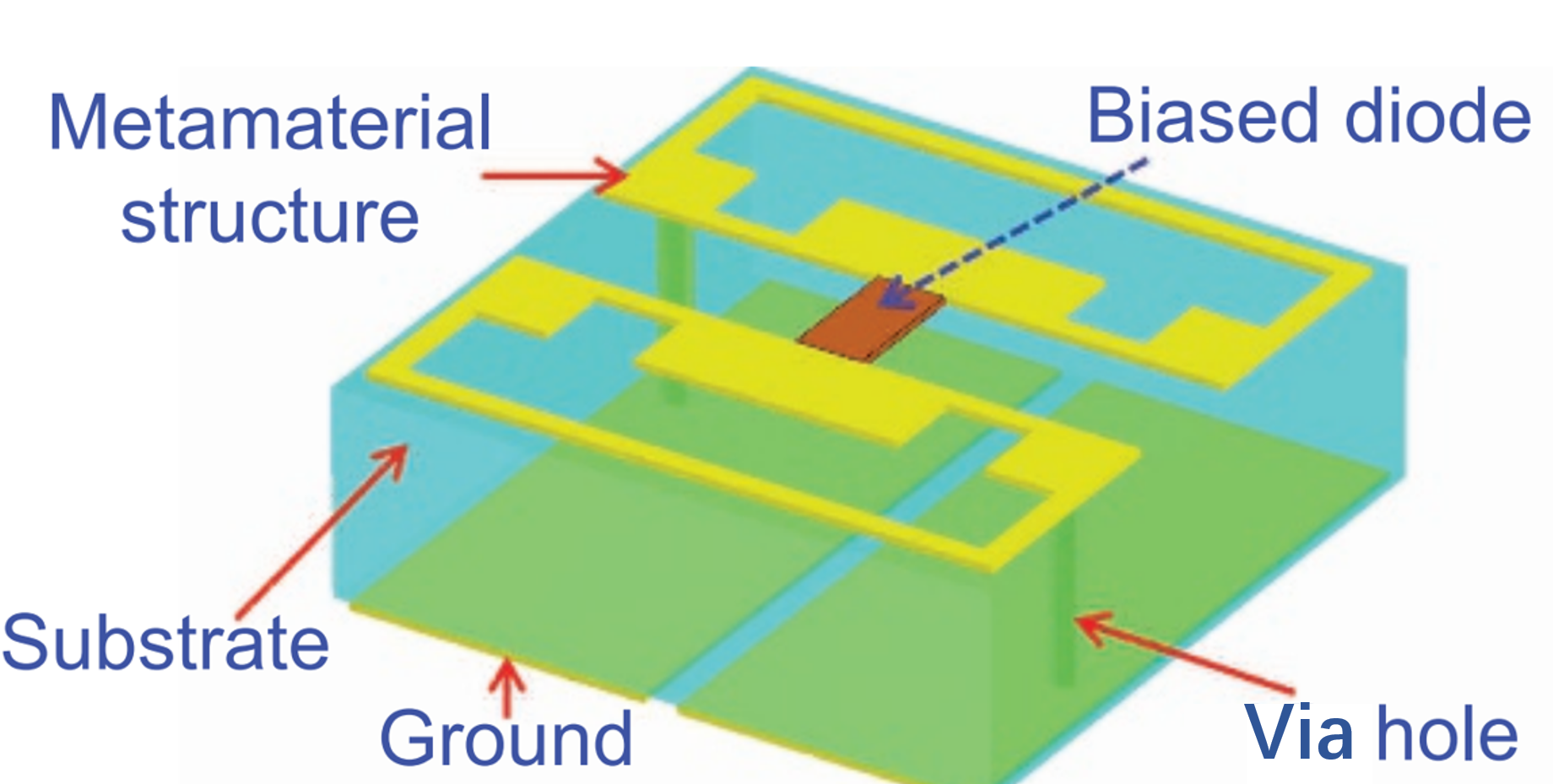}}
\caption{Programmable metamaterial unit~\cite{cui2014coding}.}
\label{F1-5}
\end{figure}

Kamoda et al. proposed to use PIN diode to realize digital phase shift~\cite{kamoda201160}. The basic model of the proposed unit is shown in Fig.~\ref{F1-4}. A rectangular microstrip patch is attached to a short stub equipped with a PIN diode. The diode acts as an RF switch so that the short-circuit stub input impedance response changes with the on/off state of the diode. When the PIN diode is switched, the reflection phase difference is determined as $180^{\circ}$ by changing the size of the stub or patch, and a 1-bit unit phase shifter is implemented in the unit. A metal via hole is added in the middle of the patch and connected to the control circuit to control the PIN diode. The annular gap (layer 2) on the ground surface and the metal disc on the third layer act to suppress the RF signal. The reflection coefficient of simulation and test shows that for 60.25 GHz incident electromagnetic wave, after RIS reflection of two states of PIN diode switch, the phase difference is approximately $180^{\circ}$. The slight difference between simulation and test is mainly due to engineering errors. The group finally produced a $160\times 160$ array with a total of 25600 units, which verified the reconstruction ability of the antenna. The measured radiation pattern under different beam directions and antenna gain was in good agreement with the calculated results. In addition, the near-field focusing ability of the proposed algorithm is verified by the near-field object imaging experiment, and the response time is much less than the requirement of the system.

Cui et al. proposed the hardware structure of programmable metamaterials for the first time\cite{cui2014coding}. As shown in Fig.~\ref{F1-5}, two planar symmetrical metal ring structures are printed on the F4B dielectric substrate and connected by PIN diodes, and each metal ring is connected to two separate ground plates through two vias for applying a direct current (DC) bias voltage. The results of the reflection coefficients obtained from the full-wave simulation using CST Microwave Studio show that the phase difference is $180^{\circ}$ at 8.2 GHz for different states of the diode, which are coded as `0' and `1', respectively. Digital control of digital metamaterials has been realized using FPGAs, which enable the manipulation of electromagnetic waves by programming different coding sequences. 
This lays the foundation for the research of the practical RIS system.

The metasurface structure proposed by Yang et al.\cite{yang2016programmable}. The structure consists of a rectangular patch and a metal ground floor. diodes connect one side of the rectangular patch to the ground floor through a metal via. The DC bias line is selected near the position of approximately zero electric fields, and a quarter-wavelength microstrip line and sector line are introduced to isolate DC and RF signals. The reflection characteristics obtained by the electromagnetic simulation software Ansys HFSS show that most of the energy is reflected under the two polarizations. When the diodes are ON and OFF, the phase difference at 11.1 GHz is $180^{\circ}$ at X-polarized incidence.

To enhance the phase dispersion capability of the reconfigurable intelligent surface (RIS) and achieve finer phase control. Huang et al proposed a 2-bit RIS unit~\cite{huang2017dynamical}. The unit of the RIS consists of a trapezoidal patch and two rectangular patches, with two PIN diodes positioned between the gaps. The metal via in the middle is connected to the metal ground floor, and the metal vias on both sides are connected to the bias line. The control circuit at the bottom of the dielectric plate provides a DC signal. The reflection coefficient magnitude and phase of the proposed design are obtained through electromagnetic simulations using the commercial software CST. The results indicated that the reflection coefficient magnitude was above 0.85, Additionally, when the PIN diodes are in four different states, namely `00' (state 1), `01' (state 2), `10' (state 3), `11' (state 4), the phase difference between adjacent states is approximately $90^{\circ}$ near 7.25 GHz. This satisfies the requirements for a 2-bit RIS unit.

The metasurface structure designed by Zhang et al. consists of multiple units, each of which consists of a rectangular metal patch and a metal stub printed on an F4B dielectric substrate~\cite {zhang2018space}.
A PIN diode is used to connect the metal patch to the floor through a metal stub, and each column (consisting of 8 units) is connected by a bias line with a width of 0.2 mm and shares a control voltage. Full-wave simulation of the metasurface unit is established using CST. The full wave simulation of the metasurface unit is established using CST. The phase and amplitude results of the reflection coefficients in the ON and OFF states of the PIN diode show that the phase difference between the two states is $180^{\circ}$ near 10 GHz.

Zhang et al. also proposed a new metasurface
structure~\cite{zhang2019breaking}, in which each column of the metasurface is composed of 8 units and shares the same control voltage. Each unit consists of a hexagonal metal patch and two bias lines, printed on the F4B with ground. The two PIN diodes connect the two bias lines and the hexagonal patch so that the two PIN diodes correspond to four coding states respectively in the OFF-OFF, ON-OFF, OFF-ON and ON-ON states. CST is used for full-wave simulation to calculate its reflection coefficient. The results show that the phase difference of adjacent 2-bit coded states is about $90^{\circ}$ at the frequency of 9.5 GHz, and the corresponding amplitude is more than 0.79.

Li et al. proposed a 2-bit intelligent metasurface unit structure~\cite{li2019machine}, each of which can be adjusted independently. The unit consists of three square metal patches of sub-wavelength scale printed on a dielectric substrate (Roger 3010). Each PIN diode has two operating states, controlled by bias voltage. Under the irradiation of an X-polarized plane wave, the unit supports four different phase responses, which are respectively represented as `00' (state 0), `01' (state 1), `10' (state 2) and `11' (state 3). By controlling the ON/OFF states of the three, the PIN diode can be determined in a suitable combination. The corresponding four digital phase differences are $0, \pi/2, \pi$ and $3\pi/2$. To isolate the DC feed port from the microwave signal, three $30$ nH inductors are used. In this unit design, the phase difference between two adjacent states is ($90^{\circ}-15^{\circ}$, $90^{\circ}+15^{\circ}$) in the range of frequency around 3.2 GHz. Thus, this unit can be viewed as a 2-bit digital coded metasurface unit at all digital states around 3.2 GHz and can achieve reflectance of more than 85\%. 

The RIS unit designed by Li et al. in 2019 consists of two layers of dielectric substrate, a square patch on the top is responsible for reflecting electromagnetic waves, and a PIN diode is integrated into the patch, which is connected to the ground plane through a metal through a via~\cite{li2019intelligent}. The chip inductance L = 33 nH is used to suppress the coupling between AC and ground. When the PIN diode is switched from ON (OFF) to OFF (ON), a $180^{\circ}$ phase difference is observed in the frequency range of 2.41 GHz to 2.48 GHz. The phase change can be accomplished by switching the external DC voltage applied to the PIN diode from 3.3 V to 0 V.

\begin{figure*}[t!]
\centering
 \subfloat {\centering
 {\includegraphics[width=.5\linewidth]{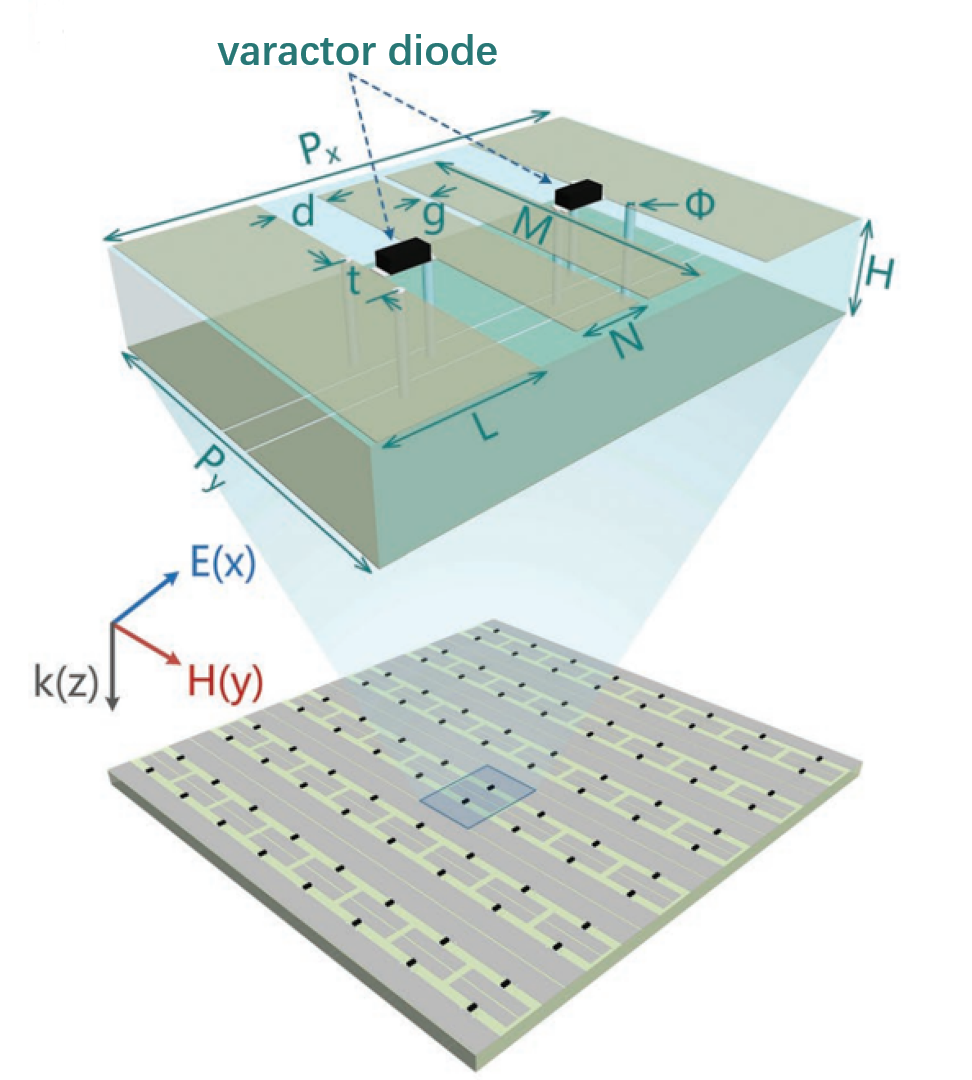}}
 \label{F1-6-1}}
 \subfloat{ \centering
 {\includegraphics[width=.45\linewidth]{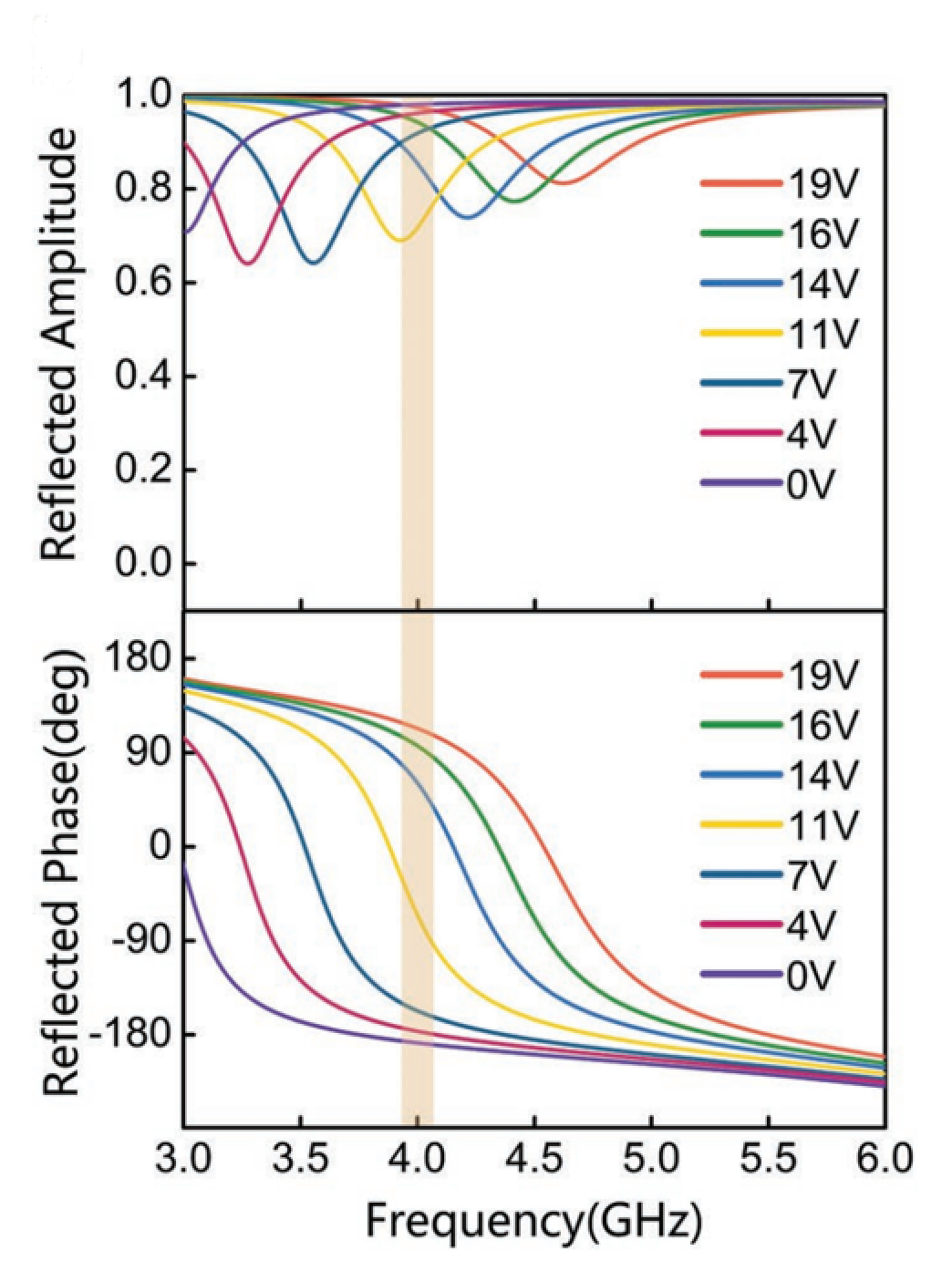}}
 \label{F1-6-2}}
 \caption{Unit design based on variable diodes~\cite{dai2019wireless}.}
 \label{F1-6}
\end{figure*}

Dai et al. designed a RIS unit structure based on varactor diodes~\cite{dai2019wireless}, as shown in Fig.~\ref{F1-6}, consisting of two pairs of rectangular patches with a slotted copper ground floor on the back of the dielectric. Each pair of metal patches is connected by a varactor diode, both ends of which are connected to the ground plate through a metal hole, and a bias voltage is applied to both ends. As can be seen from Fig.\ref{F1-6}, when the bias voltage varies from 0 to 19 V, the phase difference shows a large phase range ($\approx 300^{\circ}$). 

Lin et al. proposed a unit structure consisting of three metal layers and two dielectric layers. The PIN diode is integrated into the unit structure of the top metal layer, and the middle metal layer acts as the reflective surface and also acts as the grounding plane of the entire circuit~\cite{lin2021single}. A fan-shaped structure is designed in the bottom metal layer to suppress the high-frequency signal from the DC signal. The PIN diode is driven by two metal via holes, one metal cylinder is connected to the intermediate metal layer, and the other penetrates the intermediate metal layer to load the DC voltage. By applying different voltages, the PIN diode will be turned on or off, affecting the resonant frequency of the unit and resulting in a phase difference. The unit can achieve a $180^{\circ}$ phase difference and approximately equal reflection amplitude.

The metasurface unit proposed by Pei et al is similar to that in reference~\cite{kamoda201160}. The operating frequency is moved to 5.8 GHz by changing the unit size~\cite{pei2021ris}.

\begin{figure}[htbp]
\centering
 \subfloat {
 \centerline{\includegraphics[width=.8\columnwidth]{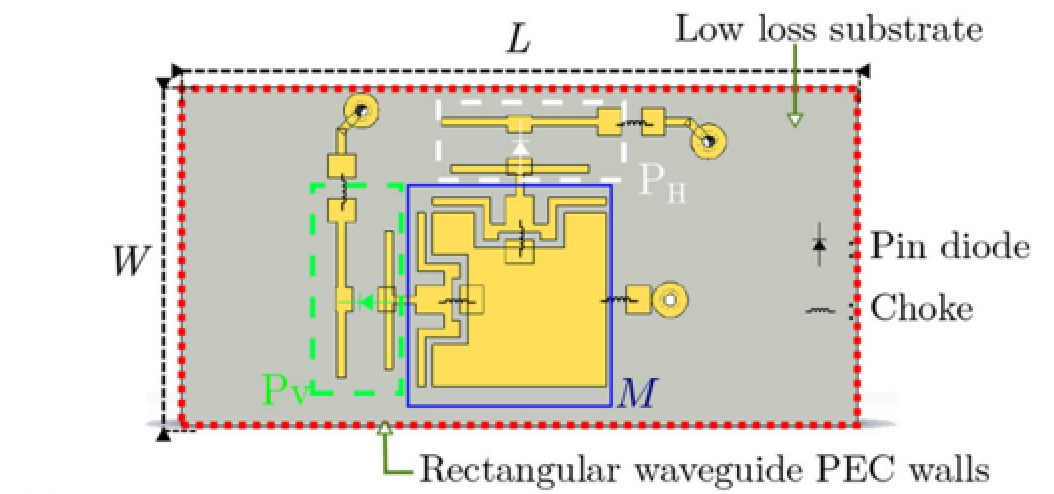}}}
 \\
 \subfloat{ 
 \centerline{\includegraphics[width=.8\columnwidth]{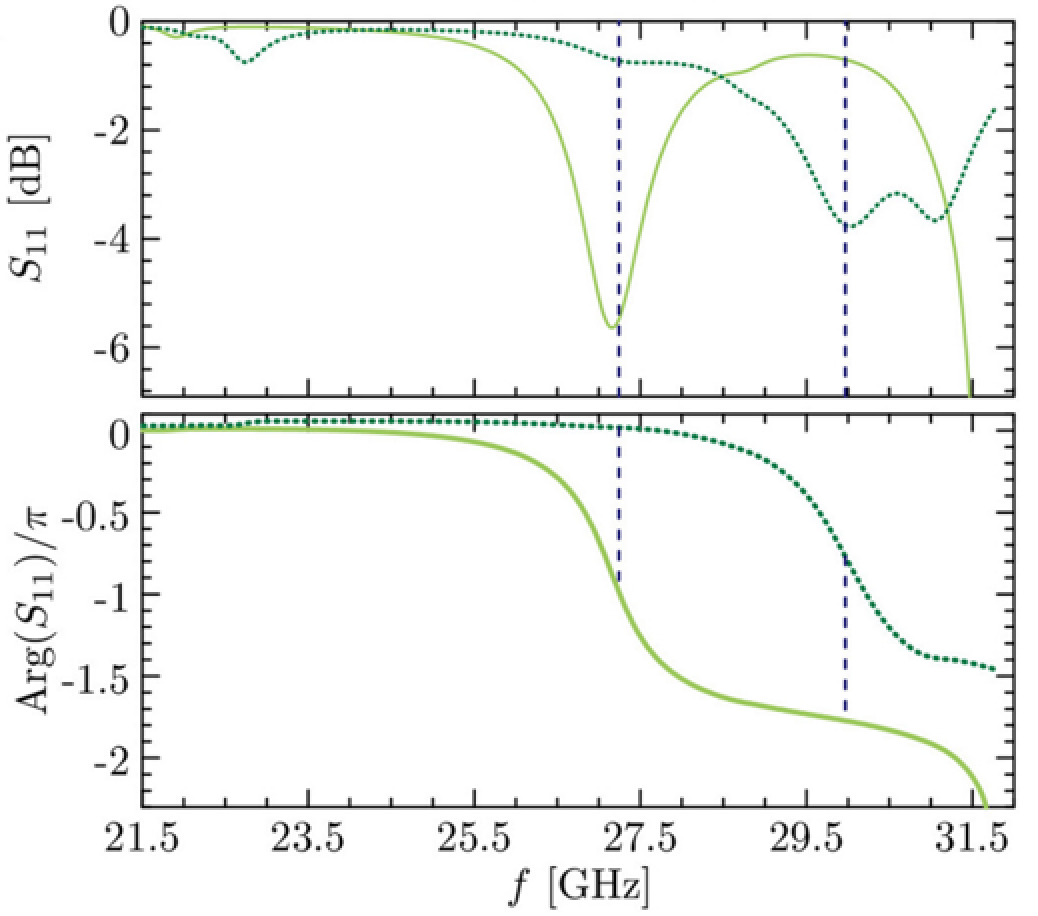}}}
\caption{Dual-polarized unit unit structure and reflection coefficient simulation~\cite{gros2021reconfigurable}.}
 \label{F1-7}
\end{figure}

Gros et al. proposed a double-polarized metasurface unit with a parasitic resonator added to the design of the unit in order to provide a $180^{\circ}$ phase shift (PH and PV in Fig.~\ref{F1-7} correspond to the horizontal and vertical polarization of the incident wave electric field vector, respectively)~\cite{gros2021reconfigurable}. The resonant frequency of the parasitic resonator is similar to that of the patch. When the coupling resonance repulsion and resonance intershift occur, the strong coupling and corresponding anti-cross behavior between the two resonators appear. By changing the electric length of the parasitic resonator, the resonant frequency of the parasitic resonator can be changed, and the phase difference can be caused. The unit structure is isolated from the RF portion of the unit by lumped inductors. Fig.~\ref{F1-7} shows the simulated reflection coefficient of the unit, with a phase difference of approximately $180^{\circ}$ in the frequency range 27.5 GHz to 29.5 GHz.

\begin{figure}[htbp]
\centering
 \subfloat {
 \centerline{\includegraphics[width=.8\columnwidth]{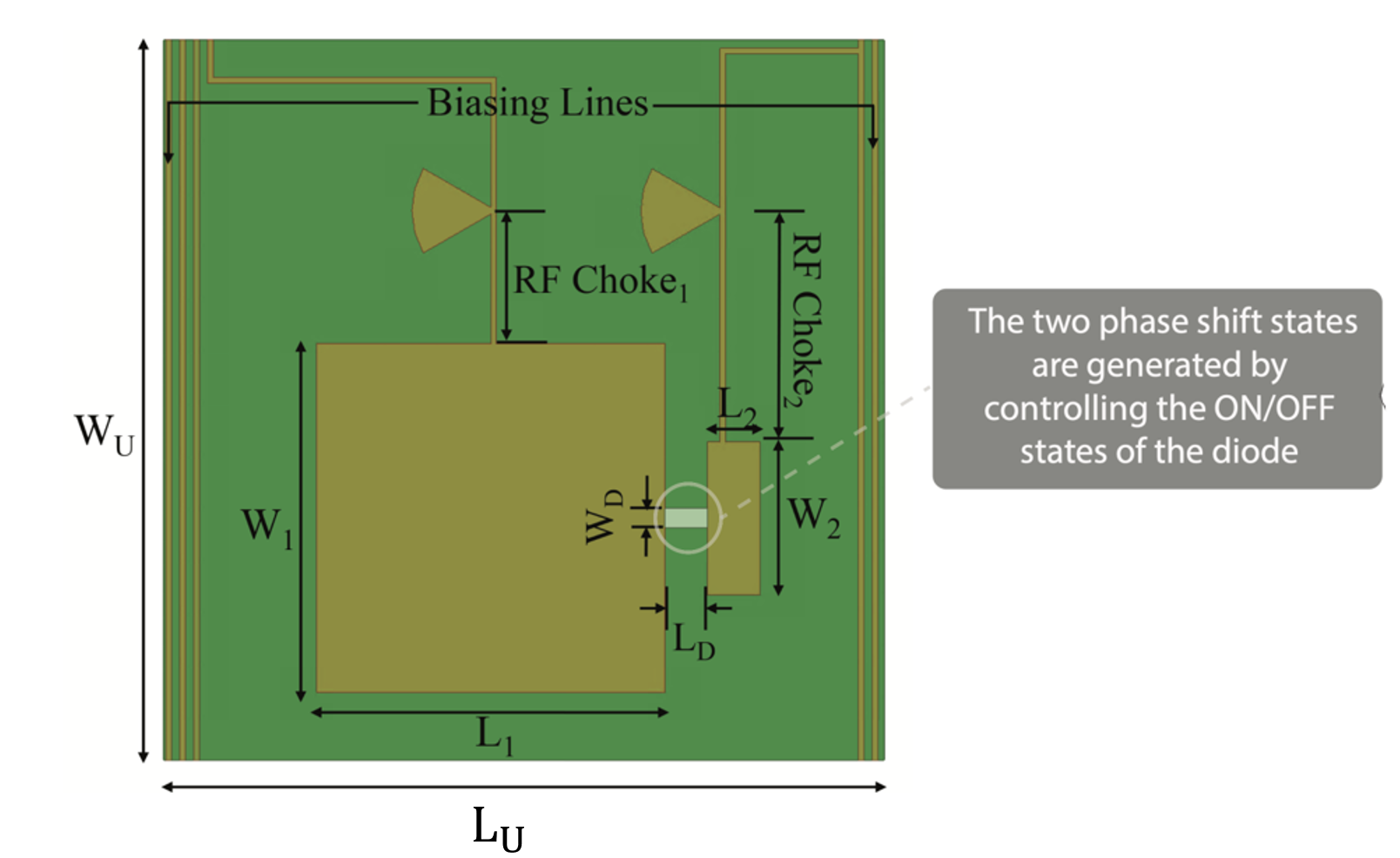}}
}
\\
 \subfloat { 
 \centerline{\includegraphics[width=.8\columnwidth]{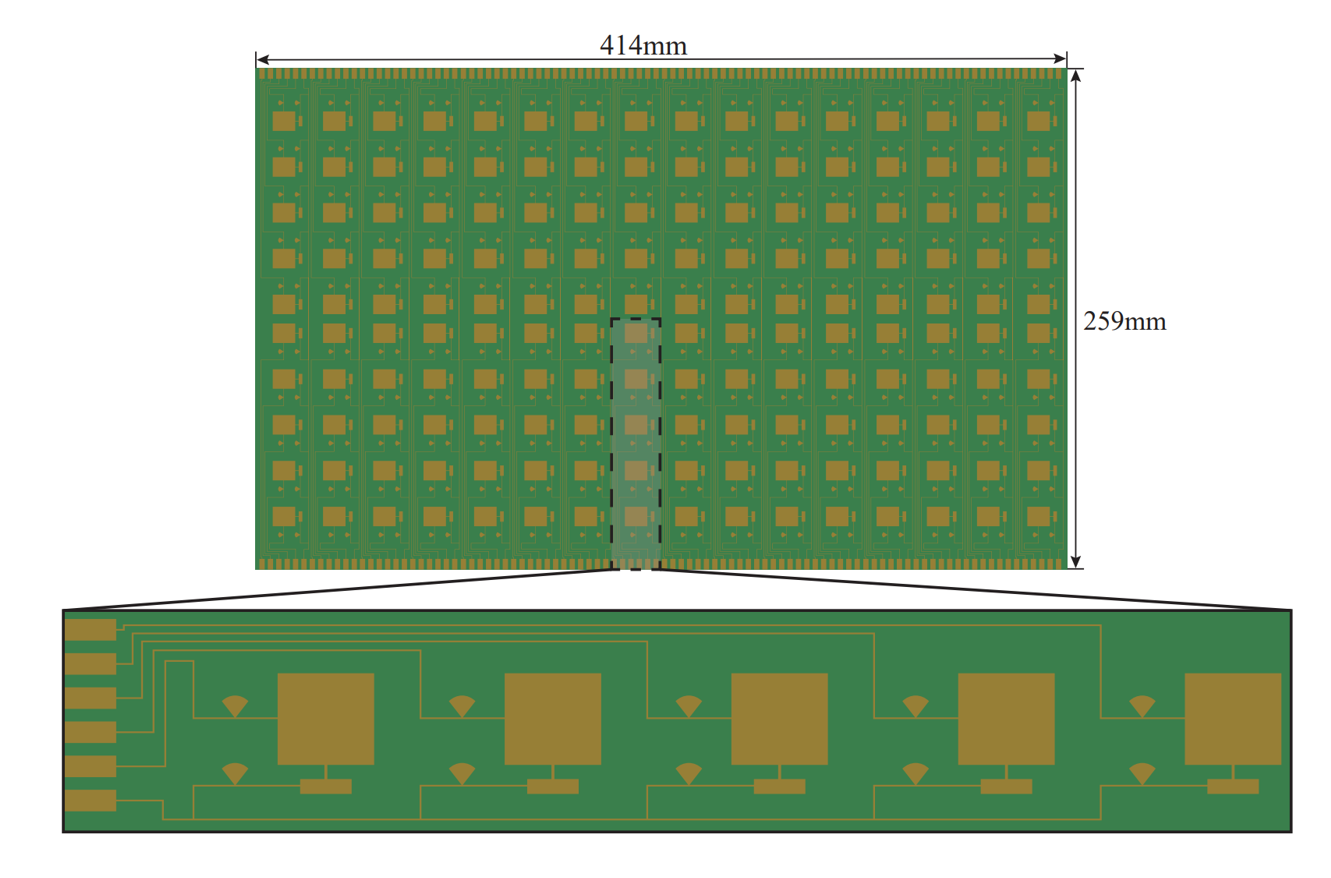}}
}
\caption{Single-layer topological metasurface structure~\cite{trichopoulos2022design}.}
 \label{F1-8}
\end{figure}

Trichopoulos et al. proposed a new RIS unit design~\cite{trichopoulos2022design}. The goal of this unit is to design a single-layer topology (other than the ground plane) that does not require the use of vertical components (such as through holes). Thus, the metasurface unit consists of a ground a dielectric substrate, and a top metal layer. The top layer consists of a rectangular patch and a parasitic patch connected by a PIN diode. To isolate the RF signal from the bias lines, a radial stub is used in each bias line. Bias lines are routed in groups of 5 units to ensure minimal wiring complexity, as shown in the full array topology in Fig.~\ref{F1-8}.

\begin{figure}[htbp]
\centering
 \subfloat 
 {
 \centerline{\includegraphics[width=.7\columnwidth]{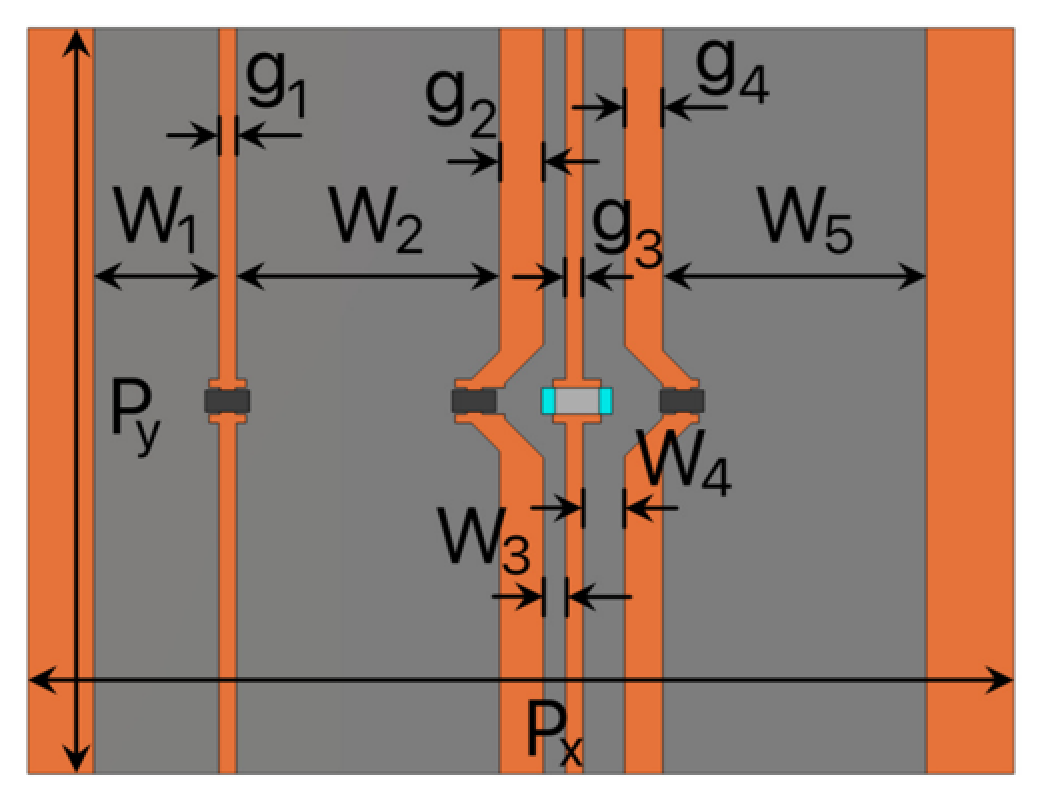}}
 \label{F1-9-1}}
 \\
 \subfloat 
{ 
 {\includegraphics[width=.45\columnwidth]{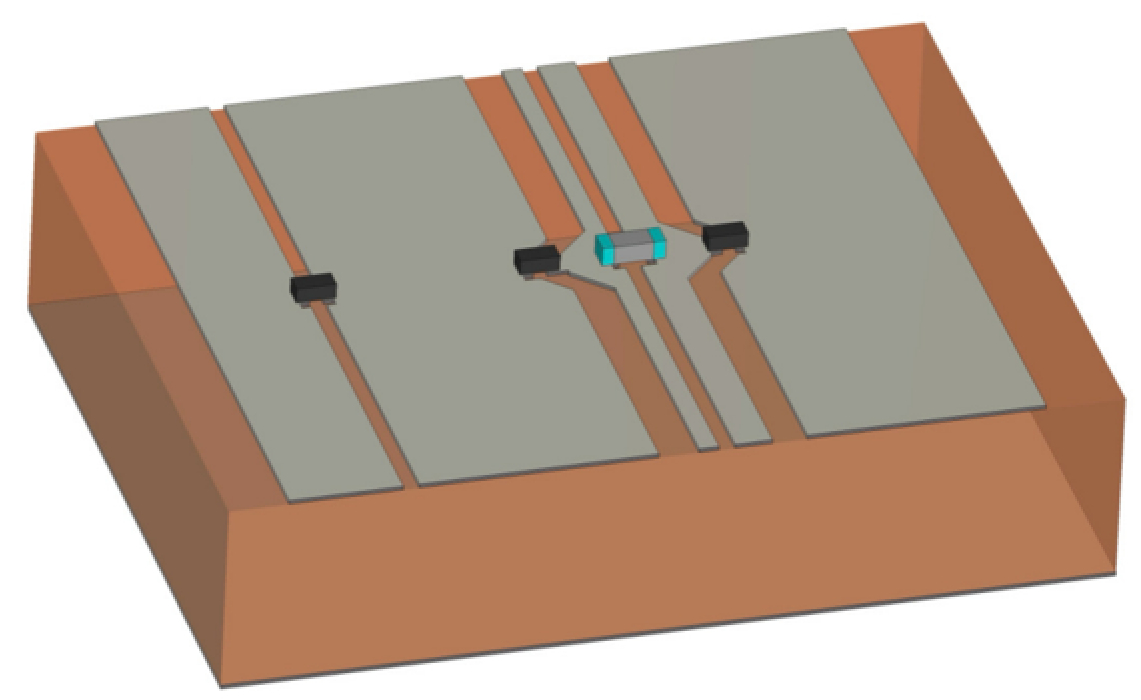}}
 \label{F1-9-2}}
  \subfloat 
{ 
 {\includegraphics[width=.45\columnwidth]{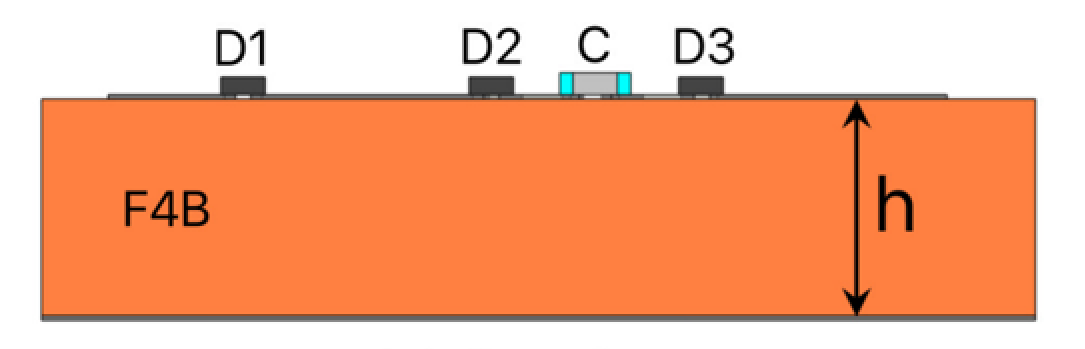}}
 \label{F1-9-3}}
\caption{Multi-bit unit unit design~\cite{rains2022high}.}
 \label{F1-9}
\end{figure}

In~\cite{rains2022high}, a multi-bit column-driven plane is designed as shown in Fig.~\ref{F1-9}. Each unit consists of 5 patches connected by 3 PIN diodes and a capacitor. The unit applies DC voltage to 3 of the patches by the side of the column, and the remaining 2 patches are connected to the ground. Three diodes in different states can achieve 3-bit.

\begin{figure}
\centerline{\includegraphics[width=.9\columnwidth]{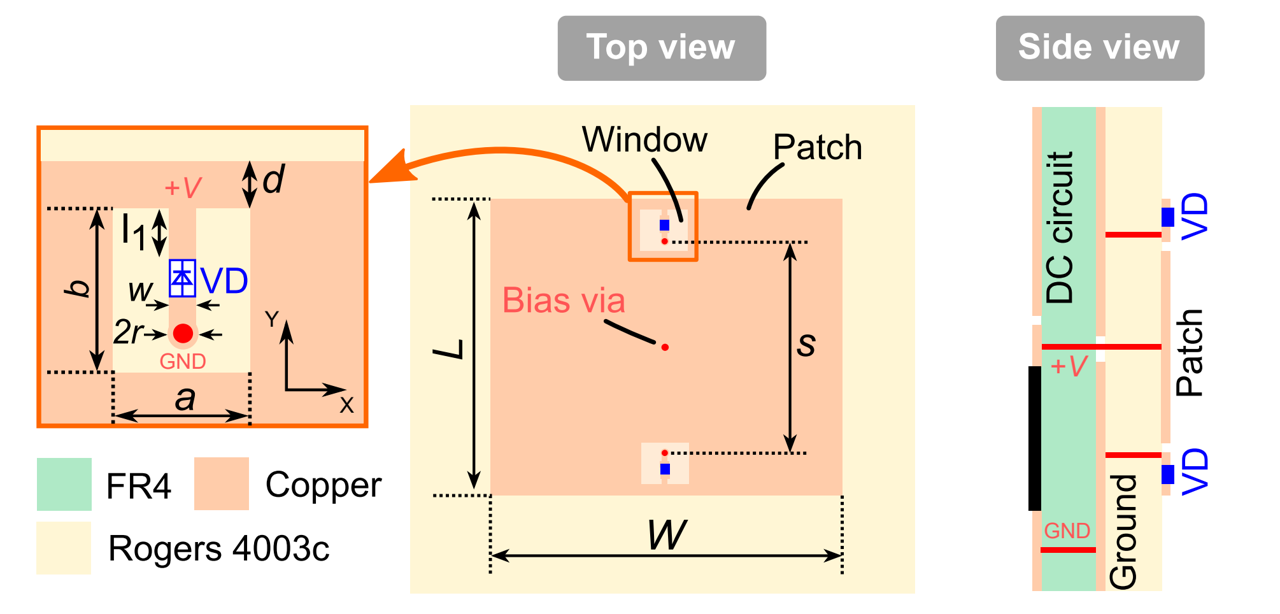}}
\caption{Unit structure based on varactor diode~\cite{sayanskiy20222d}.}
\label{F1-10}
\end{figure}

The metasurface unit designed by Sayanskiy et al. is shown in Fig.~\ref{F1-10}. The unit consists of a rectangular patch on which two varactor diodes are integrated, and the center of the patch is connected with a direct current circuit through a through hole~\cite{sayanskiy20222d}. The resonant frequency of the patch is determined by its length L and the capacitance of the varactor diode connected to the ground plane at ${\rm s}/2$ from the center of the patch. In order to connect two diodes between the patch and the ground plane, the unit made two rectangular windows of size $a \times b$ on the patch and added two small contact plates to connect to the ground through two additional through-holes. Each diode is soldered between its window edge and the corresponding contact plate. In the unit design, the ground plane and the contact piece are maintained at zero DC potential, and the patch is under a bias voltage formed by the DC circuit. As can be seen from the amplitude and phase of the reflection coefficients in Fig.~\ref{F1-10}, the selected parameters do provide 0 and 1 states, corresponding to U = 0 V and U = 3.2 V, with a difference of about  $180^{\circ}$ in the reflection phase.

Chen et al. proposed a millimeter wave band reflection unit~\cite{chen2022accurate}, and each unit was composed of two rectangular patches. A PIN diode was loaded between the two patches. The array unit operates at 27 GHz, and the phase difference of incident electromagnetic waves generated by two switching states was $180^{\circ}$.

The unit structure proposed by Fara et al. is a metal patch etched on a dielectric substrate, which is separated by annular grooves into two conducting regions, and the inner and outer regions are interconnected by four varactor diodes. By adjusting the capacitance of the varactor, the desired reflectance coefficient is provided~\cite{fara2022prototype}.

In summary, the microstrip patch-type structure is reconfigurable by adding some branches and corresponding electromagnetic switches to the first layer of the multi-layer structure and changing the current distribution of the patch through the state change of the electromagnetic switch.

\subsection*{b. Slot-coupling type unit}
\begin{figure}[htbp]
\centerline{\includegraphics[width=.9\columnwidth]{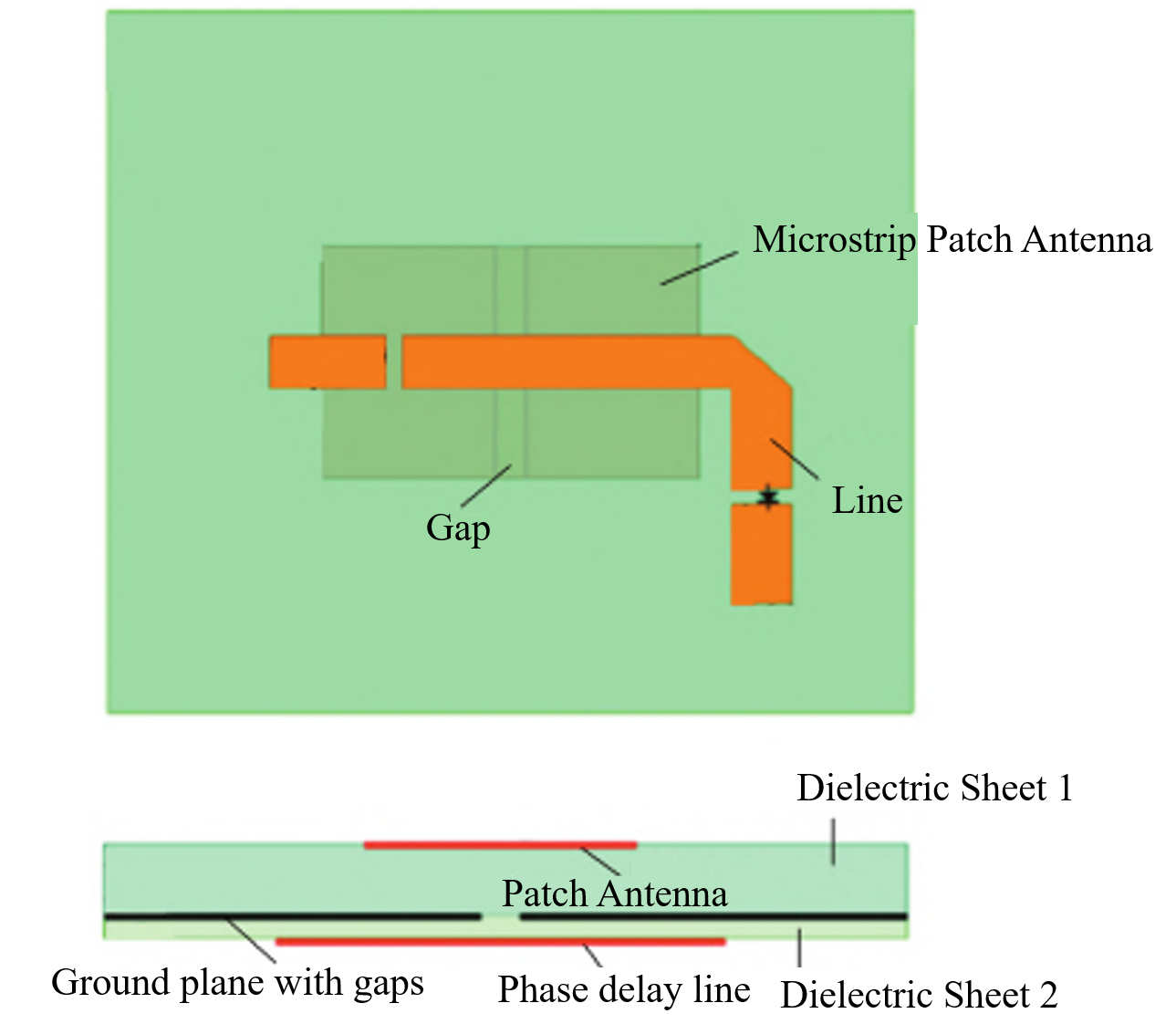}}
\caption{Slot coupled unit structure~\cite{xue2018research}.}
\label{F1-11}
\end{figure}

\begin{figure}[t!]
    \centering

    \subfloat[RIS hierarchy diagram]{
		\includegraphics[width=.8\columnwidth]{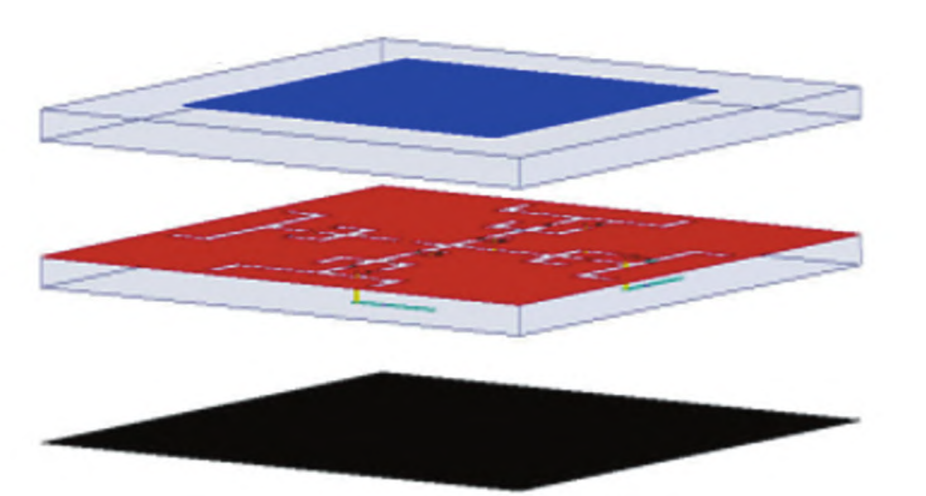}}
    \\
    \subfloat[RIS unit structure]{
		\includegraphics[width=.75\columnwidth]{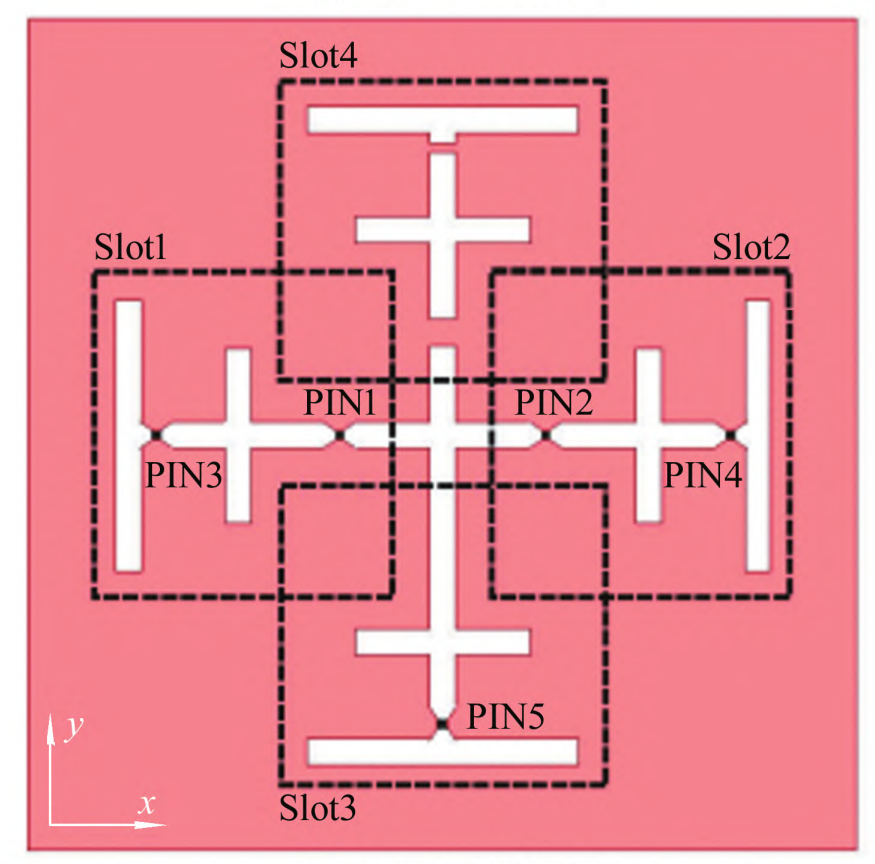}}
    \caption{Symmetrical 2-bit unit structure~\cite{dai2020reconfigurable}.}
    \label{F1-12}
\end{figure}

\begin{figure}[htbp]
\centerline{\includegraphics[width=1\columnwidth]{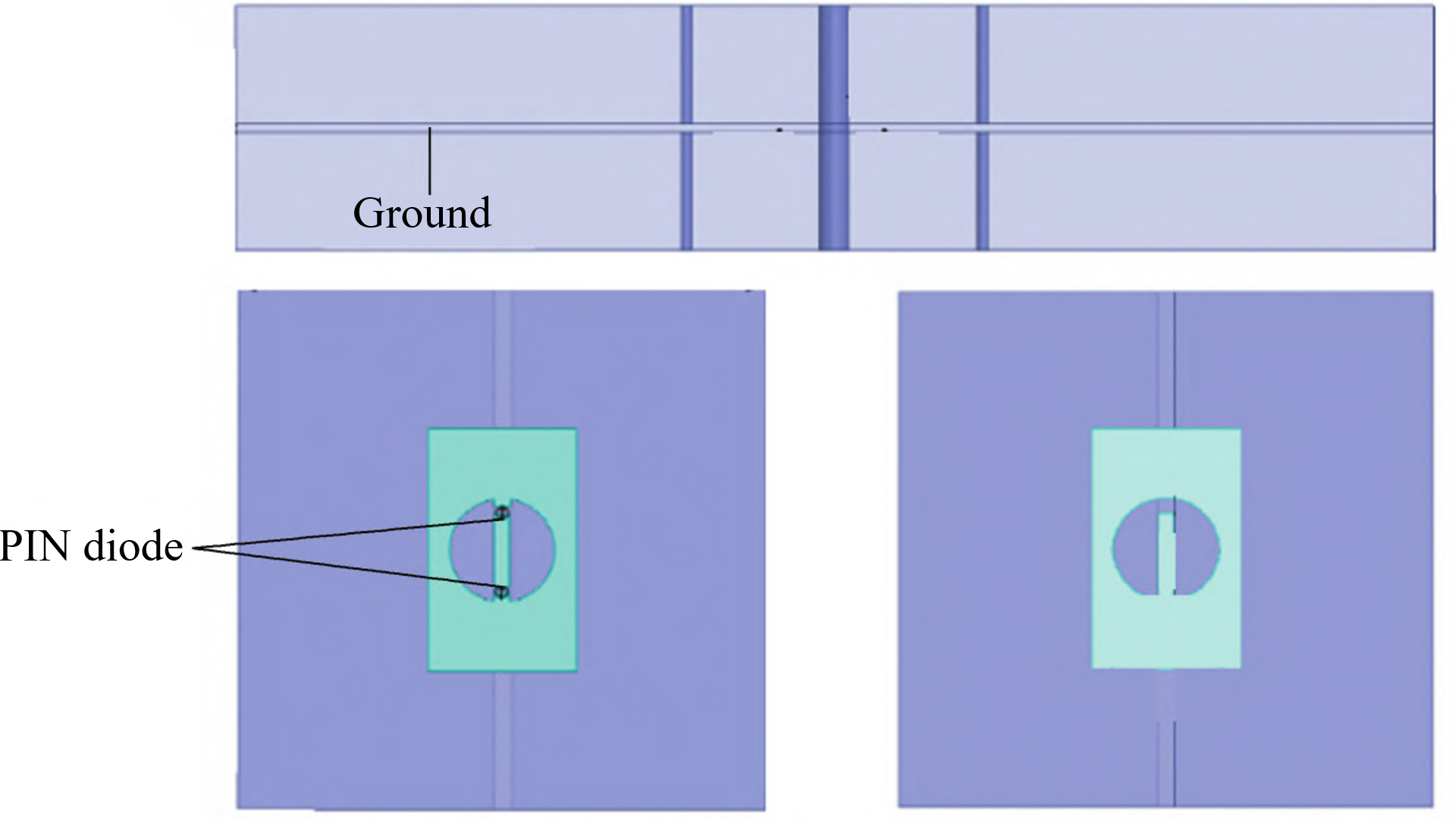}}
\caption{Unit structure of transmissive RIS~\cite{qiao2019design}.}
\label{F1-13}
\end{figure}

\begin{figure}
\centerline{\includegraphics[width=.8\columnwidth]{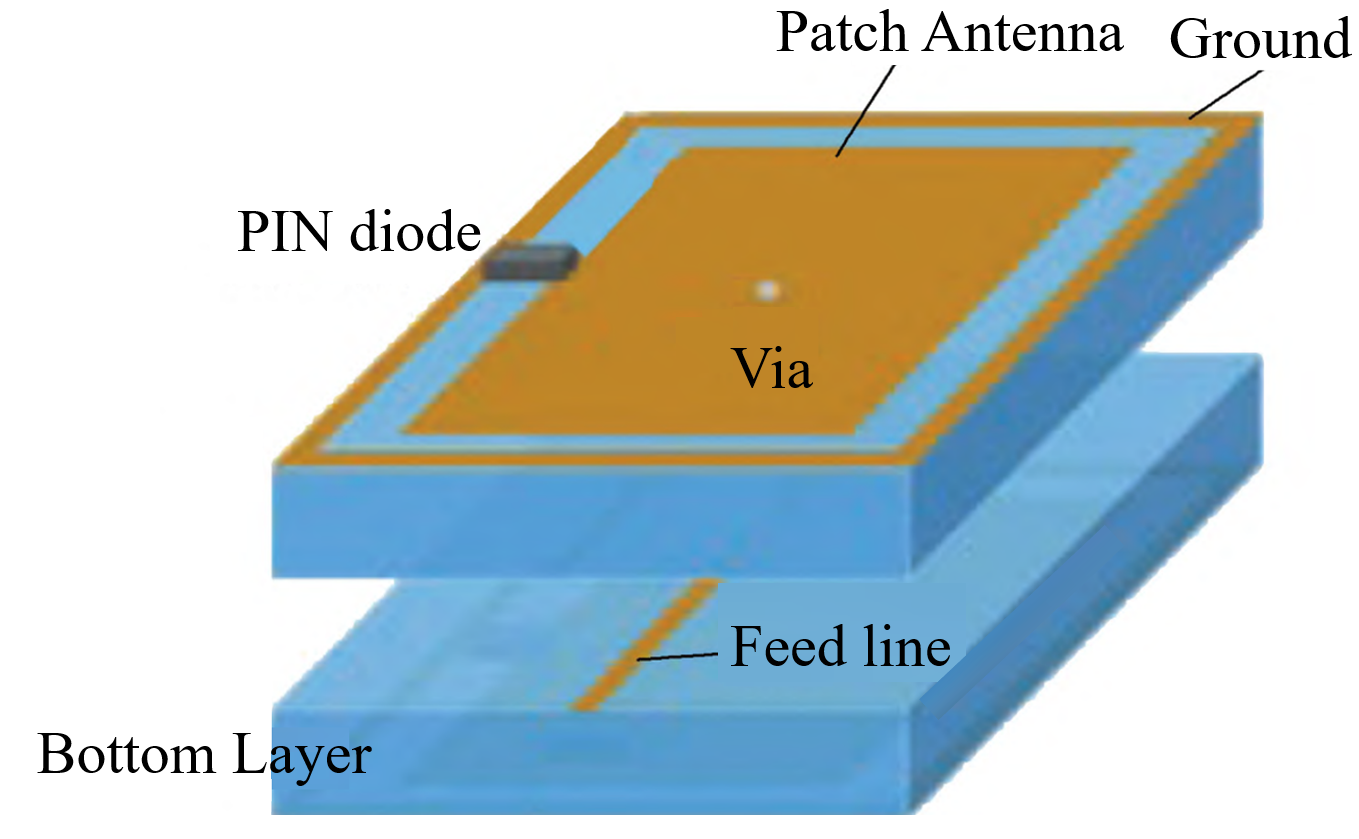}}
\caption{The structure design of a unit capable of simultaneous reflection and transmission of electromagnetic waves~\cite{zhang2022intelligent}.}
\label{F1-14}
\end{figure}

\begin{figure}[t!]
    \centering

    \subfloat[Unit structure diagram]{
		\includegraphics[width=.5\columnwidth]{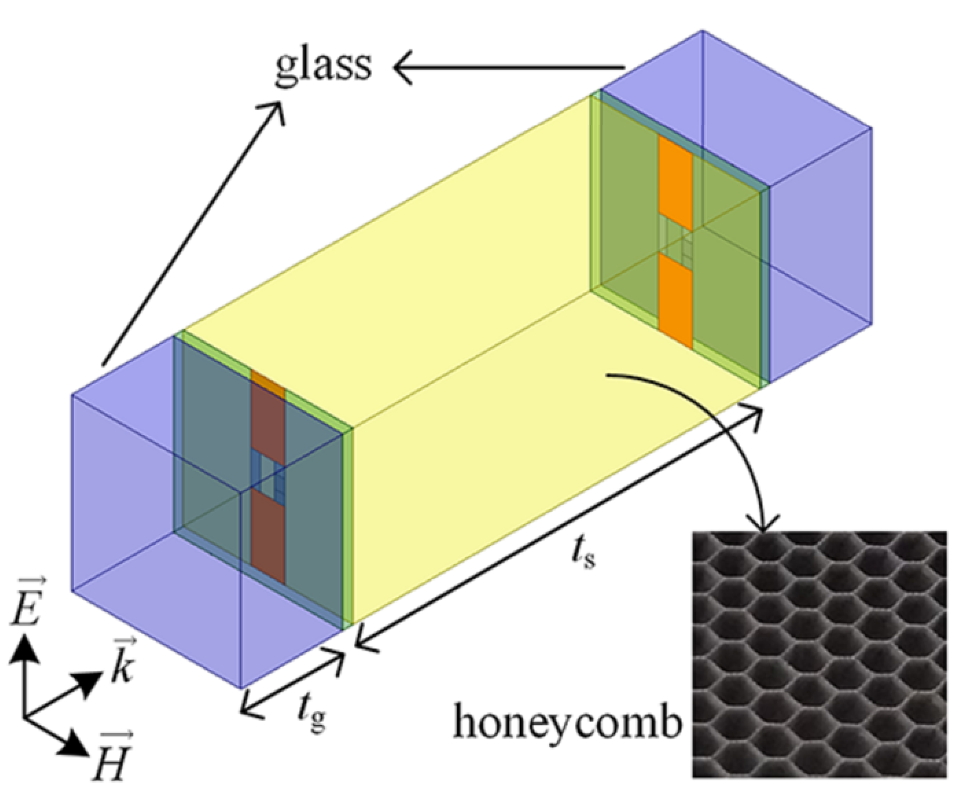}}
    \subfloat[Single layer structure diagram]{
		\includegraphics[width=.45\columnwidth]{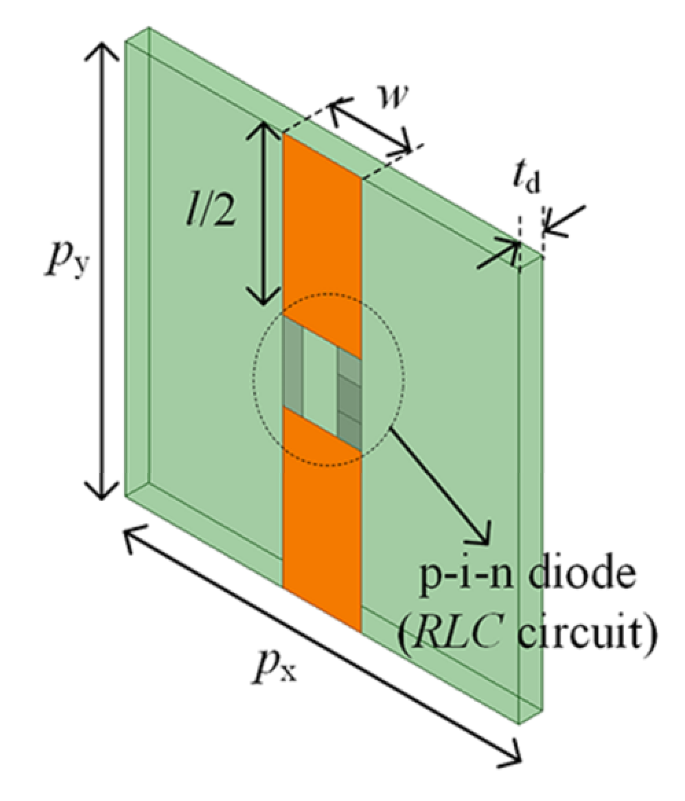}}
    \caption{A unit structure designed for simultaneous refraction, reflection, and absorption of electromagnetic waves~\cite{cai2017high,zhang2020beyond}.}
    \label{F1-15}
\end{figure}

The slot-coupling reflection unit not only achieves excellent reflection phase performance but also expands the bandwidth of the reflection array. Most importantly, the phase delay lines of slot-coupled units are located beneath the floor, making it very convenient for loading active components in reconfigurable reflection arrays, greatly simplifying bias circuit design. gap-coupled units are mainly composed of rectangular microstrip patches, slotted ground planes, phase delay lines, and two layers of dielectric material with different relative permittivity. Currently, there are not many designs based on slot-coupling.

The unit structure proposed by Xue et al.\cite{xue2018research} is shown in Fig.~\ref{F1-11}. A rectangular microstrip patch is located on the top layer dielectric substrate, a slotted ground plane is situated between the top and bottom dielectric layers, and the phase delay line is printed on the bottom layer of the dielectric substrate. The two layers of the substrate are pressed together. When PIN diodes are loaded on the phase delay line, these diodes can be controlled to be either on or off by an external bias voltage, thereby allowing the alteration of the electromagnetic wave propagation path, and enabling the reconfigurable characteristics of the RIS.

Dai et al.~\cite{dai2020reconfigurable} introduced a symmetric RIS unit structure in 2020, as shown in Fig.~\ref{F1-12}. Each unit consists of an upper square patch, a slotted ground plane, and a ground plane. The upper patch is responsible for receiving and radiating energy, while the ground plane primarily provides grounding and suppresses backward radiation. In this unit structure, the slotted plane is a key component for controlling the phase of the RIS. Its detailed configuration is depicted in Fig.~\ref{F1-12}, where PIN1/PIN2 are configured as 1, 2, 3, and 4 for ON/OFF, OFF/ON, ON/OFF, and OFF/ON, respectively. PIN3, PIN4, and PIN4 have configurations of 1 and 2 as ON, and 3 and 4 are both configured as OFF. Within each unit, four sets of slots are symmetrically etched, each integrated with five PIN diodes. Ideally, a 2-bit RIS unit provides four quantized phase states with a phase difference of 90 degrees. The states and corresponding radio frequency current paths for the four-unit configurations of the PIN diodes are illustrated in Fig.~\ref{F1-12}. The analog phase and amplitude performance of the 2-bit RIS unit under these four configurations display a phase difference of approximately $90^{\circ}$. These states remain highly stable within the frequency range of 2.0 to 2.6 GHz, with an insertion loss of less than 1.2 dB.

Due to the challenges associated with modeling equivalent circuits and structural design, the slot-coupled RIS unit structure is not commonly encountered at present. Its reconfigurability is primarily achieved through slot etching in the ground plane within multi-layered metal structures.

\subsection*{c. Transmissive type unit}

A transmissive RIS unit is shown in Fig.~\ref{F1-13}. Rectangular patch antennas are printed on the upper surface of the upper-layer dielectric substrate and the lower surface of the lower-layer dielectric substrate. There are two semicircular gaps on the rectangular patch antennas, with one side of the gaps being connected in the middle. The two patch antennas are interconnected through a metal via in the center. There is a shared ground plane in the middle. The upper-layer rectangular patch antennas are connected to the ground plane through two metal vias. The design of these two metal vias is to maintain antenna symmetry. Two PIN diodes are integrated into the lower-layer rectangular patch antennas. By controlling the bias voltage at the two ends of the PIN diodes, the opening direction of the lower-layer rectangular patch antenna can be changed, thus controlling the phase state of the unit~\cite{qiao2019design}.

Zhang and colleagues~\cite{zhang2022intelligent} introduced a unit design known as Intelligent Omni-Surface (IOS), which combines both transmissive and reflective functionalities. As shown in Fig.~\ref{F1-14}, a rectangular patch is connected to a rectangular metal frame via PIN diodes, and there's a through-hole in the middle of the rectangular patch connected to the bias line. Unlike a typical RIS unit, this unit does not have a metal ground plane. Instead, it features a bottom symmetric surface similar to the top layer's design. This unit can operate simultaneously in both reflection and transmission modes. When both diodes are either on or off, representing states 0 and 1, the phase difference is approximately $180^{\circ}$. However, it's worth noting that there is an energy loss of about half in either the reflection or transmission mode.

Similar IOS symmetric structural designs are also explored in references~\cite{cai2017high,zhang2020beyond}. Furthermore, in references ~\cite{de2021tri,wang2022transmission,song2022reconfigurable}, multi-layer unit structures have been designed, enabling the metasurface to switch between three states: transmission, reflection, and electromagnetic wave absorption. The structure of such units is depicted in Fig.~\ref{F1-15}.

The key difference between transmissive RIS and other RIS lies in their ability to not only reflect electromagnetic waves but also transmit or absorb them. This versatility enables them to be applied in a wider range of scenarios.

In summary, Reconfigurable Intelligent Surfaces (RIS) that allow for manual adjustment of their states are primarily controlled by the status of controllable electromagnetic components such as diodes. This control manipulates parameters like phase shift for incident electromagnetic waves. By integrating multiple diodes through various designs, RIS with different bit configurations, including 1-bit, 2-bit, and higher, can be achieved. The main types of these RIS include microstrip-based, gap-coupled, and transmissive RIS, with microstrip-based RIS being the most extensively researched among them.

\subsection{Channel Model}

Apart from the electromagnetic model based on the RIS units, the accuracy of the channel model also has a decisive impact on the results of experiments such as beamforming in practical test systems. 
Currently, channel modeling for RIS systems can be categorized into two types: one is statistical modeling based on statistical information, and the other is physical modeling based on physical information.

\subsection*{(1) Statistical channel model}

\subsection*{a. Cascaded model}
In RIS-assisted communication systems, the widely used baseband equivalent channel model \cite{basar2019wireless,wu2019intelligent,han2019large,tang2021securing,jung2021optimality} was initially proposed by R. Zhang et al. 
Assume that the number of antennas at the base station (BS) is $N$, the number of RIS units is $M$, and the number of user equipment (UEs) is $K$.
Considering a flat fading channel, the signal propagates through multipath from the transmitter to RIS and then arrives at the receiver after phase modulation and combination by RIS. 
The overall channel can be expressed in a cascaded form as:
\begin{equation}
{{\mathbf{y}}_{t}}=\left( {{\mathbf{h}}_{\text{d}}}+\mathbf{h}_{r}^{H}\mathbf{\Theta} \mathbf{H} \right)\mathbf{s}+{{\mathbf{n}}_{t}}
\label{Cascaded channel}
\end{equation}
where ${{\mathbf{h}}_{d}} \in \mathbb{C}^{M \times K}$ denotes the direct path (line-of-sight, LOS) between UEs and BS. ${{\mathbf{h}}_{r}} \in \mathbb{C}^{M \times K}$ and $\mathbf{H} \in \mathbb{C}^{M \times N}$ denote the channel between BS and RIS, and the channel between RIS and UEs, respectively. $\mathbf{s} \in \mathbb{C}^{N \times 1}$ is the transmitted signal. 
$\Theta =\operatorname{diag}\left( {{\left[ {{\Phi }_{1}},{\Phi }_{2},\ldots,{{\Phi }_{M}} \right]}^{T}} \right)$ denotes the reflection coefficient matrix of RIS, which is a diagonal matrix with its diagonal elements representing the reflection coefficient of each RIS unit, i.e., ${{\Phi }_{m}}={{\beta }_{m}}{{e}^{j{{\varphi }_{m}}}}$ where ${{\varphi }_{m}}$ and ${{\beta }_{m}}$ are the phase shift response and amplitude response, respectively. Without loss of generality, it is assumed that ${{\beta }_{m}}=1$. ${{\mathbf{n}}_{t}}$ is the additive white Gaussian noise.

Under this modeling, the ${\left( m,n \right)}$-th element of the channel matrix denotes the channel coefficient between the $n$-th transmitting antenna and the $m$-th receiving antenna, i.e., ${{H}_{m,n}}={{\alpha }_{{{H}_{m,n}}}}{{e}^{j{{\theta }_{{{H}_{m,n}}}}}}$, which includes the amplitude and phase of the channel.

However, since RIS is passive and not able to process baseband signals, the channel state information (CSI) of the channel between BS and RIS and between RIS and UEs cannot be obtained. To address this issue, \cite{zhang2020capacity} deployed RF links on RIS to obtain CSI. \cite{mishra2019channel} proposed an ON-OFF strategy, which divides the channel into multiple time slots and controls a certain reflection unit in the ON state while keeping other units in the OFF state so that the received signal information of that time slot can be obtained.

Based on the channel model~\eqref{Cascaded channel} mentioned earlier, an improved cascaded channel model was proposed in \cite{du2021capacity}.
The received signal expression for user $k$ at time $t$ is as follows:
\[{{\mathbf{y}}_{k}}=\left( {{\mathbf{h}}_{k}}+{{\mathbf{u}}^{H}}\text{diag}\left( {{\mathbf{h}}_{r}} \right)\mathbf{H} \right)\mathbf{s}+n_{t}\]
where the symbols ${\mathbf{h}}_{r}$, $\mathbf{H}$, and $\mathbf{s}$ have the same meanings as mentioned earlier.
${{\mathbf{u}}^{H}}=\left[ {{\Phi }_{1}},{\Phi }_{2},\cdots ,{{\Phi }_{M}} \right]$, ${\mathbf{h}}_k$ represents the $k$-th row of channel matrix ${\mathbf{h}}_d$, and $n_t$ represents the noise at time $t$.
Based on this modeling method, the CSI of the cascaded channel $\text{diag}\left( {{\mathbf{h}}_{r}} \right)\mathbf{H}$ can be acquired by BS, avoiding separate estimation for two channels.

\subsection*{b. Geometric model}
The channels of RIS-assisted mmWave or terahertz systems are generally modeled geometrically, which is also known as the Saleh-Valenzuela (SV) model \cite{saleh1987statistical,meijerink2014physical}. Due to the poor penetration of mmWaves, the scattered paths of the channel are often much less than the number of transmitting and receiving antennas, so the channel model has rich geometric characteristics.

Assuming the numbers of antennas at the transmitter, receiver, and the RIS are $N_t$, $N_r$, and $N$, the number of path clusters between the UE and the RIS is $P$, and the number of path clusters between BS and the RIS is $Q$.

The overall channel from UE to the BS can be denoted as \cite{sun20223d,ma2022modeling}:
\[{{\mathbf{H}}_{overall}}={{\mathbf{h}}_{d}}+\mathbf{H}\Theta {{\mathbf{h}}_{r}}\]
where the channel based on geometric modeling is given as:
\[{{\mathbf{h}}_{r}}=\sqrt{\frac{{{N}_{\text{r}}}{{N}_{\text{I}}}}{P}}\sum\limits_{p=1}^{p}{{{\alpha }_{p}}{{\mathbf{a}}_{\text{r}}}\left( \theta _{\text{r}}^{p},\phi _{\text{r}}^{p} \right)\mathbf{a}_{\text{t}}^{H}\left( \theta _{\text{t}}^{p},\phi _{\text{t}}^{p} \right)}\]
\[\mathbf{H}=\sqrt{\frac{{{N}_{\text{t}}}{{N}_{\text{I}}}}{Q}}\sum\limits_{q=1}^{Q}{{{\beta }_{q}}{{\mathbf{a}}_{\text{r}}}\left( \psi _{\text{r}}^{q},\varphi _{\text{r}}^{q} \right)\mathbf{a}_{\text{t}}^{H}\left( \psi _{\text{t}}^{q},\varphi _{\text{t}}^{q} \right)}\]
where ${{\alpha }_{p}}$, $\theta _{\text{r}}^{p}\left( \text{resp}.\phi _{\text{r}}^{p} \right)$ and $\theta _{\text{t}}^{p}\left( \text{resp}.\phi _{\text{t}}^{p} \right)$ represent the channel gain, azimuth angle (resp. elevation angle) of the angle of arrival (AoA), and azimuth angle (resp. elevation angle) of the angle of departure (AoD) of the p-th path between RIS and UE; similarly, ${{\beta }_{q}}$, $\psi _{\text{r}}^{q}\left( \text{resp}.\varphi _{\text{r}}^{q} \right)$and $\psi _{\text{t}}^{q}\left( \text{resp}\text{.}\varphi _{\text{t}}^{q} \right)$ represent the channel gain, azimuth angle (resp. elevation angle) of AoA, and azimuth angle (resp. elevation angle) of AoD of the $q$-th path between BS and RIS. ${{\mathbf{a}}_{\text{r}}}$ and ${{\mathbf{a}}_{\text{t}}}$ represent the receiving and transmitting array response vectors, respectively.

For example, for a uniform planar array (UPA) with units spacing of half-wavelength which contains $N_1\times N_2$ units, the steering vector is given by:
\begin{align*}
\begin{split}
       \mathbf{a}\left( \theta ,\phi  \right)=\frac{1}{\sqrt{N_1 N_2}} \left[ 1,{{e}^{j\pi \left( \sin \theta \sin \phi +\cos \phi  \right)}}, \cdots ,\right. \\
    \left. e^{j\pi ((\sqrt{N_1}-1 )\sin \theta \sin \phi +( \sqrt{N_2}-1 )\cos \phi)} 
    \right]^T
\end{split}
\end{align*}

Geometric-based channel models and hybrid channel models based on digital maps are the main models currently adopted and accepted in 3GPP and ITU standard research \cite{s1}\cite{s2}. The geometric-based channel modeling method provides a new means for mmWave channel estimation. The channels of mmWave and terahertz wireless communication systems mainly consist of LOS paths, which results in the sparsity of mmWave channels. Moreover, the sparsity of the channel within the bandwidth of the studied system almost remains unchanged \cite{ma2021intelligent}. Therefore, the channel estimation problem can be converted into a sparse recovery problem \cite{gao2016channel}\cite{gao2016channelICC}. Compared with traditional channel estimation methods, those based on channel sparsity can effectively reduce training complexity.

\subsection*{c. Rician fading model}
Due to the fact that RIS is passive and the reflected energy is related to the received energy, it is generally assumed that there exist LOS links between RIS and BS/UE. In scenarios with rich scattering, such as urban streets, the multipath components caused by the environment should be considered as well. In this case, the channel is supposed to be modeled as the Rician fading model. The channel between BS and the $m$-th unit of RIS is given as
\[{\textbf{h}_{r,m}}=\sqrt{\frac{{{K}_{1}}}{1+{{K}_{1}}}}{\bar{\textbf{h}}_{r,m}}+\sqrt{\frac{{{K}_{1}}}{1+{{K}_{1}}}}{\tilde{\textbf{h}}_{r,m}}\]
while the channel between the $i$-th unit of RIS and UE is given as \cite{yildirim2020modeling}:
\[{\textbf{H}_{m}}=\sqrt{\frac{{{K}_{2}}}{1+{{K}_{2}}}}{\bar{\textbf{H}}_{m}}+\sqrt{\frac{{{K}_{2}}}{1+{{K}_{2}}}}{\tilde{\textbf{H}}_{m}}\]
where ${{K}_{1}}$ and ${{K}_{2}}$ are the Rician factors of the two channels respectively. ${\bar{\textbf{h}}_{r,m}}$ and ${\bar{\textbf{H}}_{m}}$ represent the LOS part while ${\tilde{\textbf{h}}_{r,m}}$ and ${\tilde{\textbf{H}}_{m}}$ represent the NLOS part of the channel. The multipath component is generally assumed to be a narrowband Gaussian process, following a complex Gaussian distribution $\mathcal{CN}(\mu,\sigma^2)$.

Based on the Rician fading model, \cite{singh2021reconfigurable} derived the expression of channel capacity and the closed-form expressions for average symbol error probability for RIS-assisted single-input single-output (SISO) mmWave systems. \cite{msleh2022ergodic} obtained the ergodic capacity of RIS-assisted MIMO communication systems by analyzing the probability density functions (pdf) of the two cascaded channels.

\subsection*{d. Spatial correlation model}
In the RIS-assisted MIMO and massive MIMO scenarios, there are usually a large number of units in RIS and UE. However, with the trend of miniaturization of communication devices, it is necessary to design compact antenna arrays to solve the problems brought by multi-antenna \cite{zhou2012effect}. In the case of small inter-unit spacing, there exist spatial correlations between units, which is also known as the mutual coupling effect \cite{xu2021sum}.

Taking the channel matrix between RIS and UE as an example, the Rician fading model can be expressed as the sum of the LOS path and multipath components. With the multipath components modeled by spatial correlation, the Rician fading model based on the spatial correlation can be given as \cite{demir2021ris}:
\[\mathbf{H}=\mathbf{\tilde{H}}+\mathbf{\bar{H}}={{\mathbf{R}}^{\frac{1}{2}}}\mathbf{G}{{\mathbf{T}}^{\frac{1}{2}}}+\mathbf{\bar{H}}\]
where $\mathbf{R}$ and $\mathbf{T}$ are the spatial correlation matrices of the receiver and the transmitter, respectively. 
$\mathbf{G}$ is a Gaussian random matrix, the elements of which follow independent and identically distributed (i.i.d) complex Gaussian distribution \cite{abdullah2022impact}.

This section mainly introduces four statistical models for RIS modeling. 
Currently, the channels from the base station to RIS and from RIS to users often use Rayleigh fading or Rician fading channels. 
However, in real-world scenarios, due to the dense arrangement of RIS antennas, the channel in the spatial domain is often correlated. 
Therefore, in MIMO systems, spatial correlation modeling is typically considered.
Different from traditional discrete antenna array models, the above channel modeling cannot reflect the intrinsic electromagnetic characteristics and structure of RIS.
Therefore, it is necessary to analyze the channel response from a microscopic perspective.

\subsection*{(2) Physical channel model}
\subsection*{a. Based on free space path loss}
Most of the above channel models are based on the statistical characteristics of the channel, modeling the responses of RIS to the signal as diagonal matrices in exponential form. However, these modeling methods ignored the physical and electromagnetic characteristics of RIS, the near-field effect was not taken into consideration as well. Based on the electromagnetic response of RIS to the incident electromagnetic wave, \cite{tang2020wireless} first proposed a free-space propagation model for RIS-assisted communication systems. For example, in the far-field beamforming mode, the path loss of the electromagnetic wave reflected to the receiving antenna by RIS can be denoted as:
\[\delta_{\text{farfield}}^{\text{beam}}=\frac{64{{\pi }^{3}}{{\left( {{d}_{1}}{{d}_{2}} \right)}^{2}}}{{{G}_{t}}{{G}_{r}}G{{N_c}^{2}}{{N_l}^{2}}{{d}_{x}}{{d}_{y}}{{\lambda }^{2}}F\left( {{\theta }_{t}},{{\phi }_{t}} \right)F\left( {{\theta }_{r}},{{\phi }_{r}} \right){{A}^{2}}}\]
where ${{d}_{1}}$ is the distance between transmitter and RIS and ${{d}_{2}}$ is the distance between RIS and receiver. $N_l$  and $N_c$ are the numbers of rows and columns of the uniform planar array RIS. ${{G}_{t}}$, ${{G}_{r}}$, and $G$ are the antenna gain of the transmitter, receiver and RIS units, respectively. ${d}_{x}$ and $d_y$  are the size of each unit along the $x$-axis and $y$-axis. $F\left( {{\theta }_{t}},{{\phi }_{t}} \right)$ and $F\left( {{\theta }_{r}},{{\phi }_{r}} \right)$ represent the normalized power radiation pattern of the transmitter and receiver, $A$ represents the amplitude response of the RIS units, $\lambda$ represents the wavelength.
This paper also provides the near-field broadcasting mode's free space path loss conclusion.

The models based on physical and electromagnetic responses fully consider factors such as the distance between RIS and transceivers, the size of RIS, the near-field and far-field effects, and the antenna radiation pattern which provides further theoretical support for modeling RIS-assisted communication systems. 
Furthermore, based on the consideration of the electromagnetic and physical characteristics of RIS, Tang et al.\cite{tang2022path} established a free-space path loss model for RIS-assisted communication systems under various scenarios. They verified the accuracy of the proposed model through practical testing.

\subsection*{b. Based on electromagnetic unit response}
In this part, we propose to characterize the reflection and scattering effects of RIS through its physical and electromagnetic response, and extend the model to MIMO scenarios \cite{mi2022towards}. 

Considering rectangular units of RIS with length and width being $\left( a,b \right)$, the incident wave is uniform plane wave denoted as ${{E}_{i}}\left( {{\theta }_{i}},{{\phi }_{i}} \right)$. The reflection coefficient of the RIS unit to the electromagnetic wave is denoted as $\Gamma$. Under the assumption of far-field effects, the scattered electric field at distance ${{r}_{s}}$ with azimuth angle $\theta_s$ and elevation angle $\phi_s$ can be expressed as ${{E}_{s}}$ \cite{mi2022towards}: 
\begin{align*}
 & E_{r}^{s}\left( {{r}_{s}},{{\theta }_{s}},{{\phi }_{s}} \right)\simeq 0 \\ 
 & E_{\theta }^{s}\left( {{r}_{s}},{{\theta }_{s}},{{\phi }_{s}} \right)\simeq C\frac{A}{\lambda }\frac{{{e}^{-j2\pi {{r}_{s}}/\lambda }}}{{{r}_{s}}}{{E}_{i}}\cos {{\theta }_{i}}\cos {{\theta }_{s}} \\ 
 & \cdot \left( \cos {{\phi }_{i}}\sin {{\phi }_{s}}-\sin {{\phi }_{i}}\cos {{\phi }_{s}} \right)\text{Sa}\left( a,b,{{\theta }_{s}},{{\phi }_{s}},{{\theta }_{i}},{{\phi }_{i}} \right) \\ 
 & E_{\phi }^{s}\left( {{r}_{s}},{{\theta }_{s}},{{\phi }_{s}} \right)\simeq C\frac{A}{\lambda }\frac{{{e}^{-j2\pi {{r}_{s}}/\lambda }}}{{{r}_{s}}}{{E}_{i}}\cos {{\theta }_{i}} \\ 
 & \cdot \left( \sin {{\phi }_{i}}\sin {{\phi }_{s}}+\cos {{\phi }_{i}}\cos {{\phi }_{s}} \right)\text{Sa}\left( a,b,{{\theta }_{s}},{{\phi }_{s}},{{\theta }_{i}},{{\phi }_{i}} \right)  
\end{align*}
where $A=ab$, $C=-j(1-\Gamma )/2$. The function $Sa(\cdot )$ is defined as 
\begin{align*}
  & \text{Sa}\left( a,b;{{\theta }_{s}},{{\phi }_{s}};{{\theta }_{i}},{{\phi }_{i}} \right)=\frac{\sin \left( \frac{\pi a}{\lambda }\left( \sin {{\theta }_{s}}\cos {{\phi }_{s}}+\sin {{\theta }_{i}}\cos {{\phi }_{i}} \right) \right)}{\frac{\pi a}{\lambda }\left( \sin {{\theta }_{s}}\cos {{\phi }_{s}}+\sin {{\theta }_{i}}\cos {{\phi }_{i}} \right)} \\ 
 & \times \frac{\sin \left( \frac{\pi b}{\lambda }\left( \sin {{\theta }_{s}}\sin {{\phi }_{s}}+\sin {{\theta }_{i}}\sin {{\phi }_{i}} \right) \right)}{\frac{\pi b}{\lambda }\left( \sin {{\theta }_{s}}\sin {{\phi }_{s}}+\sin {{\theta }_{i}}\sin {{\phi }_{i}} \right)}  
\end{align*}

The modeling method based on electromagnetic unit response fully considers the influence of the wavelength of the electromagnetic waves and the size, spacing, and number of units on RIS. 
\cite{mi2022towards} first studied the multiple-input multiple-output (MIMO) behavior of RIS, and the model shows that a simple linear system of equations can describe the MIMO behavior of RIS under reasonable assumptions, which can serve as a basic model for analyzing and optimizing the performance of RIS-assisted systems in the far-field. Continuous and discrete strategies are taken to model single patch and patch array and their interactions with multiple incident electromagnetic waves. In addition, a physically accurate formula was introduced to calculate the scattered electric field of rectangular metal patches and several mathematically easy-to-handle models are proposed to characterize the input/output behavior of the array. 

Based on general scalar diffraction theory and the Huygens-Fresnel principle, \cite{di2020analytical} proposed a calculating method for the received power of RIS in closed form. \cite{gradoni2021end} proposed an end-to-end and EM-compliant electromagnetic compatibility model considering the mutual coupling between units of RIS.

In recent years, communication theory analysis based on channel modeling has garnered significant attention in the academic community. 
Based on the RIS-assisted SISO wireless communication system model proposed by Wu et al.\cite{wu2019intelligent}, references \cite{van2020coverage}\cite{van2021outage}\cite{selimis2021performance}\cite{ibrahim2021exact} have individually considered channel modeling for uncorrelated Rayleigh channels, correlated Rayleigh channels, and Nakagami-m distribution.
They have provided theoretical characterizations, both asymptotic and non-asymptotic, for coverage, outage probability, (equivalent) signal-to-noise ratio, and ergodic capacity in communication systems. 
These studies also discuss the influence of phase-shift matrix design on communication system theoretical capacity properties in certain special cases.
Based on the references mentioned above, Papazafeiropoulos et al.~\cite{papazafeiropoulos2021coverage} extended the work by considering scenarios with multiple RIS in a cooperative setting, assuming correlated Rayleigh channels. 
They conducted a theoretical comparison between two design approaches: a few large-scale RIS and a large number of small-scale RIS. 
The results indicated that having more RIS, as opposed to more RIS units, can provide better communication performance gains.
In the same system modeling framework, 
references \cite{zhang2021performance}\cite{kudathanthirige2020performance}\cite{boulogeorgos2020ergodic} extended the theoretical analysis from reference \cite{selimis2021performance} to various different channel assumptions. 
Additionally, it's worth mentioning that Li et al. 
\cite{li2023performance} considered wireless communication systems with RIS element positions close together and coupling, by assuming a special structure of channels as the Kronecker product. 
They studied group-wise joint control strategies for RIS array units in practical scenarios where RIS units are in close proximity.
References \cite{kammoun2020asymptotic}\cite{yang2020accurate}\cite{moustakas2021capacity} used probability and random matrix theory tools to investigate signal-to-noise ratio and mutual information statistics in the context of downlink multi-user and multi-RIS scenarios, as well as dual-hop scenarios. 
They discussed theoretically optimal linear coding schemes that can guide the control of RIS phase-shift matrices.
Héliot et al. \cite{heliot2022closed} considered RIS-assisted wireless communication in multi-antenna and multi-hop scenarios. They employed channel multiplication modeling to explore the trade-off between energy efficiency and spectral efficiency. Their findings indicate that, compared to increasing the number of antennas, increasing the number of hops can effectively enhance the system's energy efficiency. However, in Device-to-Device (D2D) scenarios, RIS-assisted MIMO communication does not guarantee energy efficiency superior to traditional multi-antenna solutions.

This section introduces two different physical models of RIS. The channel model, based on free-space fading, establishes the relationship between path loss in RIS-assisted wireless channels and the physical parameters of the RIS. 
In addition to modeling the fading coefficient, the unit modeling based on electromagnetic response provides a vector representation of the RIS-emitted electric field at any point in space. 
This offers a precise electromagnetic model for micro-level channel modeling. The main components of the basic RIS model are shown in Tables \ref{Table3.1} and \ref{Table3.2}.

\begin{table*}[!htbp]
\centering
\setlength{\tabcolsep}{0.2mm}
\caption{The Development and Characteristics of Electromagnetic Structure Models for RIS}
\begin{tabular}{ccll}
\toprule
\multicolumn{2}{c}{Model} & \multicolumn{1}{c}{Development} & \multicolumn{1}{c}{Characteristic} \\ \midrule
\multicolumn{2}{c}{\multirow{3}{*}[-8.5ex]{\begin{tabular}[c]{@{}c@{}}Fixed structure \\ metasurfaces\end{tabular}}}                                                               & \begin{tabular}[c]{@{}l@{}}Yu et al.\cite{yu2011light} revolutionized the conventional\\ design approach of metasurfaces by introducing\\ the generalized Snell's law. They leveraged this\\ law to devise metasurfaces with non-uniform\\ distributions.\end{tabular}                                                    

& \multirow{3}{*}[-6.0ex]{\begin{tabular}[c]{@{}l@{}}The electromagnetic structure of a fixed metasurface is inherently\\ stationary, which implies its limited phase tunability. Nonetheless,\\ researchers have harnessed this characteristic to achieve phase control\\ across various frequency bands. Additionally, they have achieved\\ multi-bit phase manipulation by altering the geometric parameters.\end{tabular}} 
\\
\multicolumn{2}{c}{}                                                                                                                                         & \begin{tabular}[c]{@{}l@{}}Li and Yao\cite{li2018manipulation} created a multi-bit digital \\ coding metasurface using PB phase engineering, \\ with identical-sized but differently oriented \\ meta-atoms.\end{tabular}           &                                                                          
\\
\multicolumn{2}{c}{}                                                           & \begin{tabular}[c]{@{}l@{}}Abdullah and Koziel\cite{abdullah2021supervised} conducted a supervised \\ learning-based investigation into the design of \\ metasurface units with broadband radar cross-\\section (RCS) reduction capabilities. Their design \\enables separate 1-bit phase modulation and\\ amplitude modulation for incident waves with\\ different linear polarizations.\end{tabular} &                                                  \\

\multirow{5}{*}[-18.5ex]{\begin{tabular}[c]{@{}c@{}}Programmable \\ metasurfaces\end{tabular}} & \multirow{3}{*}[-7.0ex]{\begin{tabular}[c]{@{}c@{}}Microstrip \\ patch-style\end{tabular}} & \begin{tabular}[c]{@{}l@{}}A Japanese research team\cite{kamoda201160} proposed using \\ PIN diodes for digital phase shifting.\end{tabular}
& \multirow{5}{*}[-15.0ex]{\begin{tabular}[c]{@{}l@{}}Programmable metasurfaces primarily utilize the control of diode\\ states to manipulate the phase of the metasurface, enabling the\\ realization of multi-bit functionality. The integration of multiple\\ diodes with different designs allows for the implementation of\\ metasurfaces with 1-bit, 2-bit, and higher bit counts. Currently,\\ the main types of metasurfaces include microstrip-patch-based, \\ gap-coupled, and transmissive designs, with microstrip-patch-based\\ metasurfaces being the most extensively studied, while the other\\ two types have received comparatively less attention.\end{tabular}} \\ &              

& \begin{tabular}[c]{@{}l@{}}Cui et al.\cite{cui2014coding} first proposed the hardware \\ structure of programmable metamaterials, laying \\ the research foundation for practical systems \\ of intelligent metasurfaces (RIS).\end{tabular}       

&                                                                              \\
&                                                                              & \begin{tabular}[c]{@{}l@{}}Reference\cite{huang2017dynamical} presents a 2-bit RIS element\\ design, where the phase difference between\\ adjacent states are approximately 90º near\\ 7.25 GHz, meeting the requirements\\ of 2-bit units.\end{tabular}                                                          
&                                                                           
\\
& Slot-coupled                                                                 & \begin{tabular}[c]{@{}l@{}}Reference\cite{xue2018research} introduces a gap-coupled design \\ that achieves reconfigurability by controlling \\ diode switching through an external bias voltage, \\ allowing for the alteration of electromagnetic \\ wave propagation paths.\end{tabular}                        &                                                                              \\
& Transmission-type                                                                  & \begin{tabular}[c]{@{}l@{}}Reference\cite{zhang2022intelligent} presents a unit design known as \\ the Intelligent Omnidirectional Surface (IOS), \\ which is capable of operating in both reflection \\ and transmission modes. However, it suffers from \\ an energy loss of approximately half the power in\\ either the reflection or transmission mode.\end{tabular}            
&                                                                             \\ \bottomrule
\end{tabular} \label{Table3.1}
\end{table*}

\begin{table*}[!htbp]
\centering
\setlength{\tabcolsep}{0.2mm}
\caption{The Fundamental Characteristics of an RIS Channel Model}
\begin{tabular}{ccll}
\toprule
\multicolumn{2}{c}{Model} & \multicolumn{1}{c}{Development and characteristics}                                                               \\ \midrule
\multirow{8}{*}[-9.5ex]{Statistical model} & \begin{tabular}[c]{@{}c@{}}Cascaded channel\\ model\end{tabular}              & \begin{tabular}[c]{@{}l@{}}The cascaded channel model is obtained by partitioning and modeling the channel \\ step by step under flat fading channel conditions. This channel model is commonly \\ used, but it has certain limitations due to the inherent uncertainty in channel \\ information.\end{tabular}                            

\\
& Geometric model                                                               & \begin{tabular}[c]{@{}l@{}}Geometric modeling is commonly used in scenarios involving millimeter-wave and \\ terahertz systems. Due to the limited penetration capability of millimeter waves \\ and the relatively fewer scattering paths in the channel, often much fewer than \\ the number of transmit and receive antennas, the channel model exhibits rich \\ geometric characteristics. By geometrically modeling each part of the channel \\ and then cascading them, a comprehensive geometric model can be obtained.\end{tabular} \\

 & \begin{tabular}[c]{@{}c@{}}Rician fading\\ model\end{tabular}                 & \begin{tabular}[c]{@{}l@{}}The Rician fading model is employed to account for scenarios where multipath \\ effects are significant. It models the BS-RIS and RIS-UE segments of the channel \\ separately using the Ricean fading model.\end{tabular}                                                           
 
\\
& \begin{tabular}[c]{@{}c@{}}Spatial correlation\\ model\end{tabular}           & \begin{tabular}[c]{@{}l@{}}Spatial correlation modeling is derived from the combination of the Rician fading\\ model and spatial correlation, which is a result of the closely spaced units in\\ MIMO and massive MIMO scenarios.\end{tabular}                                                                                                                                       \\
\multirow{2}{*}[-5.0ex]{Physical model}    & \begin{tabular}[c]{@{}c@{}}Free-space\\ path loss based\end{tabular}          & \begin{tabular}[c]{@{}l@{}}By taking into account the physical and electromagnetic properties of the RIS, array\\ aperture, near-field effects, and other influencing factors on the electromagnetic response\\ of incident waves, a physical model is proposed. This model is more realistic compared\\ to statistical models.\end{tabular}                                                                                     \\
& \begin{tabular}[c]{@{}c@{}}Electromagnetic unit\\ response based\end{tabular} & \begin{tabular}[c]{@{}l@{}}Compared to physical models based on free-space fading, a physical model based \\ on the electromagnetic unit response further focuses on individual electromagnetic \\ units, rather than considering the overall electromagnetic effects of the RIS. \\ It takes into account the micro-scale effects of RIS, including factors such as \\ the size of array units, spacing, wavelength, and the number of units, \\ and their influence on the response of the RIS.\end{tabular}

\\ \bottomrule
\end{tabular} \label{Table3.2}
\end{table*}

\section{Prototypes and testbeds}\label{Section3}
A typical RIS board generally consists of three layers and a controller module. In the first layer, electromagnetic design principles are employed to create distinct units for different bits. The second layer incorporates a copper board to prevent signal energy leakage. The third layer serves as the control circuit board, responsible for adjusting the reflection coefficients (amplitude and phase shift) of each reflection unit by modulating the voltage. An FPGA module functions as the controller, connecting with both the host computer and the control circuit board. The host computer inputs the control codebooks to the FPGA module, which then translates these codebooks into corresponding reflection coefficients and forwards them to the control circuit board. Upon receiving the commands, the control circuit board alters the state of each reflection unit accordingly. This multi-layered structure and control mechanism enables precise control of the RIS's reflective properties.

\begin{figure}
\centering{\includegraphics[width=1\columnwidth]{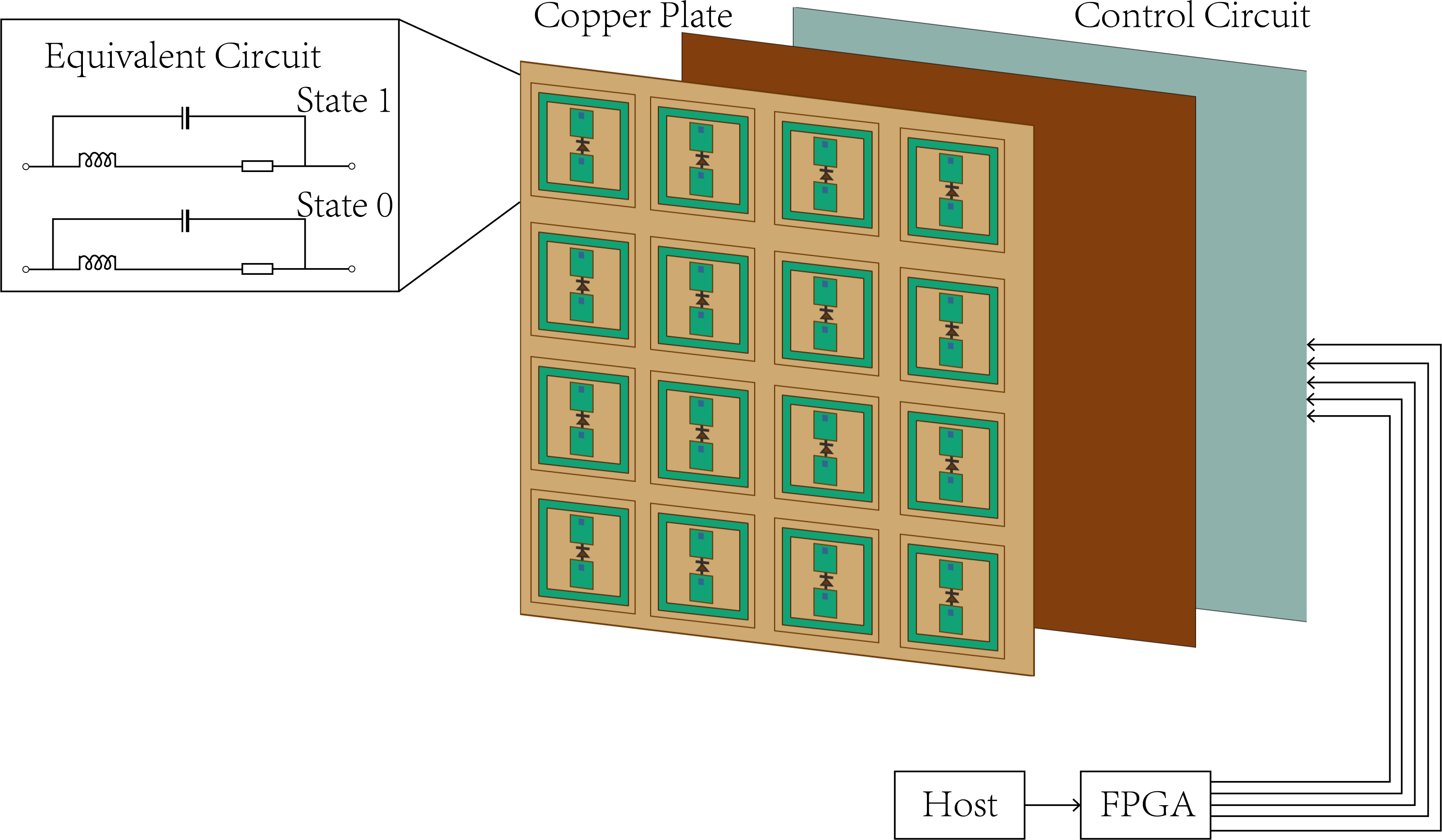}}
\caption{The structure of RIS board.}
\label{F2-1}
\end{figure}

Within each unit of the RIS, there exist one or more diodes. Fig. \ref{F2-1} illustrates the equivalent circuit diagrams corresponding to two distinct states of RIS reflection units controlled by PIN diodes. By adjusting the bias voltage of the PIN diode through a direct current (DC) feedline, the PIN diode can be set to either a `0' or `1' state, thereby inducing a phase shift.

In the design of RIS units, the primary focus revolves around the applicable frequency range and the number of discrete bits. From the perspective of RIS prototype design, the design objectives can broadly be categorized into two main directions: low power consumption and high precision. For RIS prototype systems aimed at communication purposes, the currently achievable fundamental functionalities primarily encompass signal coverage and spatio-temporal modulation, among others.

\subsection{Low Power Consumption Prototypes}
RIS unit components are a type of shape-adaptable surface with varactor diodes or PIN diodes. They can be controlled through bias circuits to achieve real-time control over the direction of electromagnetic wave propagation. As a nearly passive artificial device, the deployment of RISs needs to be spatially unrestricted, implying that a wireless power supply is often necessary. When employing wireless power sources such as batteries, minimizing power consumption becomes a critical concern.

Dai et al. implemented a RIS with 256 2-bit units operating at a frequency of 2.3 GHz using PIN diodes\cite{dai2020reconfigurable}. In the 2.3 GHz frequency band, the RIS achieves a gain of 21.7 dBi, with a power consumption of approximately 153 W. To attain the same gain, conventional phased-array antennas would require at least 64 antenna units with a power consumption of approximately 370 W. The RIS prototype designed in this work achieved a 58.6\% reduction in power consumption while delivering performance similar to traditional phased-array antennas.

Gros et al. designed a 10 cm $\times$ 10 cm RIS containing 400 1-bit units, comprising 800 PIN diodes, operating in the millimeter-wave frequency range from 27.5 to 29.5 GHz\cite{gros2021reconfigurable}. Millimeter-wave experiments demonstrated that this RIS system achieved a gain of 25 dBi at the cost of only 8 W of power consumption through the utilization of PIN diodes. When all PIN diodes are in the ON state, the peak power consumption reaches 16 W. Subsequently, Popov et al. used this unit to design a RIS measuring 20 cm $\times$ 20 cm, which included 1600 independently controlled units\cite{popov2021experimental}. In a non-line-of-sight (NLOS) scenario with a transmitter-receiver distance of 10 meters, enabling the RIS results in a 30 dB increase in signal strength.

In addition to PIN diode-based RIS designs, varactor diodes are also commonly employed in low-power designs. Pei et al. employed varactor diodes to design a RIS board operating at a frequency of 5.8 GHz, comprising 1100 1-bit units\cite{pei2021ris}. In the 5.8 GHz frequency band, the RIS yields an antenna power gain of 17.1 dBi, with power consumption primarily attributed to electronic components on the RIS. Specifically, the power consumption includes 0.0138 W for bidirectional voltage converters, 0.918 W for level adjusters, and 0.076 W for varactor diodes, totaling 0.934 W. In addition to the RIS's power consumption, the FPGA controller consumes 1.5 W in this design.

Similarly, Tang et al. were the first to implement a prototype of a RIS-assisted MIMO-QAM (Quadrature Amplitude Modulation) wireless communication system, with a RIS board composed of 256 units designed using varactor diodes and operating at a frequency of 4.25 GHz\cite{tang2020mimo}. Experimental results demonstrated that this prototype achieved 2 $\times$ 2 MIMO transmission based on RIS and 16-QAM modulation, with a data transmission rate of up to 20 Mbit/s. Furthermore, the power consumption of the RIS and control circuit boards is merely 0.7 W.

In fact, for frequency ranges exceeding 30 GHz, the quality factor of varactor diodes is relatively poor\cite{nasserddine2016millimeter}, rendering them unsuitable for modulating spatially high-frequency waves. Therefore, it is crucial to reduce the power consumption of RIS designs based on PIN diodes. Building upon the foundation of PIN diode-based designs, the power consumption of the RIS has been reduced by implementing a series resistance current-limiting operation while ensuring minimal impact on the gain.

\begin{figure}
\centerline{\includegraphics[width=.7\columnwidth]{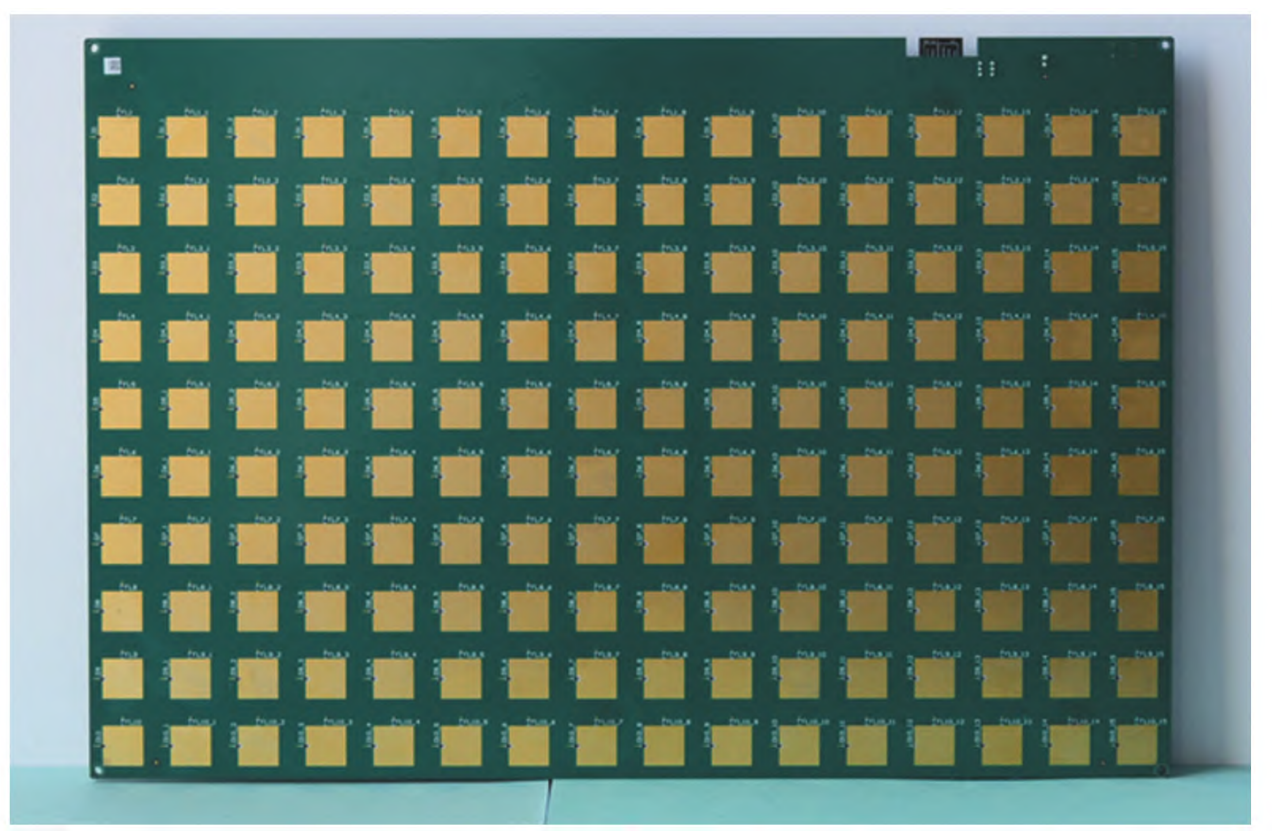}}
\caption{$10\times 16$ 1-bit RIS prototype.}
\label{F2-2}
\end{figure}

\begin{figure}[t!]
    \centering

    \subfloat[Amplitude response]{
		\includegraphics[width=.7\columnwidth]{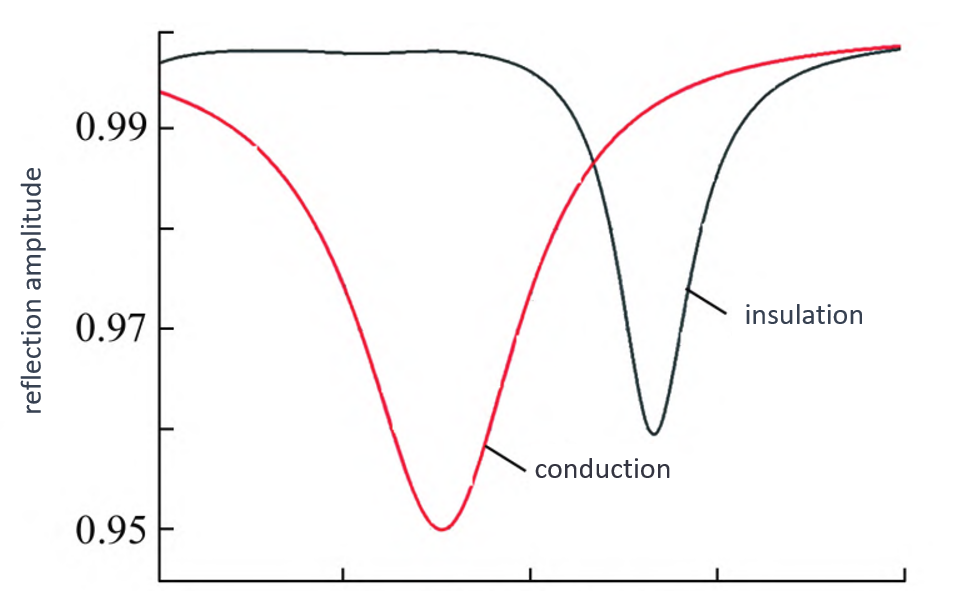}}
    \\
    \subfloat[Phase response]{
		\includegraphics[width=.7\columnwidth]{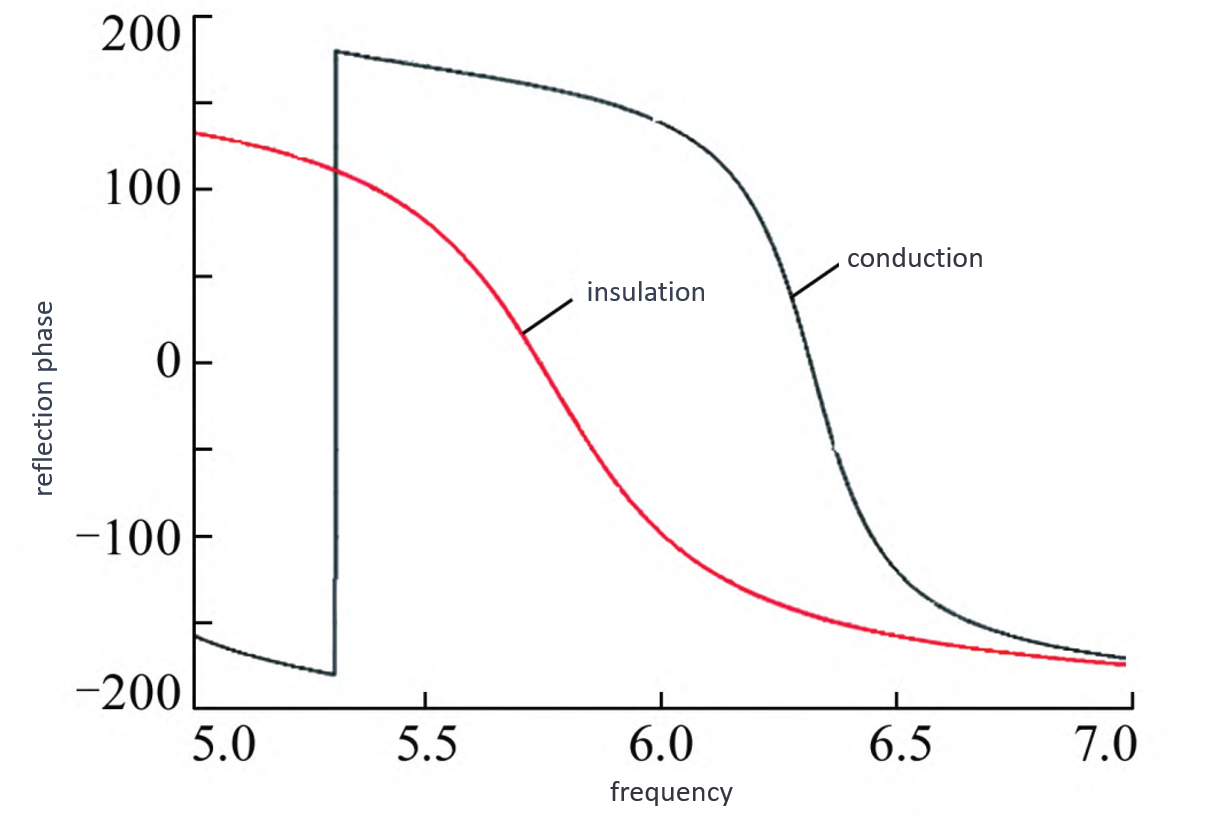}}
    \caption{Amplitude response (top) and phase response (bottom).}
    \label{F2-3}
\end{figure}

\begin{figure}
\centerline{\includegraphics[width=.9\columnwidth]{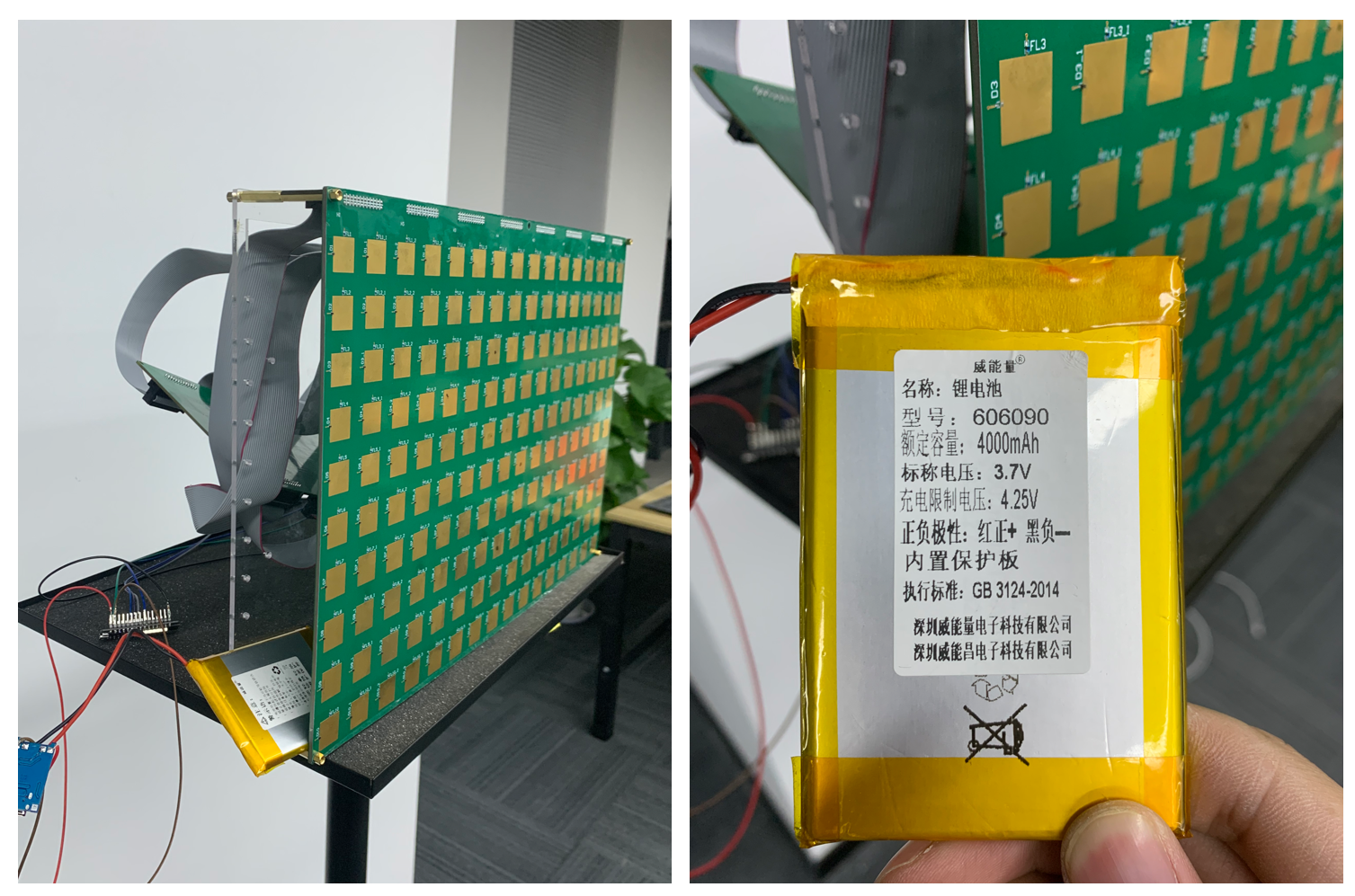}}
\caption{Lithium battery powered RIS.}
\label{F2-4}
\end{figure}

Fig. \ref{F2-2} presents the designed prototype of the RIS, which operates at a frequency of 5.8 GHz and comprises 10 $\times$ 16 1-bit units. The entire RIS and its control system operate at a full-load power consumption of only 0.2 W\cite{xiong2023ris}. Fig. \ref{F2-3} illustrates the amplitude and phase responses of the designed RIS prototype. It is evident that when the RIS operates at its working frequency, the amplitude responses in both ON and OFF states remain above 0.95, and the phase response difference between the two states is close to $180^\circ$. Furthermore, benefiting from its low power consumption, the RIS is powered by a lithium battery for the first time, as depicted in Fig. \ref{F2-4}.

Power consumption can be reduced by using PIN diodes or varactor diodes. However, when considering the limitations of varactor diodes in modulating spatially high-frequency waves, PIN diodes may be a better choice as high-frequency electromagnetic switches.

\subsection{High Precision Prototypes}
Tan et al. proposed the use of varactor diodes in designing RIS units to meet the requirement of phase continuity\cite{tan2016increasing}. However, varactor diodes typically have longer response times and lower precision in achieving continuous phase variations. Consequently, some researchers have shifted their focus towards RIS unit designs based on PIN diodes\cite{tan2018enabling, carrasco2012x,zhang2016design}.

A 1-bit RIS can only provide two phase states, namely 0 and $\pi$. This coarse phase resolution can reduce the antenna aperture efficiency, resulting in higher sidelobes and a decrease in antenna gain by more than 3 dB\cite{wu2019beamforming}. To mitigate the performance degradation caused by 1-bit phase quantization, some scholars have designed multi-bit RIS boards. However, multi-bit solutions come with increased system complexity and hardware costs. A 2-bit phase quantization is considered a compromise between system complexity and unit performance\cite{wu2019beamforming,wu2008selection}. In this scenario, antenna gain losses can be reduced to below 1 dB\cite{pereira2010four}, and there are significant improvements in sidelobe patterns\cite{cheng2006study,tang2020design}.

Rains et al. designed a 3-bit RIS, which, in the 3.75 GHz frequency band, provided a gain of 21.13 dB compared to an aluminum board\cite{rains2022high}. However, the gain improvement is only 0.68 dB compared to the 2-bit design, but the system's design complexity significantly increases. Therefore, when designing an RIS, a balance must be struck between the high-precision gain achievable with multi-bit solutions and the complexity of system design.

TABLE~\ref{Table3.3} illustrates the number of bits, the number of RIS array units, and the corresponding gains reported in some studies. Experimental results demonstrate that increasing the number of RIS units from 160 to 320 leads to an increase in signal strength at the receiver end from -40 dBm to -27 dBm\cite{xiong2023ris}.

RIS unit designs are mainly categorized into two types based on the type of electromagnetic switch employed: some based on varactor diodes and some based on PIN diodes. Varactor diodes can be used to create RIS units that achieve continuous phase modulation, but this approach suffers from long response times and low precision in achieving continuous phase variations. The use of PIN diodes can partially address this limitation, enabling the design of discrete-phase RIS units. In the context of phase-discrete RIS units, 2-bit RIS unit designs are considered a compromise solution that balances system complexity and phase resolution.

Furthermore, the number of units on the RIS has an impact on system performance, and a larger number of units can enhance the precision of RIS control.

\begin{table*}[!htbp]
\centering
\setlength{\tabcolsep}{0.2mm}
\caption{Comparison of Existing Work}
\begin{tabular}{ccccccc}
\toprule
Year     & Reference     & Diode Type     & Frequency Band/Center Frequency     & Number of Bits     & Number of RIS Component units     & System Gain\\
\midrule
2017 &   \cite{zhang2017realization}      & -              & 9.5-10.5 GHz   & 2-bit      & 9 $\times$ 9      & 19.8 dB \\
2019 &   \cite{al2019design}              & -              & 10-14 GHz      & 2-bit      & 20 $\times$ 20    & 6 dB \\
2021 &   \cite{tang2021channel}           & -              & 2.4 GHz        & 1-bit      & 64 $\times$ 50    & 9.8 dB \& Channel Capacity Doubling \\
2021 &   \cite{fu2021combining}           & -              & 9-19 GHz       & 1-bit      & 6 $\times$ 6     & 10 dB \\
2022 &   \cite{al2022design}              & -              & 60-120 GHz     & 1-bit      & 21 $\times$ 21    & 10 dB \\
2011 &   \cite{kamoda201160}              & PIN diode      & 60.25 GHz      & 1-bit      & 160 $\times$ 160  & 10 dB \\
2016 &   \cite{zhang2016design}           & PIN diode      & 13.5 GHz       & 1-bit      & 10 $\times$ 10    & 16.5 dB \\
2019 &   \cite{zhang2019breaking}         & PIN diode      & 9.5 GHz        & 2-bit      & 16 $\times$ 8     & 10 dB \\
2019 &   \cite{zhang2019dynamically}      & PIN diode      & 9.5 GHz        & 1-bit      & 16 $\times$ 8     & Achieved Throughput 2.5 Mbps \\
2020 &   \cite{dai2020reconfigurable}     & PIN diode      & 2.3/28.5 GHz   & 2-bit      & 16 $\times$ 16    & 21.7/19.1 dB \\
2021 &   \cite{gros2021reconfigurable}    & PIN diode      & 27.5-29.5 GHz  & 1-bit      & 20 $\times$ 20    & 25 dB \\
2022 &   \cite{trichopoulos2022design}    & PIN diode      & 5.8 GHz        & 1-bit      & 16 $\times$ 10    & 20 dB \\
2022 &   \cite{rains2022high}             & PIN diode      & 3.75 GHz       & 3-bit      & 20 $\times$ 20    & Ave. 16 dB, Max. 40 dB \\
2022 &   \cite{keykhosravi2022leveraging} & PIN diode      & 2.64 GHz       & 1-bit      & 16 $\times$ 32    & 10 dB \& Achieved Throughput 10 Mbps \\
2022 &   \cite{ouyang2022computer}        & PIN diode      & 2.6 GHz        & 1-bit      & 32 $\times$ 16    & 9.9 dB \& Achieved Throughput 78.61 Mbps \\
2022 &   \cite{chen2022accurate}          & PIN diode      & 27 GHz         & 1-bit      & 56 $\times$ 20    & 256-QAM Modulation \\
2019 &   \cite{dai2019wireless}           & varactor diode & 4 GHz          & 2-bit      & 8 $\times$ 16     & 480p Video Streaming \\
2021 &   \cite{pei2021ris}                & varactor diode & 5.8 GHz        & 1-bit      & 55 $\times$ 20    & 1080p Video Streaming \\ 
2022 &   \cite{araghi2022reconfigurable}  & varactor diode & 3.5 GHz        & Continuous & 30 $\times$ 87    & 15 dB \\
\bottomrule
\multicolumn{7}{l}{$-$ represents a diode-free, fixed-structure unit design.}
\end{tabular}\label{Table3.3} 
\end{table*}

\subsection{Experimental Test - Signal Coverage}
To validate the effectiveness of RIS prototypes in wireless communication systems, researchers often design experiments for wireless coverage enhancement. These experiments aim to utilize RISs to extend the coverage area of wireless signals. Additionally, some scholars conduct Over-the-Air (OTA) tests to assess the air interface performance of RISs. The evaluation metrics for these experiments typically include received signal strength and data transmission rates.

Kamoda et al. conducted tests on the beamforming of RIS in both near-field and far-field scenarios\cite{kamoda201160}. Through measurements of radiation patterns, they verified the reconfigurability of the reflective array. Additionally, they employed an alternative method using a standard gain horn antenna to measure the antenna gain at 60.25 GHz, which yielded a gain of 41 dBi for the reflective array, in close agreement with the estimated results. These experiments validated the significant role of RISs in signal gain enhancement in wireless communication.

Cui et al. fabricated a prototype of RIS in 2014, pioneering real-time phase control of scattered electromagnetic waves using PIN diode switches to achieve digital signal transmission\cite{cui2014coding}. Tang et al. achieved real-time data transmission in SISO and MIMO systems based on RIS in 2019\cite{dai2019wireless,tang2020mimo,zhao2019programmable}. They also conducted measurements and modeling of the free-space path loss of RIS and verified channel reciprocity\cite{tang2020wireless,tang2022path,tang2021channel}. In 2020, Arun et al. constructed a prototype RIS with 3200 units, resulting in a ninefold increase in single-user signal power gain and a 2.5-fold increase in median channel capacity\cite{arun2020rfocus}. In the same year, Dunna et al. harnessed RIS to enhance scattering effects in the environment, leading to improved MIMO spatial reuse gains\cite{dunna2020scattermimo}. Through experimentation, they demonstrated that Scatter MIMO, when compared to configurations without RIS, doubled throughput, increased signal-to-noise ratio by 2 dB, extended the coverage range from 30 m to 45 m, and achieved beamforming accuracy within 0.5 dB of the ideal continuous beamforming.

Dai et al. designed a RIS prototype system operating at 2.3 GHz for indoor OTA testing, with a transmitter-receiver separation of 20 meters\cite{dai2020reconfigurable}. Experimental results demonstrated that the designed RIS prototype achieved a 21.7 dBi antenna gain in the 2.3 GHz frequency band, and the receiver was capable of real-time high-definition virtual reality (VR) video playback. Similarly, Pei et al. conducted relevant OTA experiments in both indoor and outdoor environments\cite{pei2021ris}. In the indoor setting, the experiment considered an NLOS scenario, where a 30 cm thick concrete wall separated the transmitter and receiver. Introducing the RIS resulted in an approximately 26 dB increase in received power when the transmitter power was set to -16 dBm, affirming the effectiveness of RIS in NLOS scenarios. In the outdoor environment, the authors selected two types of transmission test scenarios, specifically at 50 meters and 500 meters. In the outdoor experiment at 50 meters, the RIS provided a 27 dB power gain to the receiver compared to copper plates. Likewise, in the outdoor experiment at 500 meters, despite the challenges of long distances and low power, the RIS board yielded a 14 dB gain. Moreover, during RIS-assisted communication, it was possible to stream 1920 $\times$ 1080 resolution videos in real-time and with smooth playback.

\begin{figure}
\centerline{\includegraphics[width=.9\columnwidth]{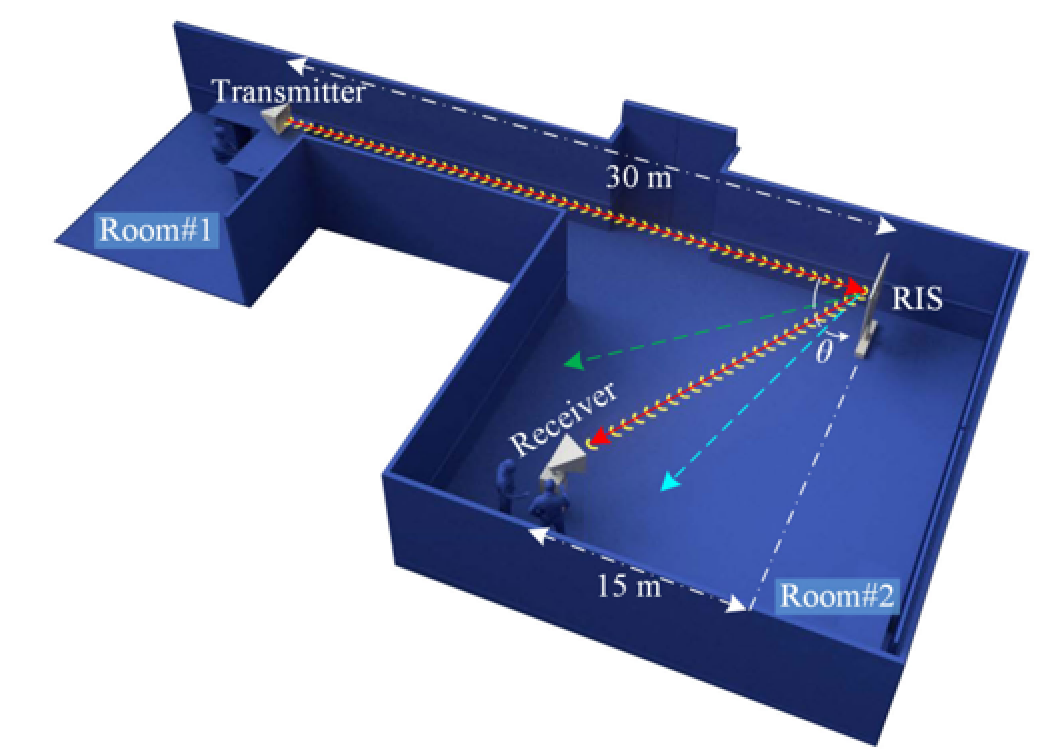}}
\caption{Utilizing RIS for blind spot coverage~\cite{araghi2022reconfigurable}.}
\label{F2-5}
\end{figure}

Araghi et al. designed an experiment in which they positioned the receiver in the blind spot of the transmitter's signal\cite{araghi2022reconfigurable}. They utilized RIS to establish the connection, thereby confirming the role of RIS in providing connectivity in coverage blind spots. The experimental setup is depicted in Fig.\ref{F2-5}.

Gros et al. reported two important experiments\cite{gros2021reconfigurable}. The first experiment compared their designed RIS with a typical reflective antenna in a near-field configuration. In this scenario, the RIS was highly effective for millimeter-wave beamforming, achieving directional gains of nearly 30 dBi with an effective angle of up to 60°. The second experiment utilized RIS to provide millimeter-wave wireless transmission in a far-field configuration, where two horn antennas were connected to a Vector Network Analyzer (VNA) without a line-of-sight link. This experiment demonstrated how RIS could create a line-of-sight link, resulting in a 25 dB gain in received energy.

Furthermore, Rains et al. conducted three indoor coverage enhancement experiments to validate the effectiveness of RIS in addressing indoor signal blockage issues using their designed RIS prototype\cite{rains2022high}. In the first experiment within a hall scenario, the RIS was positioned in the corner of the room, establishing a LOS link between the transmitting horn antenna connected to a Universal Software Radio Peripheral (USRP) and the RIS. However, there was an NLOS link between the transmitting and receiving horn antennas. Consequently, the RIS provided an additional link between the transmitting and receiving horn antennas to ensure the quality of wireless communication. The experimental results demonstrated that when the receiving horn antenna was approximately 12 meters away from the RIS, the received signal strength increased by approximately 17 dB. The second experiment, conducted at a corridor intersection, resembled the first experiment in that the RIS acted as a passive relay. In this scenario, the wireless signal received at a distance of about 18 meters from the RIS exhibited a gain of 30.6 dB. In the third experiment, RIS was employed for signal relay between floors. In this setup, the receiving horn antenna and RIS were both placed in a common lounge on the first floor, with a LOS link between them. The transmitting horn antenna was located on an intermediate level above the second floor, directly facing the RIS through two windows. In this scenario, RIS provided a 21.13 dB signal gain to the receiving horn antenna.

Keykhosravi et al. conducted experiments to address RIS positioning issues in a SISO system\cite{keykhosravi2022leveraging}. They used two transceiver modules, BFM6010, operating at 60 GHz, to simulate two RIS units. These simulated RIS units were equipped with 8 $\times$ 3 array antennas, and the user's position was determined at the intersection point of the two array antennas. 

\begin{figure}[t!]
    \centering

    \subfloat[]{
		\includegraphics[width=.4\columnwidth]{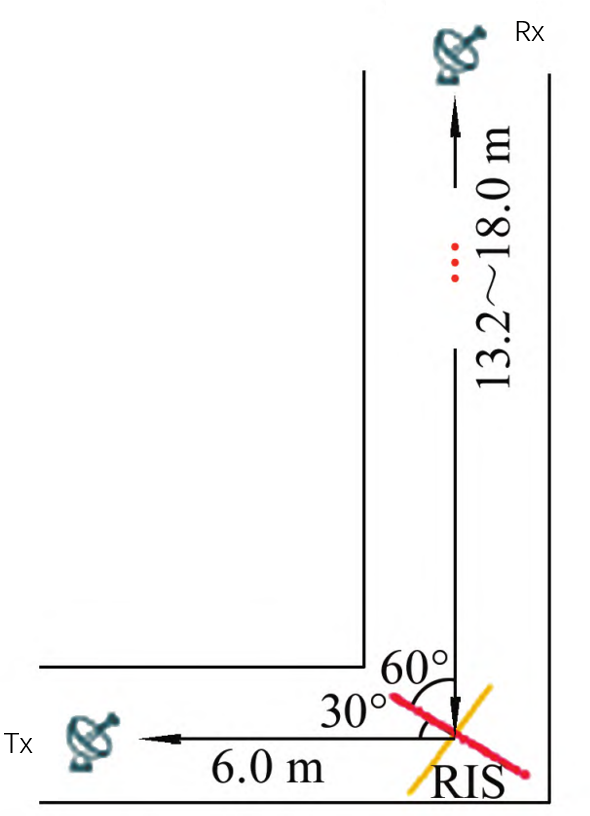}}
    \subfloat[]{
		\includegraphics[width=.4\columnwidth]{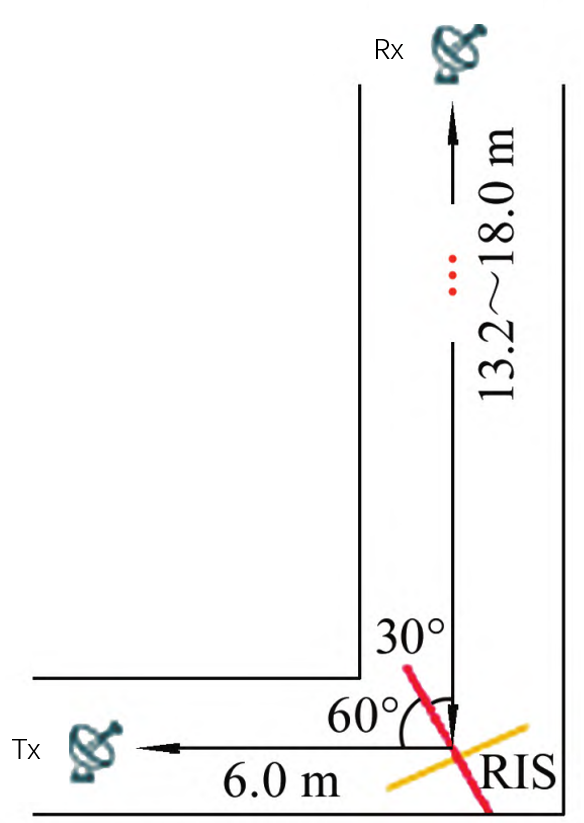}}
    \caption{Signal coverage in the `L' shaped corridor~\cite{fei2022research}.}
    \label{F2-6}
\end{figure}

Cui et al. reported on the establishment of an NLOS scenario in an L-shaped corridor\cite{fei2022research}. Through their experiments, they demonstrated that RIS could provide a maximum signal gain of up to 30 dB, thereby confirming the coverage enhancement capability of RIS. The scene diagram is depicted in Fig.\ref{F2-6}.

Ouyang et al. introduced a novel approach by incorporating computer vision into the RIS system\cite{ouyang2022computer}. They installed cameras on the RIS to capture visual information from the surrounding environment. The RIS then utilized this information to identify the desired direction of reflected beams and adjusted the reflection coefficients based on pre-designed codebooks. In contrast to traditional methods that rely on channel estimation or beam scanning to obtain reflection coefficients, this approach not only reduces beam training overhead but also eliminates the need for additional feedback links. Experimental validation with a prototype system demonstrated the system's ability to rapidly adjust reflection coefficients using visual information, enabling dynamic beam tracking.

Trichopoulos et al. integrated a 160-element RIS operating at 5.8 GHz into a wireless communication system and evaluated beamforming gains, path loss, and coverage improvements in real outdoor communication scenarios\cite{trichopoulos2022design}. When both the transmitter and receiver employed directional antennas, the RIS provided signal-to-noise ratio (SNR) gains in the range of 15 to 20 dB within a $\pm 60^{\circ}$ angular range. In terms of coverage, considering the presence of obstructions between the base station and mobile users in the far-field experiments, with an average signal path of 35 meters, the RIS offered an average 6 dB (up to 8 dB maximum) SNR improvement over an area exceeding 75 square meters.

Sang et al. reported the world's first field trials of RIS enhancing coverage performance in current 5G commercial mobile networks\cite{sang2022coverage}. These trials marked the first time that manufactured RIS was deployed in urban areas. The results of the field trials showed a significant improvement in user experience by deploying RIS into the existing China Mobile (Jiangsu) 5G communication network. This improvement was observed in terms of expanded coverage and increased throughput, potentially providing evidence for further widespread applications.

In summary, researchers have designed multiple experiments to validate the signal coverage enhancement role of RIS in wireless communication, as well as its integration with other technologies to achieve additional functionalities. Researchers have successfully achieved real-time data transmission in MIMO systems based on RIS, non-line-of-sight communication and signal coverage in blind spots, and beamforming using RIS to achieve directional gain. Furthermore, researchers have explored the application of RIS to address various extended challenges, such as solving SISO positioning problems and combining RIS with visual technology for dynamic beam tracking. Experimental results have demonstrated that RIS can provide high antenna gains, increase signal strength, and have significant potential to improve signal-to-noise ratio, mitigate path loss, and extend coverage. Through the deployment and control of RIS, wireless communication performance can be enhanced, coverage expanded, and user experience improved. These research findings provide compelling evidence for the promising future applications of RIS.

\begin{figure*}[h!]
    \centering

    \subfloat[Theoretical calculation results of\\ codebook  `00000000']{
		\includegraphics[width=.6\columnwidth]{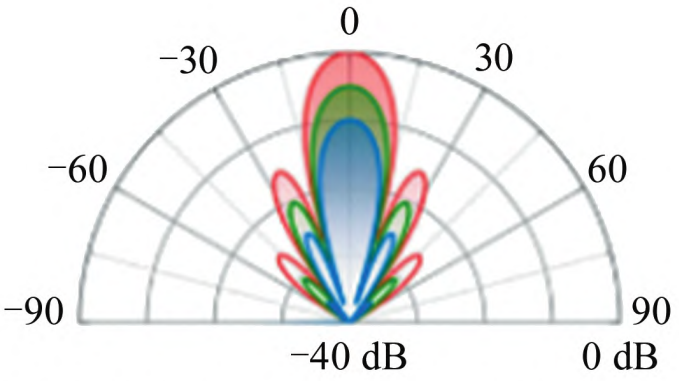}}
    \subfloat[Theoretical calculation results of\\ codebook  `00001111']{
		\includegraphics[width=.6\columnwidth]{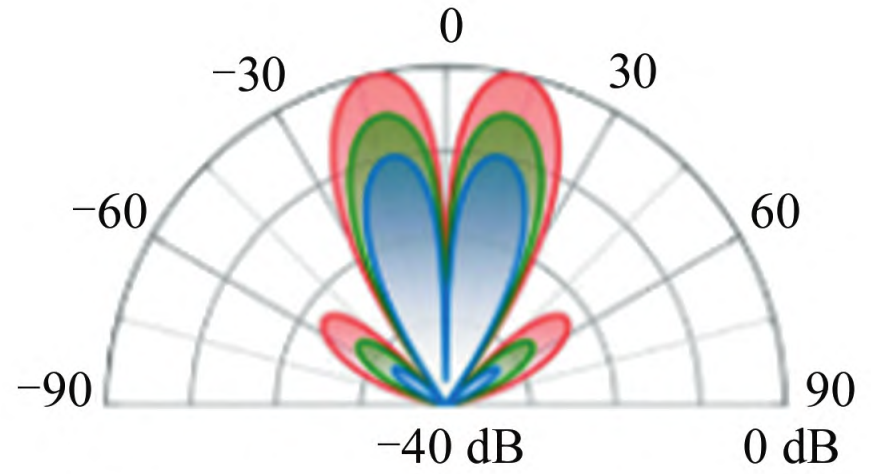}}
    \subfloat[Theoretical calculation results of\\ codebook  `00110011']{
		\includegraphics[width=.6\columnwidth]{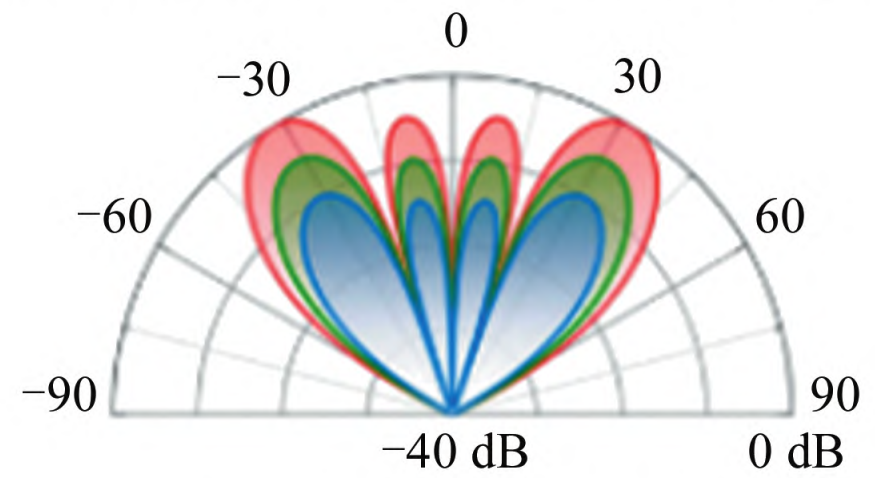}}
    \\
    \subfloat[Actual measurement results of\\ codebook  `00000000']{
		\includegraphics[width=.6\columnwidth]{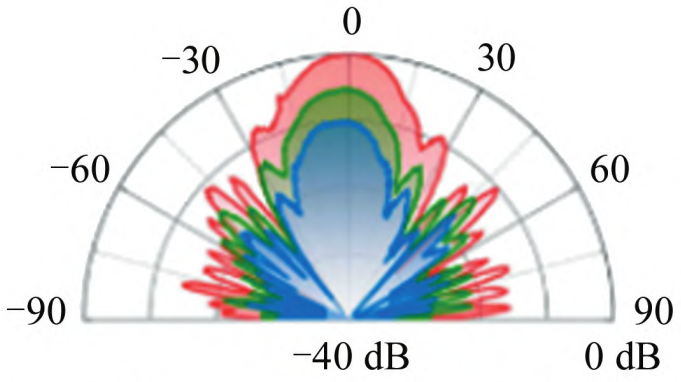}}
    \subfloat[Actual measurement results of\\ codebook  `00001111']{
		\includegraphics[width=.6\columnwidth]{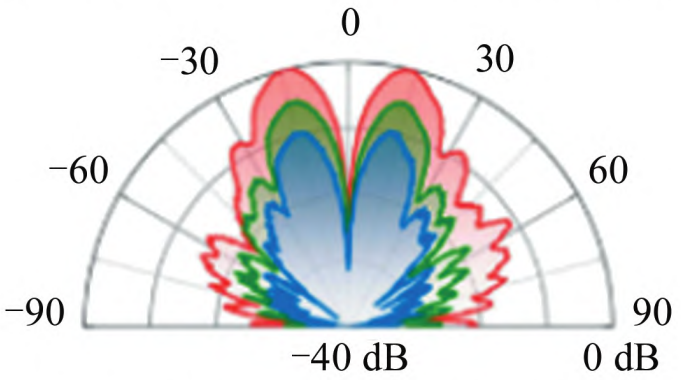}}
    \subfloat[Actual measurement results of\\ codebook  `00110011']{
		\includegraphics[width=.6\columnwidth]{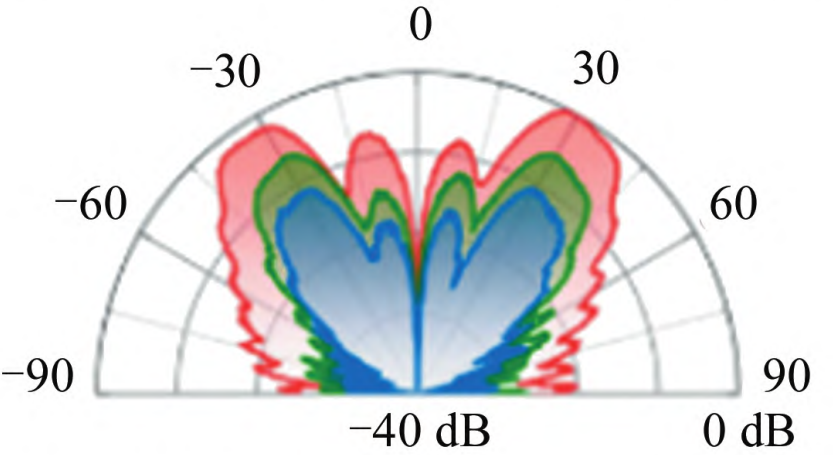}}
    \caption{Scattering of positive first-order harmonic harmonics~\cite{dai2018independent}.}
    \label{F2-7}
\end{figure*}
\subsection{Experimental Test - Space-time Modulation}
In addition to signal coverage, prototype systems for space-time modulation have also been studied in recent years, and related experimental tests have been conducted. 

Dai et al.~\cite{dai2018independent} conducted two experiments to utilize RIS for independent control of harmonic amplitude and phase. The experiments are based on an  $8\times8$ array of time-domain digitally encoded RIS units, with each unit equipped with a variable-capacitance diode. By adjusting the bias voltage of the diodes, precise adjustments in the phase response of the array could be achieved over a wide phase range (~\textasciitilde $270^{\circ}$).

Both experiments are conducted at 3.7 GHz. In the first experiment, two horn antennas are used to illuminate the RIS and receive reflected signals. Microwave signal generators and a spectrum analyzer are connected to the horn antennas via phase-stable cables to monitor the nonlinear characteristics of the metasurface under different bias voltages and modulation periods across a broad spectral range. The experimental results are compared to theoretical calculations. The results showed that different bias voltages influenced the metasurface's harmonic amplitude and phase. Different modulation periods affected the frequency spacing between harmonics. In the second experiment, time delays of 0 (0 $\mu$s) and T/2 (3.2 $\mu$s) represented the phase-opposite coding elements 0 and 1. The reflection phase of all columns in the metasurface can be described using binary coding sequences of 0 and 1. For three different metasurfaces with codes 00000000, 00001111, and 00110011, the first-order harmonic of the scattered beam was measured along the normal direction. The results(results shown in Fig.~\ref{F2-7}) confirm that different space-time codes could change the scattering pattern of harmonics. By changing the combinations of bias voltages, it is possible to achieve the attenuation of the scattered amplitude, as shown by the red, green, and blue lines in Figure ~\ref{F2-7}, without altering the scattering pattern of harmonics.

Zhang et al.~\cite{zhang2019breaking} proposed a 2-bit space-time encoding RIS capable of breaking Lorentz reciprocity to achieve spatial and spectral separation of wave reflections. They use a well-designed space-time coding sequence to demonstrate that the RIS had both spatial and temporal phase gradients, breaking time-reversal symmetry and inducing nonreciprocal wave reflection. To validate the nonreciprocal effect, the authors designed a $16 \times 8$ units 2-bit space-time encoding programmable metasurface (RIS). Each column of the metasurface consisted of 8  units, sharing a control voltage. Each unit uses two PIN diodes to connect hexagonal patches to two bias lines. The unit's parameters were designed to achieve a $90^{\circ}$ phase difference when switching between OFF-OFF, ON-OFF, OFF-ON, and ON-ON states, corresponding to 0, 1, 2, and 3 encoding states. Reflectivity tests show that at a frequency of 9.5 GHz, the phase difference between adjacent 2-bit encoding states is approximately $90^{\circ}$, and the response amplitude exceeded 0.79. Verification experiments are conducted in a microwave anechoic chamber. The prototype is initially illuminated by a transmitting (Tx) horn antenna placed along the $\theta_1$ = $34^{\circ}$ (port 1) direction, and the excitation signal has a frequency of 9.5 GHz. The result displays the scattering pattern in the forward scenario, with the primary reflected beam having a frequency of 9.50125 GHz and an angle of approximately $0^{\circ}$. For the time-reversed scenario, where the prototype is positioned with respect to the $\theta_2$=$0^{\circ}$ direction (port 2) and illuminated by the Tx horn antenna with an excitation signal at 9.50125 GHz, the main reflected beam has a frequency of 9.50250 GHz and an angle of around $34^{\circ}$. The measurement setup is designed to inspect the spectral power distribution, allowing the nonreciprocal effect to be clearly observed in the frequency domain.

\begin{figure}
\centerline{\includegraphics[width=.8\columnwidth]{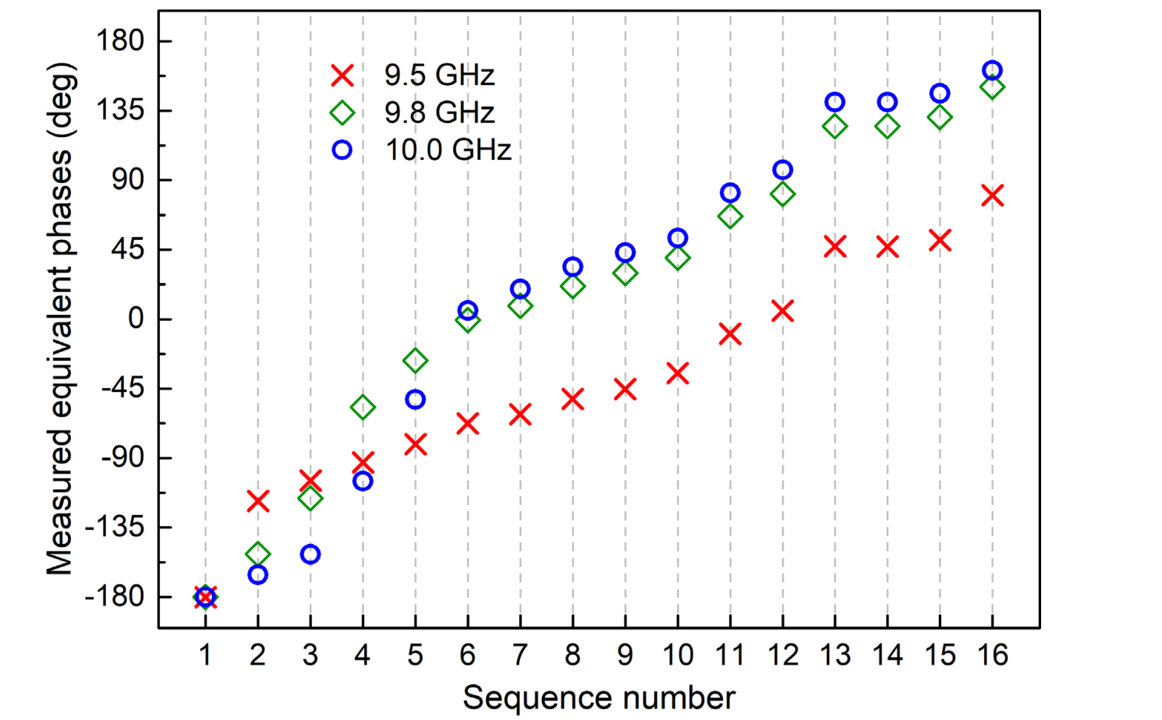}}
\caption{Measured equivalent phase~\cite{zhang2019dynamically}.}
\label{F2-8}
\end{figure}

\begin{figure}
\centerline{\includegraphics[width=.8\columnwidth]{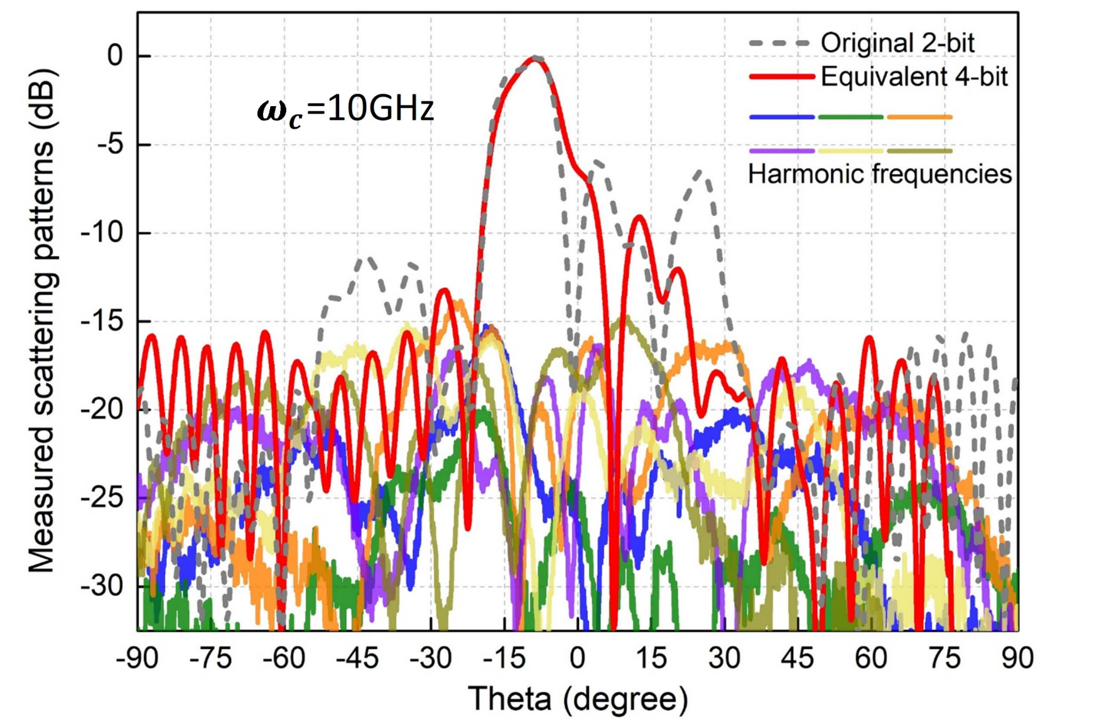}}
\caption{Electromagnetic wave scattering pattern~\cite{zhang2019dynamically}.}
\label{F2-9}
\end{figure}

Zhang et al.~\cite{zhang2019dynamically} introduced a method based on 2-bit time-domain coding for RIS, allowing for multi-bit or even near-continuous phase modulation at the central frequency or harmonic frequency. By employing vector synthesis techniques and designing appropriate 2-bit time coding sequences, the equivalent phase can be synthesized to achieve any value with a $360^{\circ}$ phase coverage. Furthermore, in beamforming applications, the use of 16 sets of time coding sequences can generate an equivalent 4-bit phase, demonstrating the advantages of reducing quantization levels. This literature describes two experiments. 
In the first experiment, by configuring the time coding sequences, a 2-bit programmable intelligent surface (RIS) can achieve 4-bit digital phases at the central frequency. This results in an ideal set of $16$ states: $-180^{\circ}$, $-157.5^{\circ}$,$-135^{\circ}$, $-112.5^{\circ}$, $-90^{\circ}$,$-67.5^{\circ}$, $-45^{\circ}$, $-22.5^{\circ}$, $0^{\circ}$,$22.5^{\circ}$, $45^{\circ}$, $67.5^{\circ}$, $90^{\circ}$, $112.5^{\circ}$, $135^{\circ}$ and $157.5^{\circ}$).
In Experiment Two, the original 2-bit encoded phase and the synthesized equivalent 4-bit encoded phase are used for beamforming at the central frequency, and the scattering patterns of these two cases are compared. The unit cells on the prototype consisted of an irregular hexagonal metal piece and two metal strips printed on a grounded F4B substrate, with a thickness of 1.5 mm. Two PIN diodes are used to connect the hexagonal patches to the two rectangular strips, serving as the bias lines for the diodes. When the two diodes are switched between the four states: OFF-OFF(00), OFF-ON(01), ON-OFF(10), and ON-ON(11), the phase difference between adjacent encoding states was approximately $90^{\circ}$, at around 9.5 GHz, and the corresponding amplitudes are all above 0.79. The prototype used in the experiment consisted of two independent samples, each with 8 columns and 8 connected units. The results of Experiment One regarding the equivalent phase are shown in Fig.~\ref{F2-8}, where the equivalent 4-bit phase still covered a range from $-180^{\circ}$ to $180^{\circ}$ at the central frequency of 10 GHz.As seen in Fig.~\ref{F2-9}, the time-domain encoded RIS with the equivalent 4-bit phase exhibits better beamforming performance at 10 GHz compared to the original 2-bit situation, with significantly lower sidelobes. Additionally, scattering power at harmonic frequencies was well suppressed. (Where $\omega_c$=10 GHz is the central frequency).

\begin{figure*}[t!]
    \centering

    \subfloat[STC codebook matrix M3]{
		\includegraphics[width=.45\columnwidth]{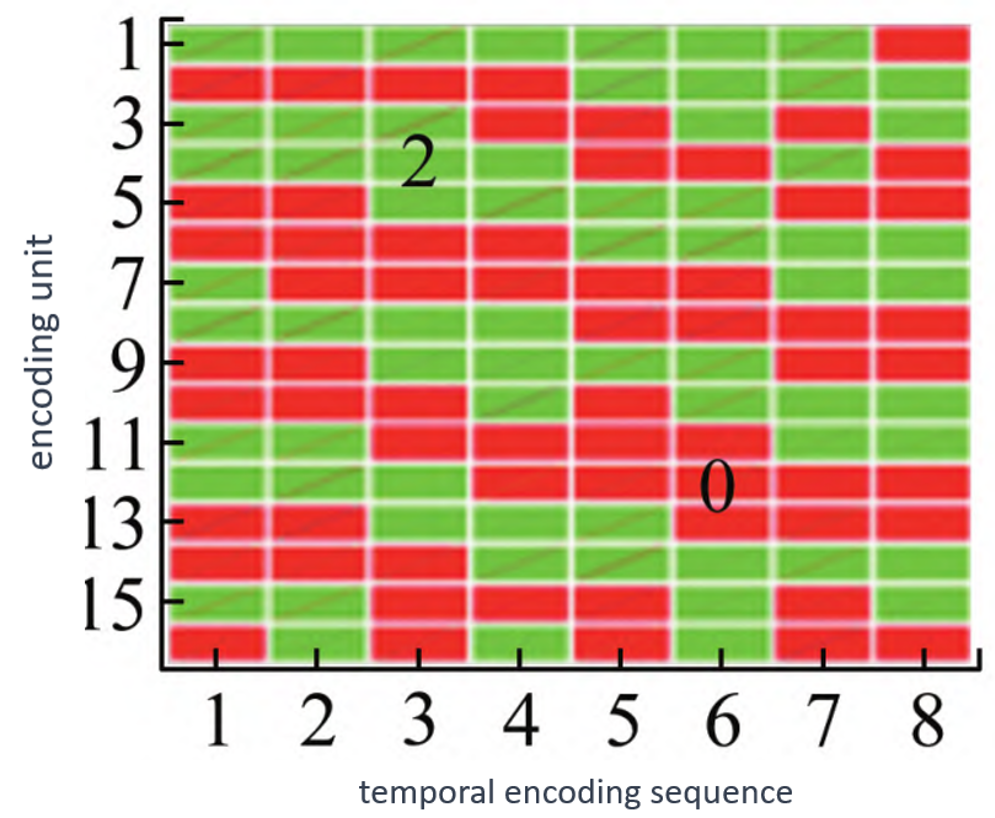}}
    \subfloat[STC codebook matrix M2]{
		\includegraphics[width=.478\columnwidth]{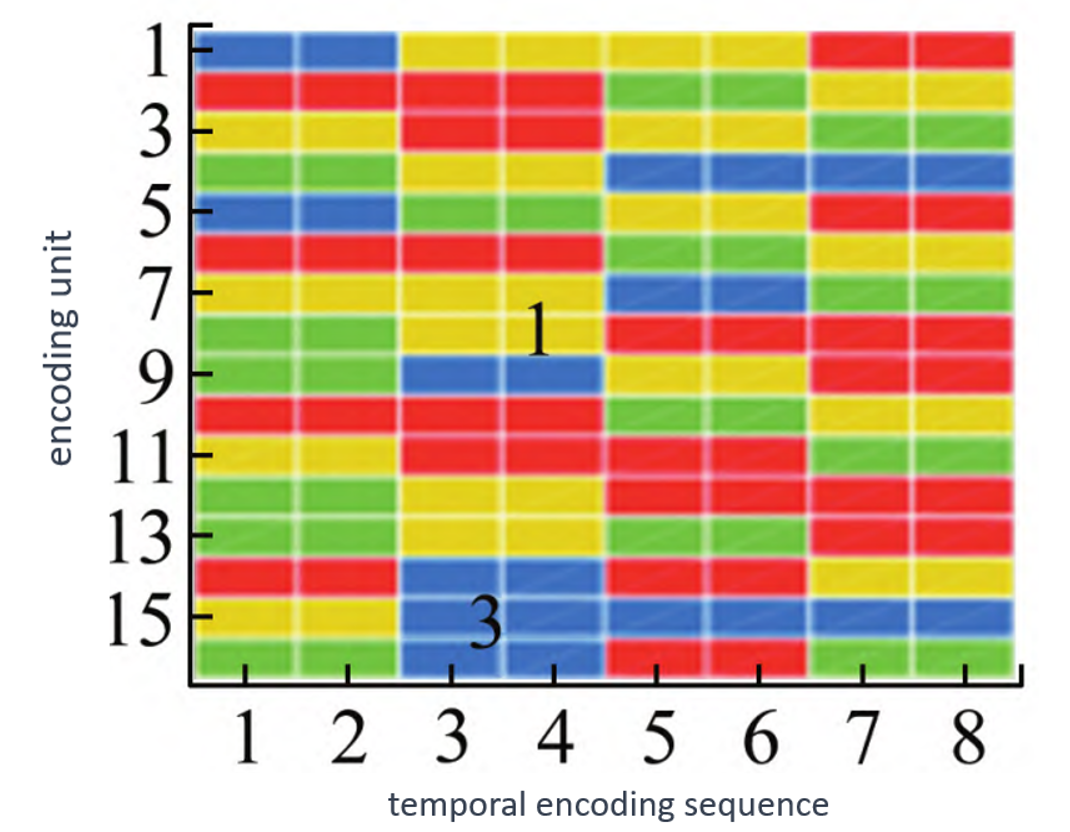}}
    \subfloat[STC codebook matrix M1]{
		\includegraphics[width=.45\columnwidth]{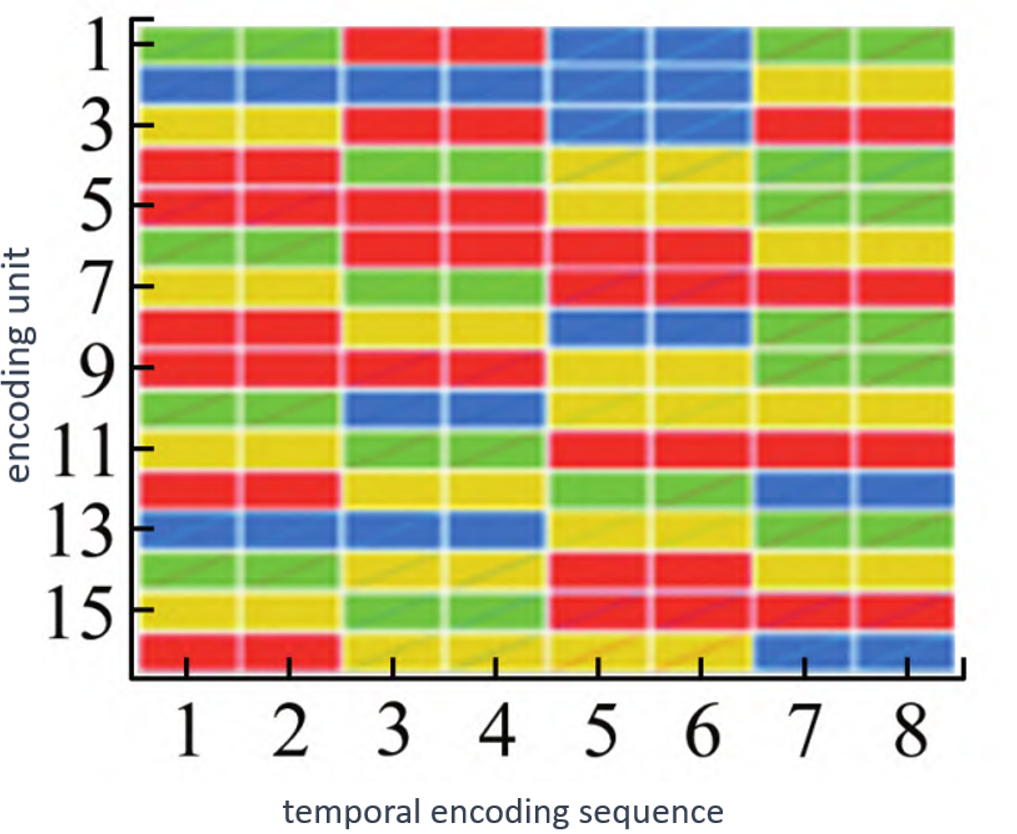}}
    \subfloat[STC codebook matrix M0]{
		\includegraphics[width=.46\columnwidth]{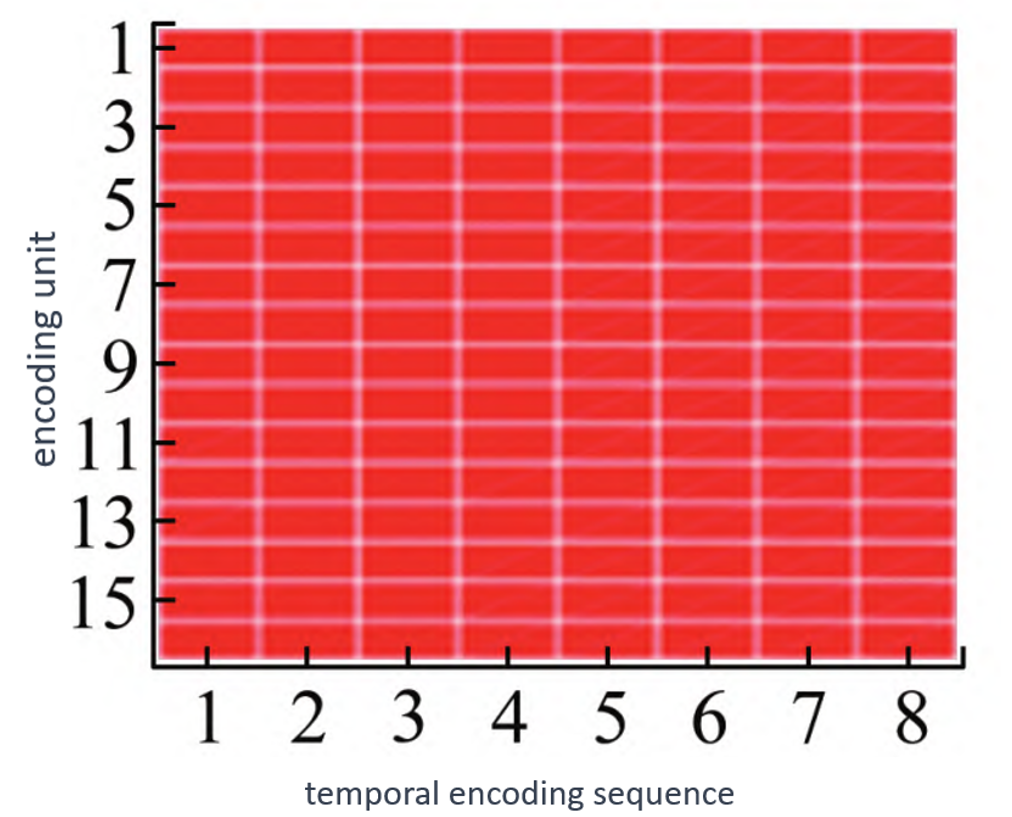}}
    \\
    \subfloat[Scattering pattern of M3]{
		\includegraphics[width=.47\columnwidth]{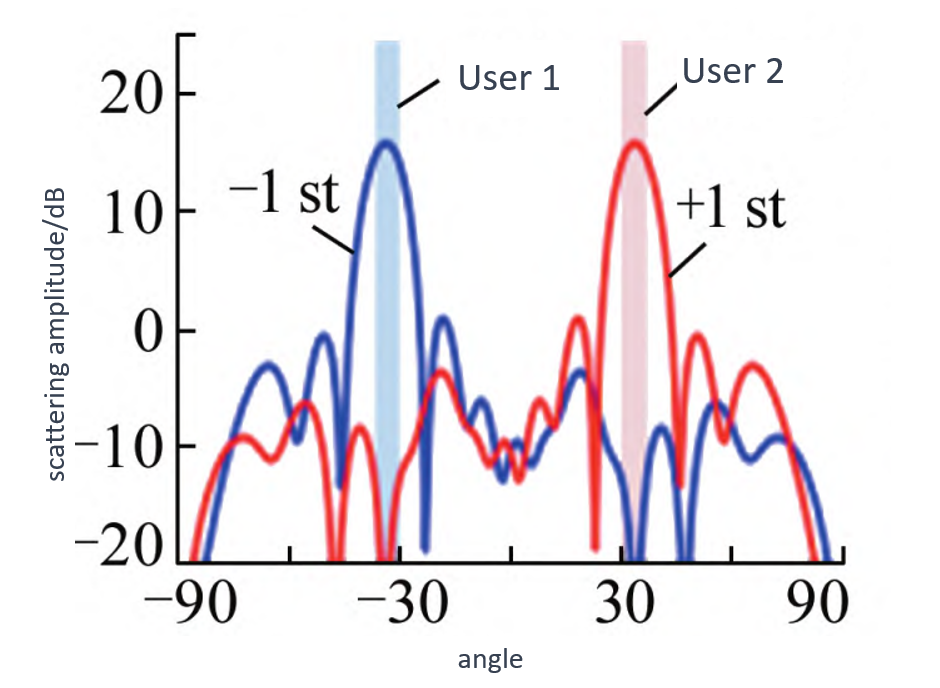}}
    \subfloat[Scattering pattern of M2]{
		\includegraphics[width=.43\columnwidth]{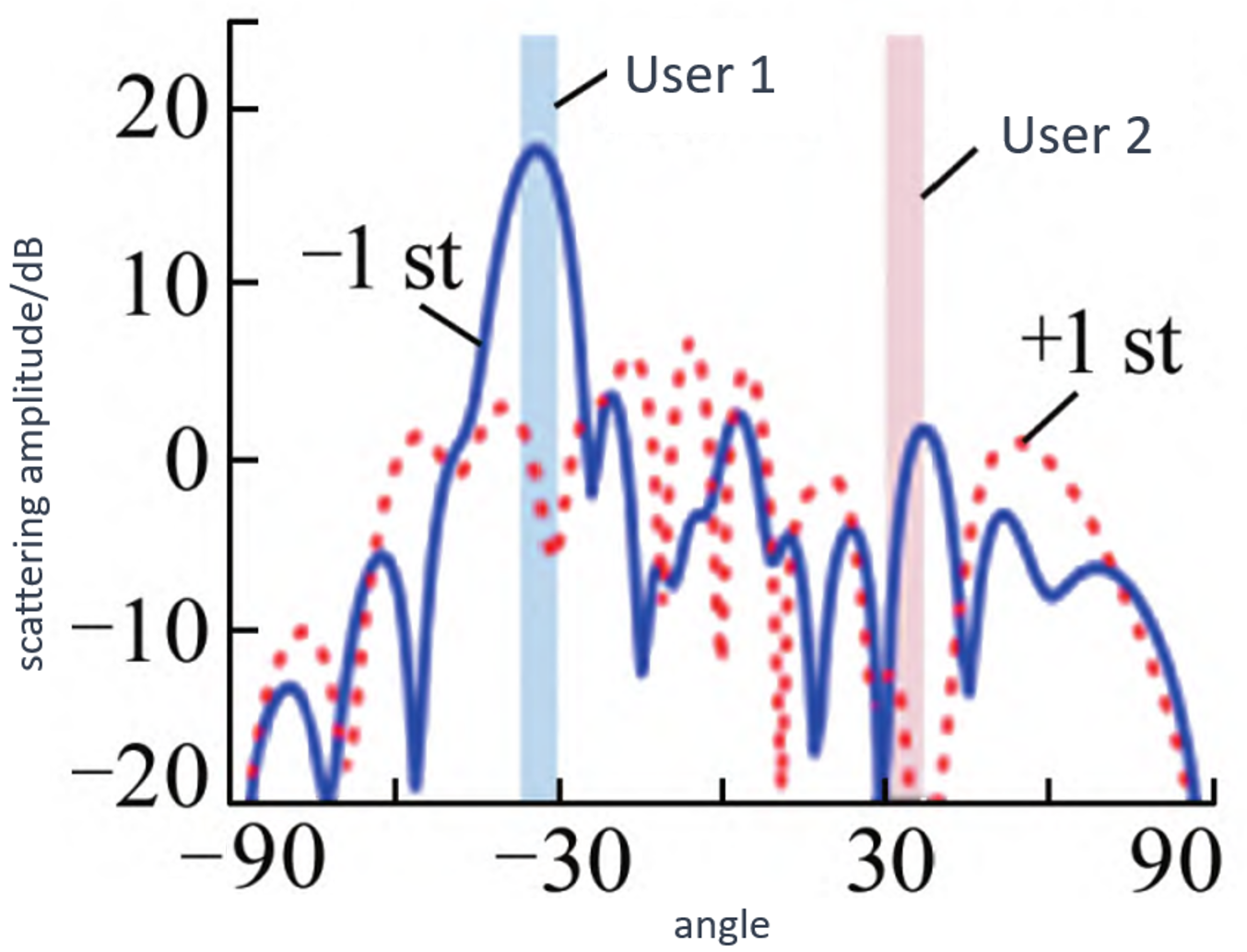}}
    \subfloat[Scattering pattern of M1]{
		\includegraphics[width=.45\columnwidth]{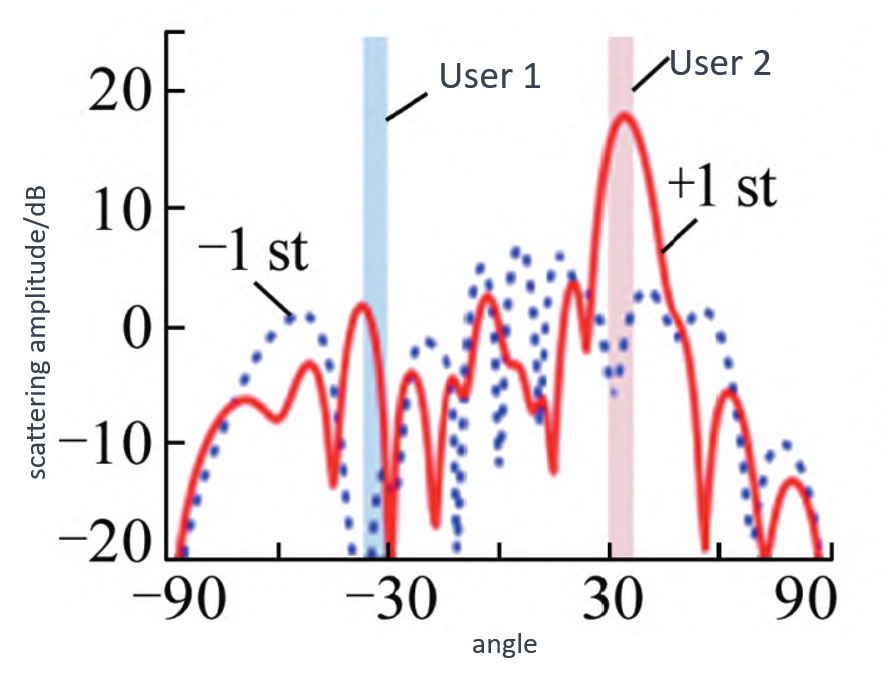}}
    \subfloat[Scattering pattern of M0]{
		\includegraphics[width=.45\columnwidth]{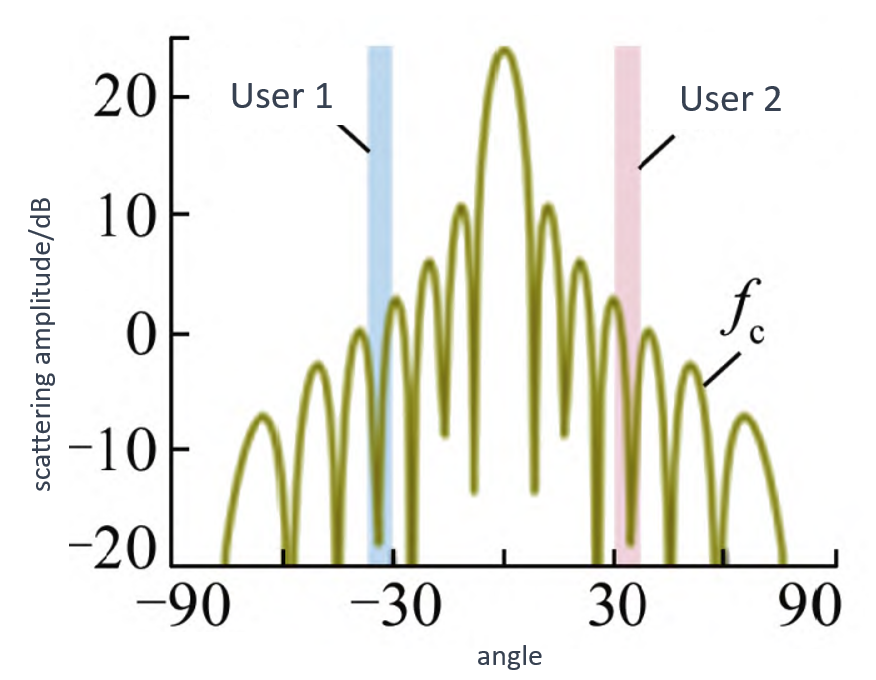}}
    \\
    \subfloat[Reception strength of M3]{
		\includegraphics[width=.45\columnwidth]{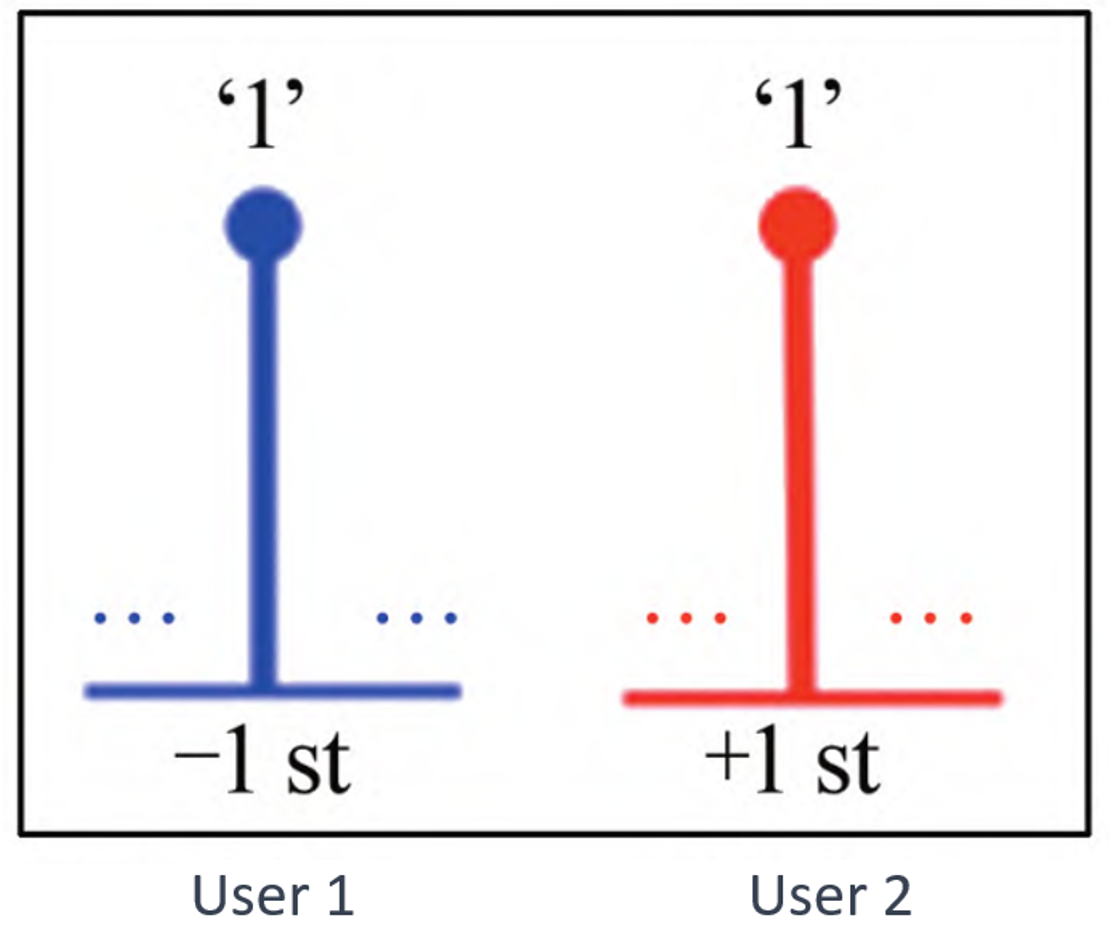}}
    \subfloat[Reception strength of M2]{
		\includegraphics[width=.465\columnwidth]{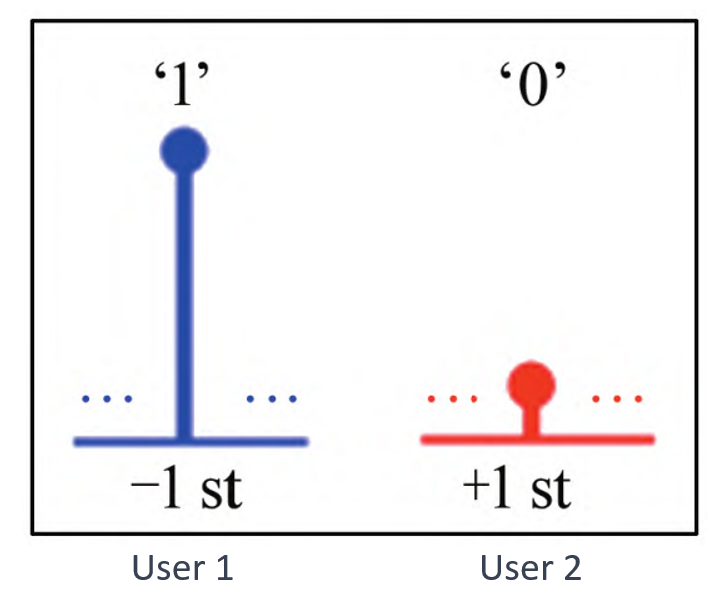}}
    \subfloat[Reception strength of M1]{
		\includegraphics[width=.45\columnwidth]{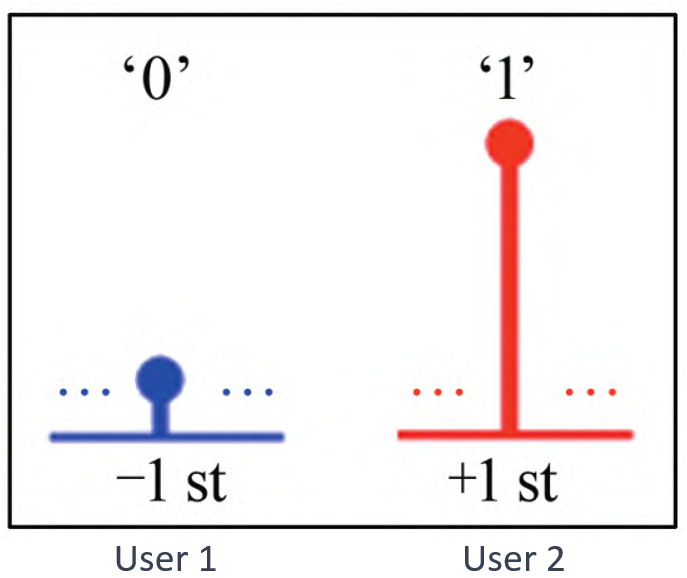}}
    \subfloat[Reception strength of M0]{
		\includegraphics[width=.455\columnwidth]{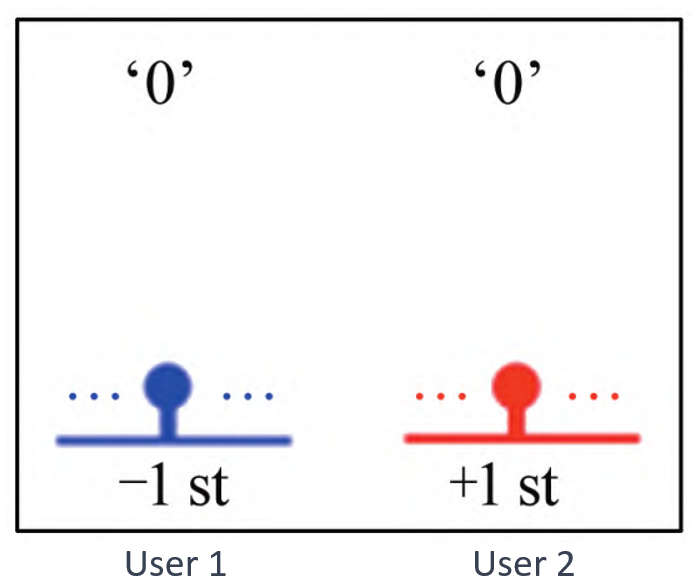}}
    \caption{4 types of space-time encoding matrices and corresponding first-order harmonic amplitudes~\cite{zhang2021wireless}.}
    \label{F2-10}
\end{figure*}

Zhang et al.~\cite{zhang2021wireless} introduced an information encoding scheme based on space-time encoding with RIS, which enables the control of electromagnetic waves in both spatial and frequency domains. In their experiments, On-Off Keying (OOK) modulation was utilized. They define that a carrier with low power, representing the first harmonic, indicates binary symbol 0, while a carrier with high power represents binary symbol 1. Using this approach, they independently transmit different images to two users in different directions.Fig.~\ref{F2-10} illustrates four space-time encoding matrices designed for modulation and provides theoretical simulations of the first harmonic amplitude. The experimental results demonstrate the capability of RIS to simultaneously and independently transmit two different images to users located in different positions.

Rajabalipanah et al.~\cite{rajabalipanah2021analog} designed a reprogrammable space-time digital metasurface (RIS) processor and established a simulation platform based on the RIS method. This platform, through the application of suitable temporal coding signals, enables the RIS to dynamically achieve the required phase and amplitude of the transfer function related to the selected operator at specific harmonic frequencies. They conducted experiments to provide several examples demonstrating the versatility of the proposed RIS processor in performing various mathematical operations and functions, such as integration, differentiation, and real-time edge detection.

\begin{figure}[htbp]
\centerline{\includegraphics[width=1\columnwidth]{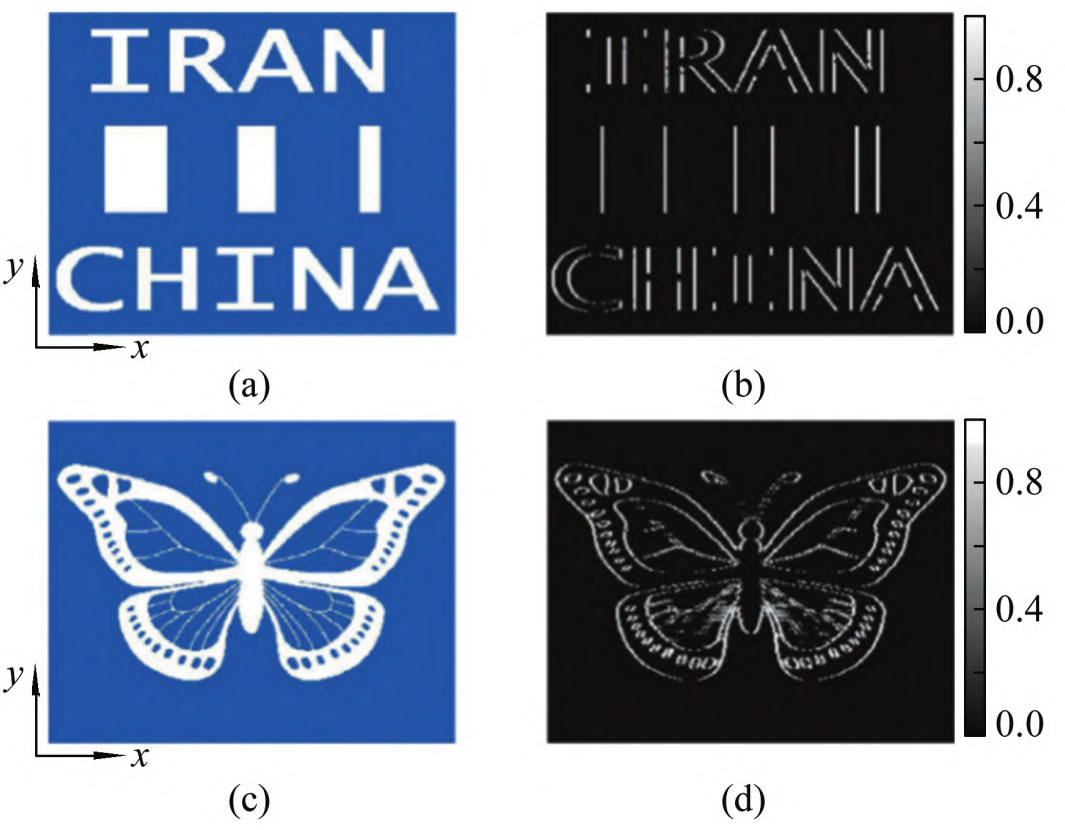}}
\caption{Use simulation functions to restore incident images~\cite{rajabalipanah2021analog}.}
\label{F2-11}
\end{figure}

They utilized an RLC model (R = 0.8 $\Omega$, L = 0.7 nH, and controllable capacitance) to represent the variable-capacitance diode in circuit simulations around the central frequency.
When different voltage levels were applied to the variable-capacitance diode, it exhibited four units with a constant $\pi/2$ phase difference, representing the reflective responses of the 00, 01, 10, and 11 encoding states. The experiment compares the actual experimental results with theoretical calculations, finding a close match between the experimental and theoretical outcomes. 
Fig.~\ref{F2-11} presents the experiment's use of simulate transfer functions for one-dimensional spatial filtering of the text `Iran China' and three rectangular shapes (Fig.~\ref{F2-11}(a)) as well as a butterfly image (Fig.~\ref{F2-11}(c)), illuminating the space-time coding RIS as two different images. As shown in Fig.~\ref{F2-11}(b) and (d), the experiment successfully achieves a normalized output field, completely displaying the contours of the incident images with the same intensity in the horizontal direction.

\begin{figure*}[h]
\centerline{\includegraphics[width=.8\linewidth]{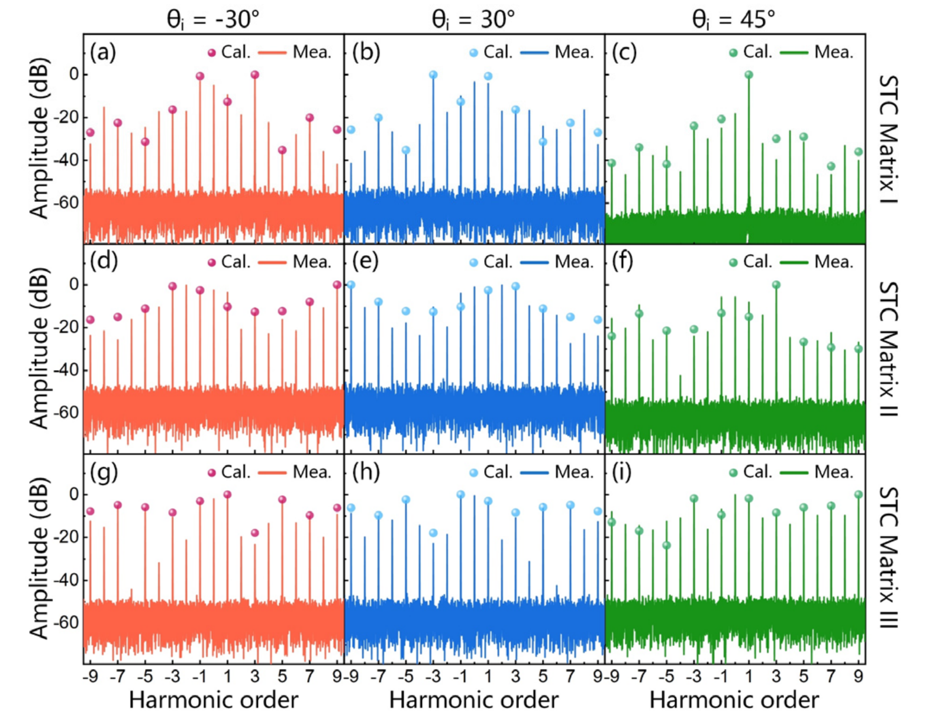}}
\caption{Direction finding harmonic signal spectrum~\cite{dai2022simultaneous}.}
\label{F2-12}
\end{figure*}

A mmWave space-time coding RIS system~\cite{dai2022simultaneous} integrates functions for the angle of arrival estimation and beamforming. It can estimate the incident angle and perform space-time encoding modulation. The authors first create a 1-bit space-time coding programmable metasurface (RIS) operating in the K-band. They use three different space-time encoding matrices to evaluate its directional capability. The experiments demonstrated that the designed space-time encoding matrices had an estimation error in the incident angle of less than 3 degrees. Fig.~\ref{F2-12} provides the spectral content of the measured directional harmonics in the reflected signal during the experiment, indicating that the measurements of odd harmonics for direction-finding are in good agreement with the calculated results. Furthermore, in all cases, the estimated angles and values are very close to the actual values, with absolute errors less than 3 degrees. The results suggest that for small incident angles, the estimated results are in good agreement with the actual values for all three space-time coding matrices. However, when the incident angle is larger, estimation errors rapidly increase. Additionally, the results indicate slight variations in accuracy among the three space-time coding matrices.

In summary, research in space-time modulation based on RIS is flourishing. The combination of RIS with space-time modulation extends the degrees of freedom of RIS from the spatial domain to the frequency domain. This allows control not only over the scattering or radiation characteristics of electromagnetic waves in the spatial domain but also over the spectral distribution characteristics in the frequency domain. With the aid of space-time modulation, RIS can achieve various pioneering tasks, such as breaking Lorentz reciprocity, achieving arbitrary phase synthesis with a $360^{\circ}$ phase coverage, and demonstrating generality in various mathematical operations and functionalities. These research achievements provide robust experimental evidence for the application of space-time modulation and showcase the potential value of RIS in the fields of communication and computation.

\subsection{Our Design and Experimental Tests}

\begin{figure}
  \centering
  \subfloat[]{
  \includegraphics[width=.9\columnwidth]{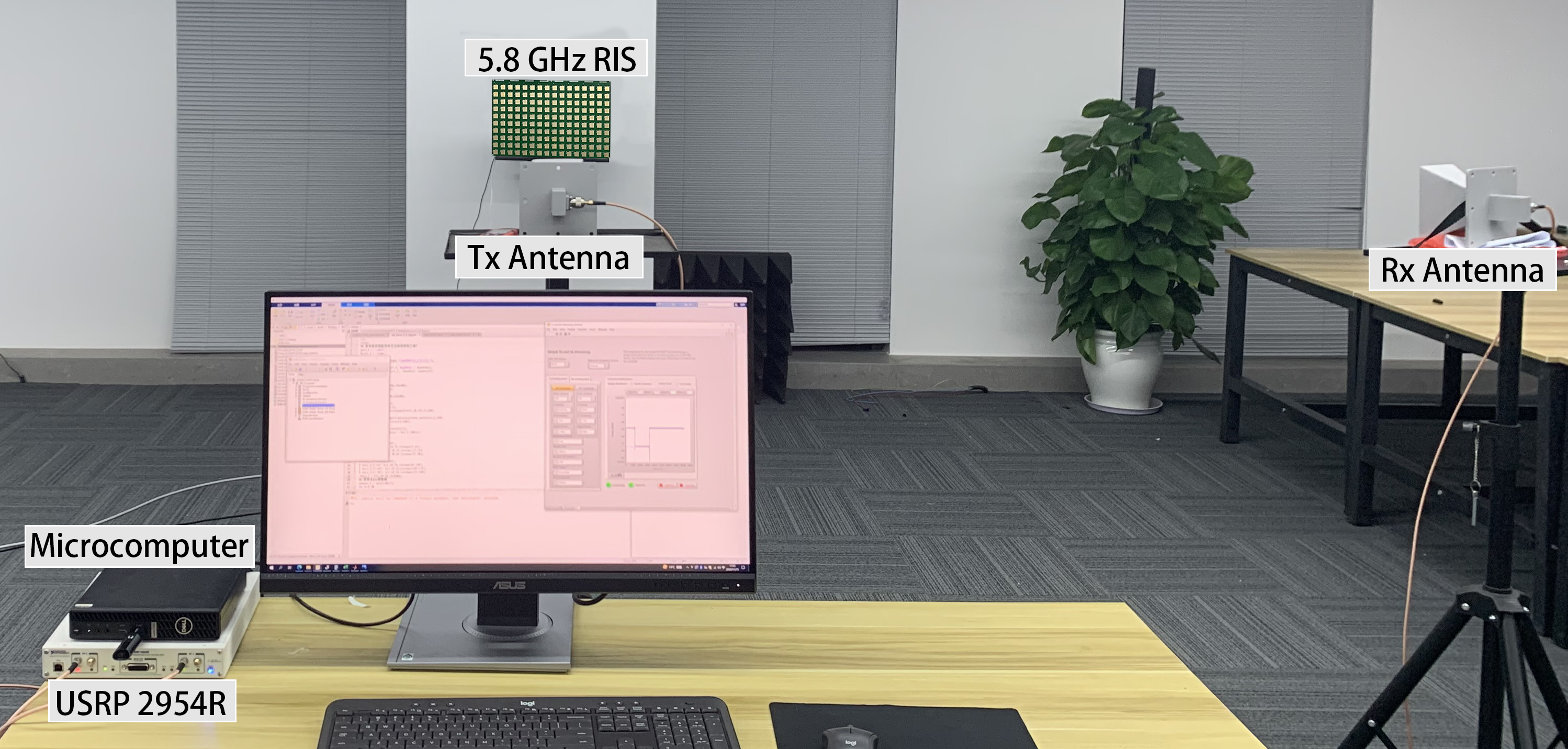}}
  \\
  \subfloat[]{
  \includegraphics[width=.9\columnwidth]{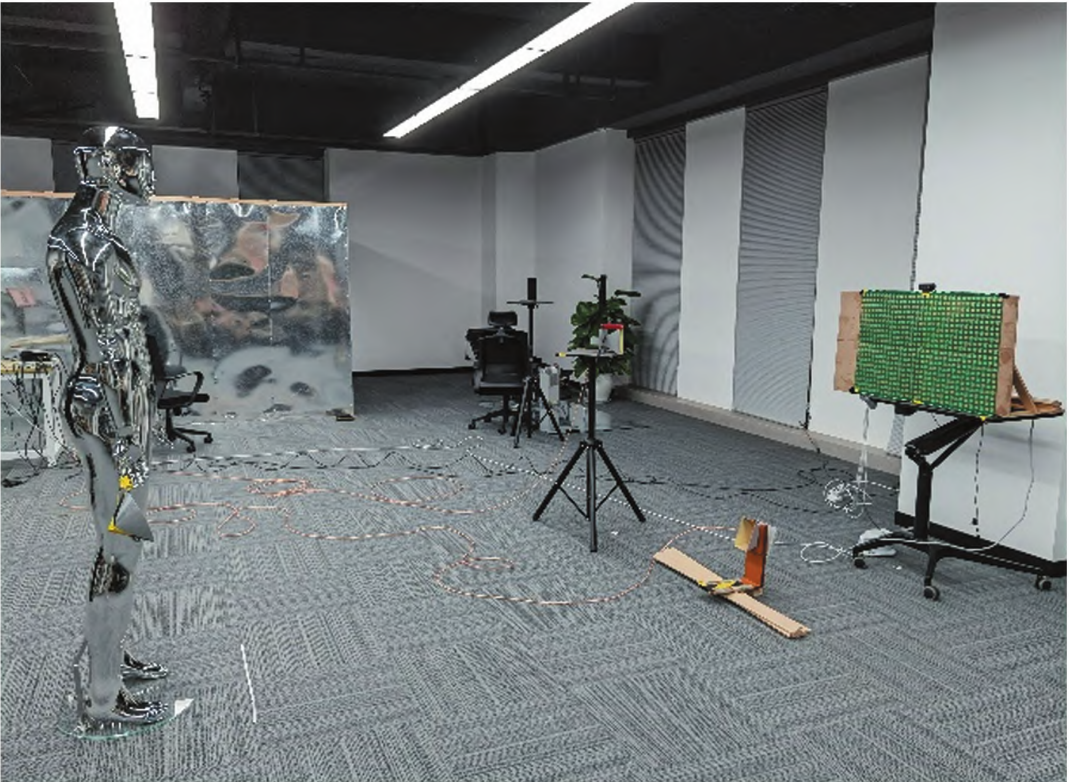}}
  \caption{The RIS prototype operating at a frequency of 5.8 GHz}
  \label{F2-13}
\end{figure}

\begin{figure*}
\centerline{\includegraphics[width=0.8\linewidth]{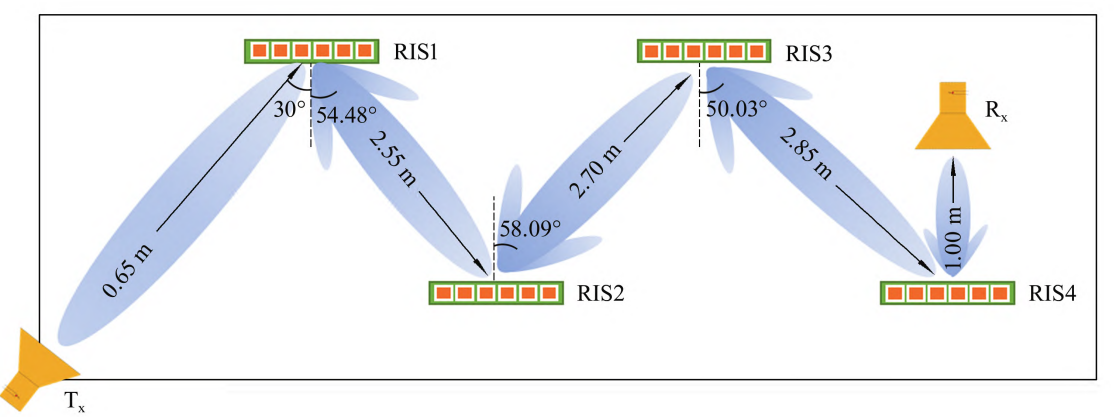}}
\caption{Multi-hop RIS}
\label{F2-14}
\end{figure*}

\begin{figure*}
  \centering
  \subfloat[-66 dBm]{
  \includegraphics[width=.35\linewidth]{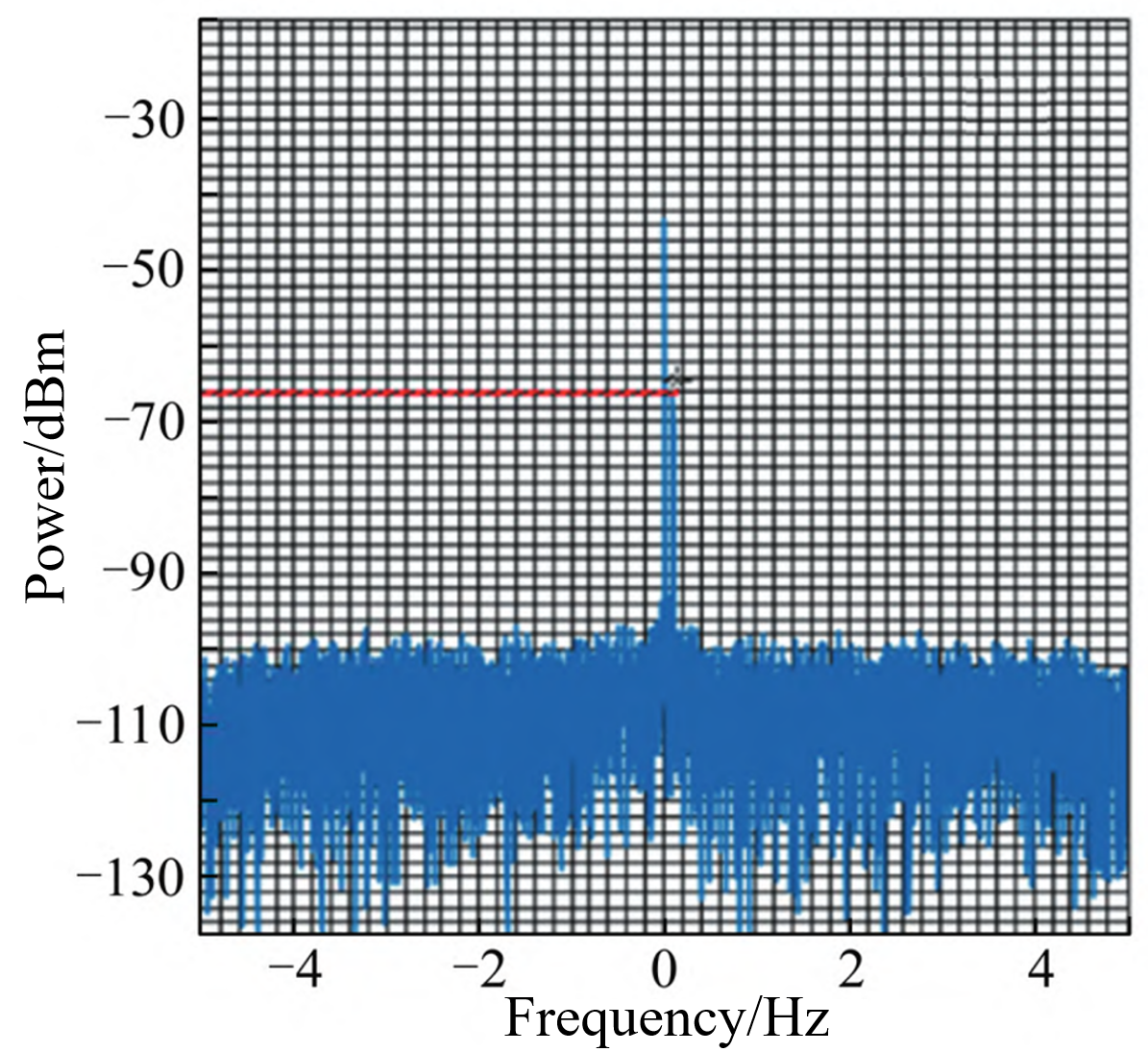}}
  \subfloat[-37dBm]{
  \includegraphics[width=.35\linewidth]{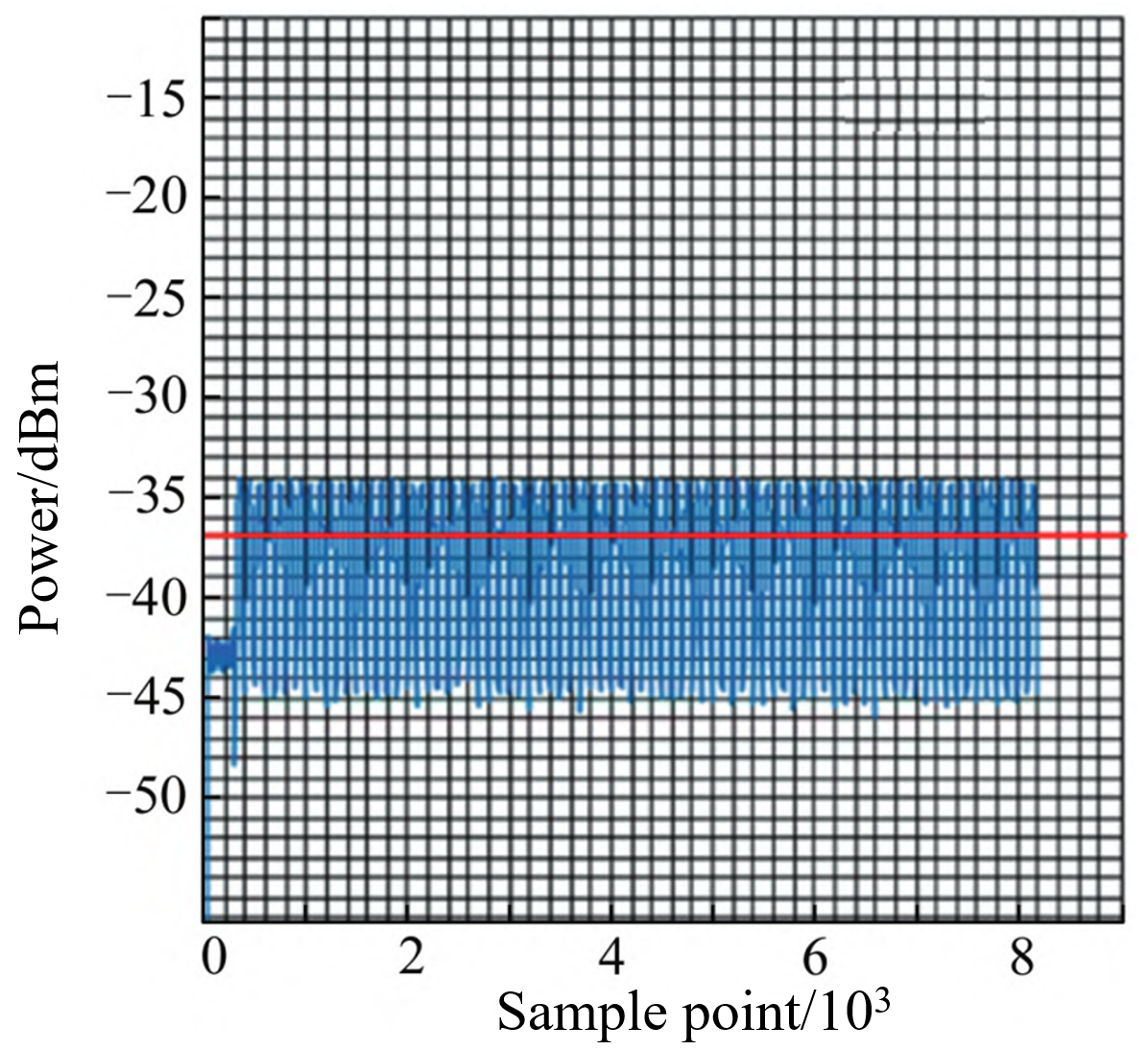}}
  \caption{Received signal strength without RIS-assisted (-66 dBm) and with 4 RIS-assisted (-37 dBm)}
  \label{F2-15}
\end{figure*}

\begin{figure*}
  \centering
  \subfloat[Signal gain measurement 1]{
  \label{F2-16-1}
  \includegraphics[width=0.81\linewidth]{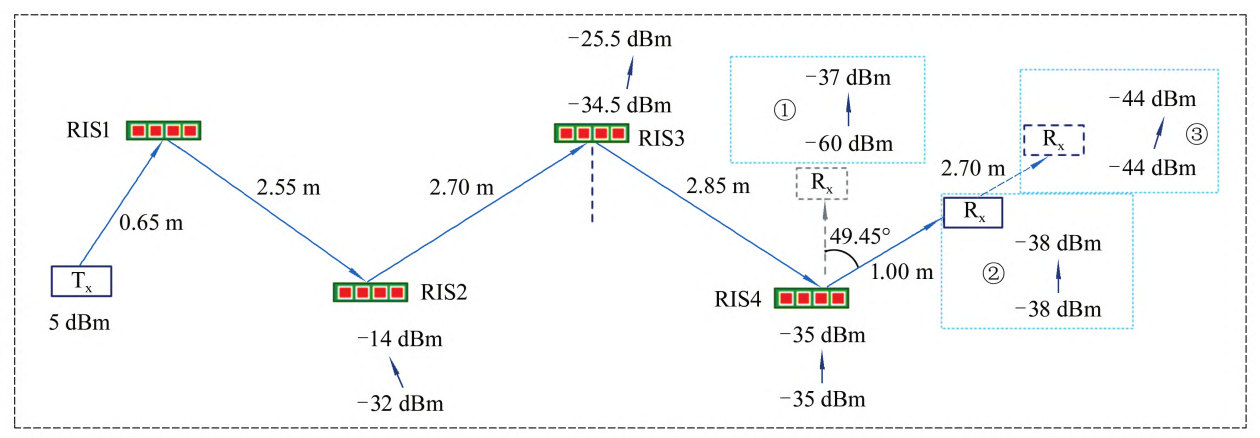}}
  \\
  \subfloat[Signal gain measurement 2]{
  \label{F2-16-2}
  \includegraphics[width=0.8\linewidth]{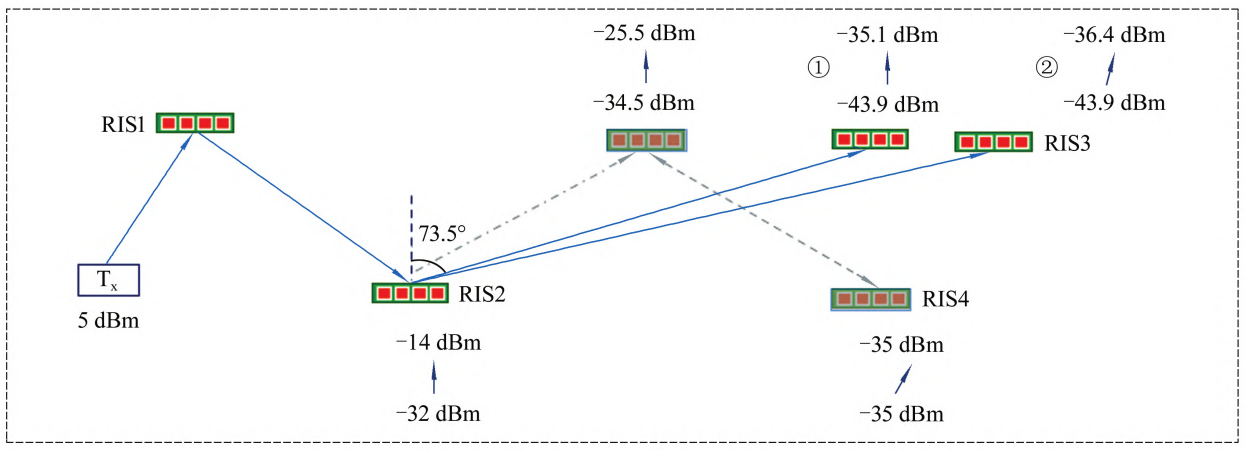}}
  \caption{Change in received signal power at each node.}
  \label{F2-16}
\end{figure*}

\begin{figure}
\centerline{\includegraphics[width=0.95\linewidth]{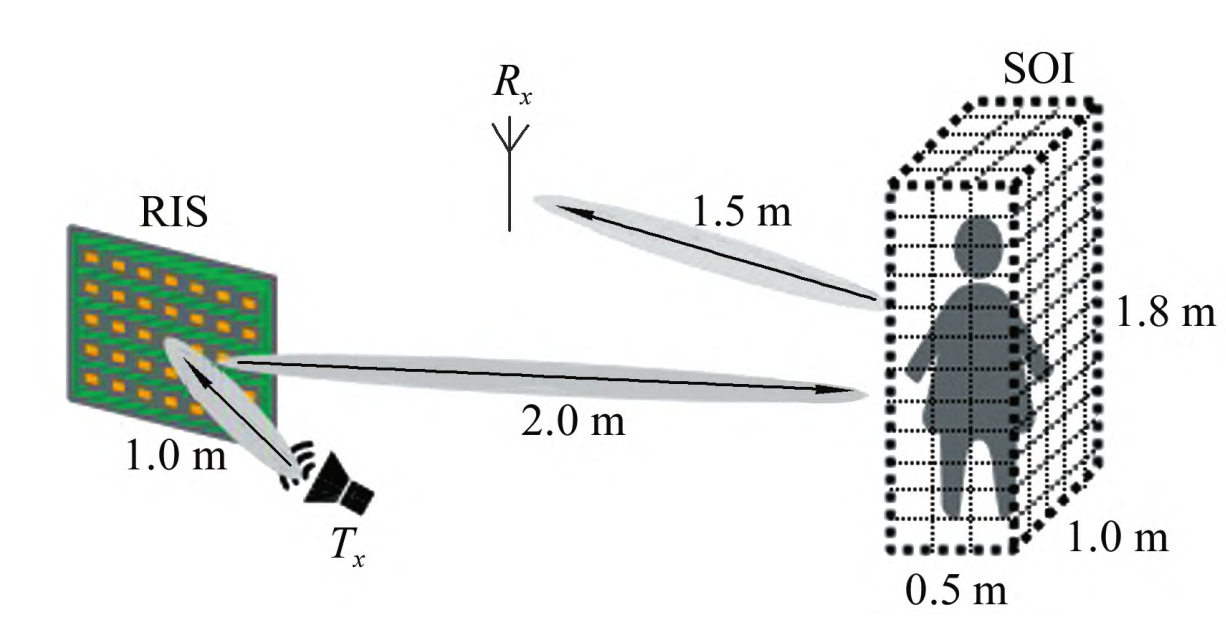}}
\caption{The gain in received signal power before and after the RIS working, assuming normal operation of the RIS's preceding node.}
\label{F2-17}
\end{figure}

To verify the effectiveness of the RIS-assisted wireless experimental system, we designed a microstrip antenna-based RIS prototype as shown in Fig.\ref{F2-13}. In the communication scenario testing, we use a beamforming method called Discrete Optimization Algorithm (DaS) proposed in reference \cite{xiong2022optimal}. This algorithm is specifically designed for the RIS system's discreteness, and for the first time, a globally optimal solution can be obtained in linear computational complexity under low-bit conditions. We conduct experiments on multi-hop RIS, multi-board RIS-assisted imaging, and indoor signal coverage enhancement. The single RIS board used in the experiment has 160 units in a $10\times16$ array, with an area of 400 mm$\times$ 265 mm.

\subsection*{(1) Dual and multi-hop trials}

Signal quality is poorer in long and winding spaces such as tunnels and mines. Employing RIS-based multi-hop propagation, multiple reflections can enhance signal coverage in such confined and curved spaces. The experimental concept is illustrated in Fig.\ref{F2-14}, where `Tx' represents the transmitting antenna, and `Rx' represents the receiving antenna. In this experiment, USRP is used for signal transmission and reception, and multiple RIS units are utilized to achieve relay-like functionality.

The experimental setup is as follows: the signal transmitting horn is positioned at a distance of 0.65 meters from the center of RIS\#1, while the distances between RIS\#1, RIS\#2, RIS\#3, and the receiving horn are 2.55, 2.70, 2.85, and 1 meter, respectively. Each RIS panel is oriented parallel to one another, with RIS\#1 and RIS\#3 on one side and RIS\#2 and RIS\#4 on the other. The angles formed between the centerlines of the four RIS panels and the normal plane of the RIS are $54.48^{\circ}$, $58.09^{\circ}$, and $50.30^{\circ}$, respectively. The direction of signal transmission from the horn forms a $30^{\circ}$ angle with respect to RIS\#1, while the direction of signal reception by the horn is perpendicular to RIS\#4.

According to DaS\cite{xiong2022optimal}, the beamforming phase shift matrix for each RIS panel is simulated and generated. Using wireless communication, the calculated optimal phase shift matrix is assigned to each RIS panel, enabling beamforming from each RIS panel to the center of the next hop RIS. Due to the directional characteristics of the horn antennas, prior to the placement of RIS panels, there is no direct line of sight path between the transmitting and receiving ends, resulting in a weak received signal with a strength of only -66 dBm. After deploying multi-hop RIS panels for signal transmission, an effective signal link is established, increasing the signal strength to -37 dBm, with a signal gain of 29 dB, as illustrated in Fig.\ref{F2-15}.

In addition, variations in signal strength before and after placing each RIS panel are also tested (RIS\#1 does not require measurement as there is no RIS gain on the link from the signal source to RIS\#1). The specific results are shown in Fig.\ref{F2-16}. The signal strength at the location of RIS\#2 increases from -32 dBm to -14 dBm, with a gain of 18 dB, after the optimal assignment of RIS\#1. Following the optimal assignment of RIS\#2, RIS\#3, and RIS\#4, the signal strengths at RIS\#3, RIS\#4, and the receiving end changes from -34.5 dBm to -25.5 dBm, from -35 dBm to -35 dBm, and from -60 dBm to -37 dBm, with gains of 9 dB, 0 dB, and 23 dB, respectively. The above results indicate that in most cases, RIS can provide significant signal gain after optimal assignment. It is worth noting that in the tests mentioned above, the signal at RIS\#4 does not experience any gain after RIS\#3 completed beamforming. To investigate whether RIS\#3 is causing the absence of gain due to specular reflection, further testing of signal strengths at the points shown in Fig.\ref{F2-16} is conducted.

In Fig.\ref{F2-16}(a), when the emission angle of RIS\#4 is continuously adjusted from $0^{\circ}$ to $49^{\circ}$ while keeping the distance from the receiving end to the center of RIS\#4 constant, the received signal strength gain decreases from 23 dB to 0 dB. At this point, the received signal strength is -38 dBm. If the distance is fixed at 2.70 meters, the received signal strength decreases to -44 dBm. However, when RIS\#4 completes beamforming at a reflection angle of $49.45^{\circ}$, there is no additional gain in received signal power.

When RIS\#4 is replaced with RIS\#3 while maintaining the reflection angle at $73.50^{\circ}$ and controlling RIS\#2 beamforming to the new RIS\#3 position, the signal is boosted from -43.9 dBm to -35.1 dBm. This is close to the gain of -35 dBm achieved by the original RIS\#4 configuration when adjusted for equivalent distance. Therefore, by reasonably setting the signal reflection angle to achieve the same signal power gain at an equivalent distance, it is possible to reduce the number of RIS units. In summary, multi-hop RIS can be employed to enhance wireless communication in confined spaces like tunnels and mines, significantly increasing received signal power gain. Starting from the third hop, the RIS panels offer zero signal gain in the specular reflection angle direction or near the specular reflection angle direction. With the same gain, the number of RIS panels can be reduced by increasing the reflection angle.

\subsection*{(2) The experiment of multiple RIS-assisted imaging}

In the context of smart homes and healthcare, people increasingly desire remote monitoring of the behavior and physiological conditions of care recipients without invading their privacy. Wireless imaging offers secure privacy while compensating for the limitations of optical imaging in low-light conditions or when a line of sight is obstructed. In recent years, wireless sensing technologies have enabled location tracking, gesture recognition, posture identification, and respiration monitoring, among other applications~\cite{xu2023enhanced}. However, traditional wireless sensing is constrained by the need for individuals to wear sensing devices, making it challenging to achieve precise sensing of specific body parts.

RIS-assisted wireless imaging systems can detect passive objects in space while focusing on any specific part of the body, presenting clearer sensing images. This experiment sets up an imaging validation system based on RIS, as shown in Fig.\ref{F2-17}. The system operates in the 5.8 GHz Wi-Fi frequency band for sensing imaging. The directional antenna Tx and the RIS panel center have a consistent height of 1 meter. The directional antenna Tx is positioned 1 meter from the RIS panel center at an incident angle of $45^{\circ}$. The space of interest (SOI) is in front of the RIS two meters and has an area of 0.5 m$\times$1.0 m$\times$1.8 m. The directional antenna Rx is placed 1.5 meters in front of the SOI at a height of 0. Using RIS beamforming, the received signal is reflected into the SOI region, and Rx detects reflected or scattered signals in the SOI region, determining the presence of objects in that area. The RIS consists of 10$\times$32 units.

\begin{figure}
\centerline{\includegraphics[width=0.95\linewidth]{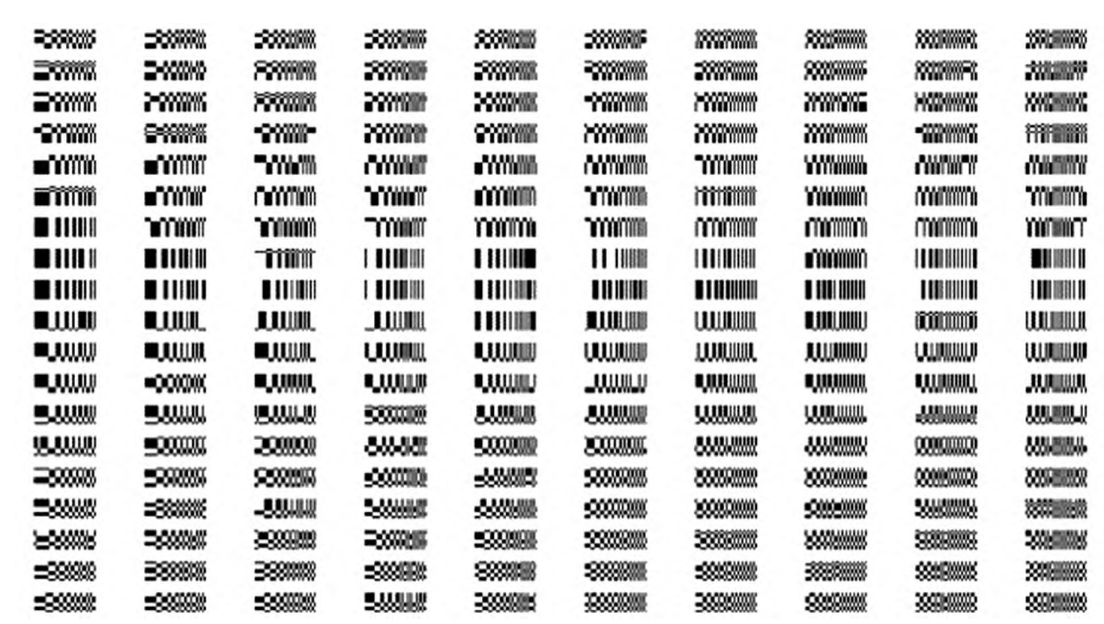}}
\caption{Experiment of RIS-assisted imaging}
\label{F2-18}
\end{figure}

\begin{figure}
\centerline{\includegraphics[width=0.9\linewidth]{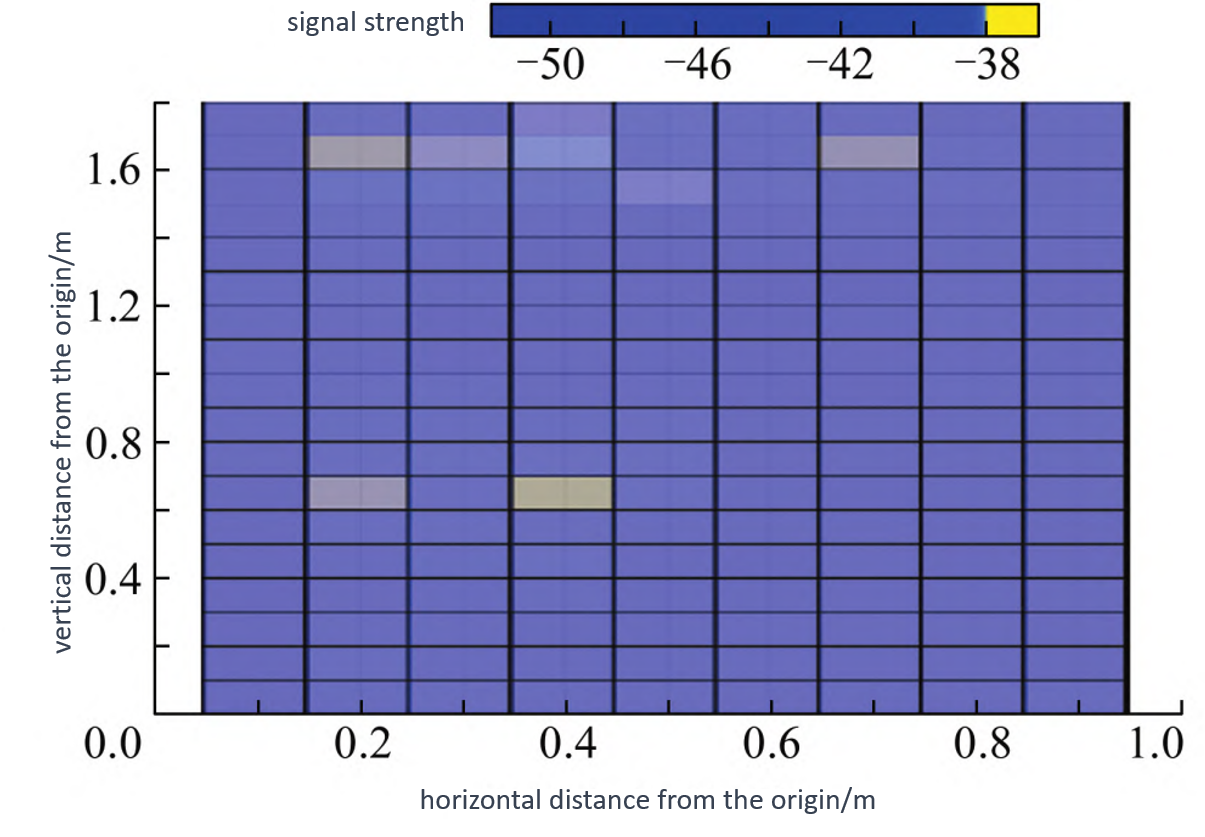}}
\caption{The optimized phase shift matrix that corresponds to the SOI discrete point.}
\label{F2-19}
\end{figure}

\begin{figure}
\centerline{\includegraphics[width=1\linewidth]{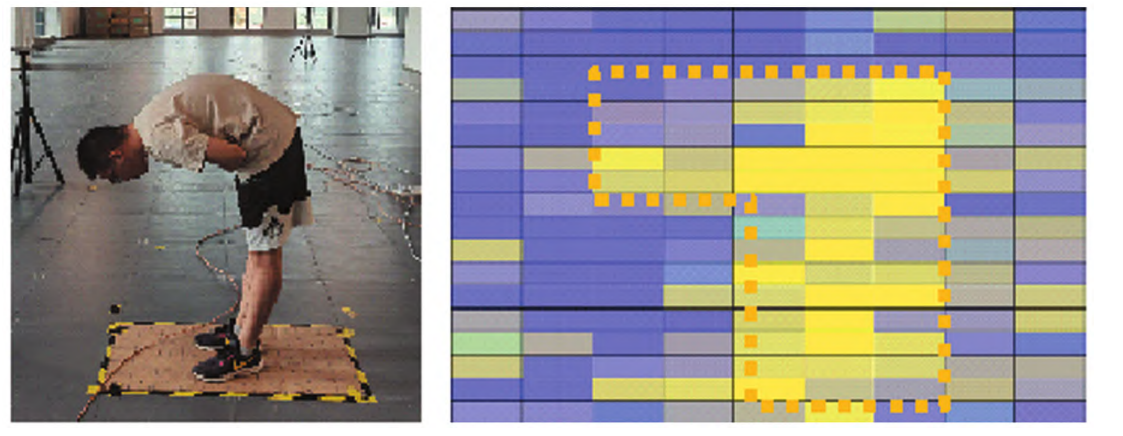}}
\caption{Imaging of a bent-over posture.}
\label{F2-20}
\end{figure}

Firstly, the SOI space is discretized into X$\times$Y$\times$Z=N points, and the index of each point is denoted as n. The RIS beams are assigned to the discretized points in the SOI space, and the encoding of the designed beam assignment phase-shift matrix is assigned to the RIS to control the maximum signal strength at that point. 
In the experiment, we use the greedy algorithm to search the RIS phase-shift matrix and obtain the optimal phase-shift matrix corresponding to each spatial point, which is denoted as $A_n$. According to the experimental setup in Fig.~\ref{F2-17}, the Rx is placed 1.5 meters in front of the SOI, and the optimal phase-shift matrix is called for the beam forming in the SOI space, and the signal strengths of the reflections back from each point in the SOI are obtained by the Rx for each point to be imaged. 
The SOI points are discretized into a 5$\times$10$\times$19 space. Fig.\ref{F2-18} shows the optimal phase shift matrix corresponding to each discrete surface of the first vertical. Fig.\ref{F2-19} shows the imaging in the SOI region without the presence of a human body, when there is no human body in the SOI region, the spatially reflected signal intensity is lower. Fig.\ref{F2-20} shows the signal strength received by the Rx antenna after traversing the optimal phase shift matrix once. Because of the existence of the human body in the SOI region, the spatial reflection of the signal strength is higher, so the yellow area in Fig.\ref{F2-20} corresponds to a stronger reflected signal.

RIS-assisted imaging can better detect objects in space in more detail. Achieving scene reconstruction, limb recognition or imaging of radio waves is usually a typical inverse problem of nonlinear electromagnetic fields, and modeling and analysis of complex electromagnetic environments using conventional methods is also a very difficult task. In subsequent studies, this task can be combined with deep learning networks for super-resolution enhancement of images to achieve clearer imaging.

\subsection*{(3) Coverage enhancement test in the chamber}

\begin{figure}
\centering
\subfloat[Plan view.]{
\includegraphics[width=.52\columnwidth]{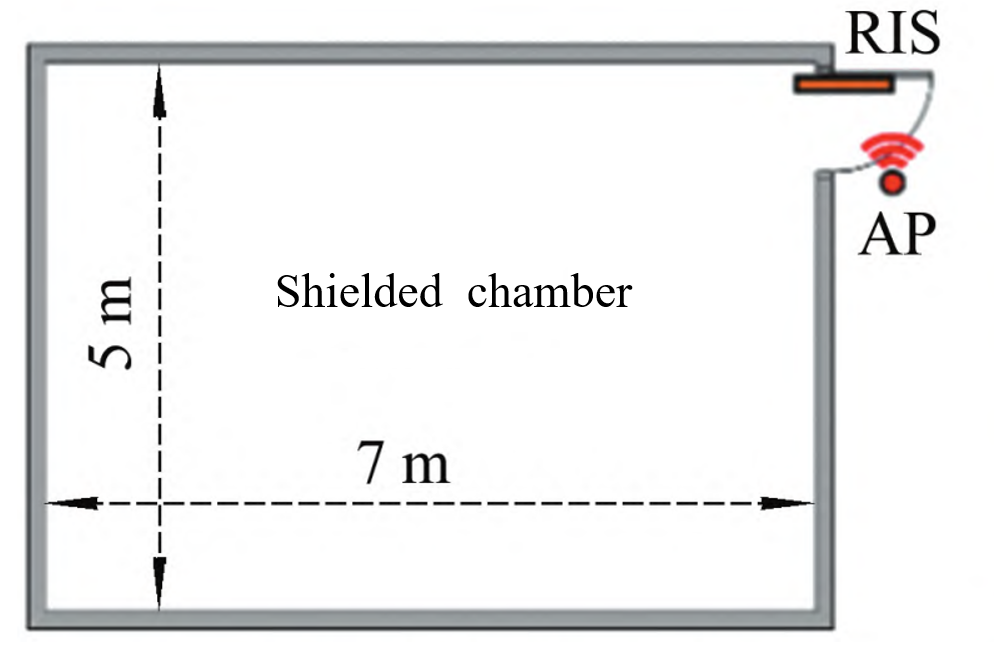}}
\subfloat[Actual scene.]{
\includegraphics[width=.42\columnwidth]{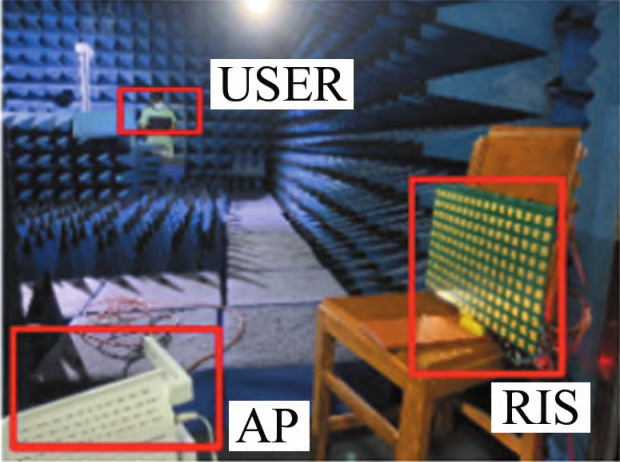}}
\caption{Single RIS-assisted signal coverage enhancement in the chamber.}
\label{F2-21}
\end{figure}

\begin{figure*}
\centering
\subfloat[Without RIS.]{
\includegraphics[width=.3\linewidth]{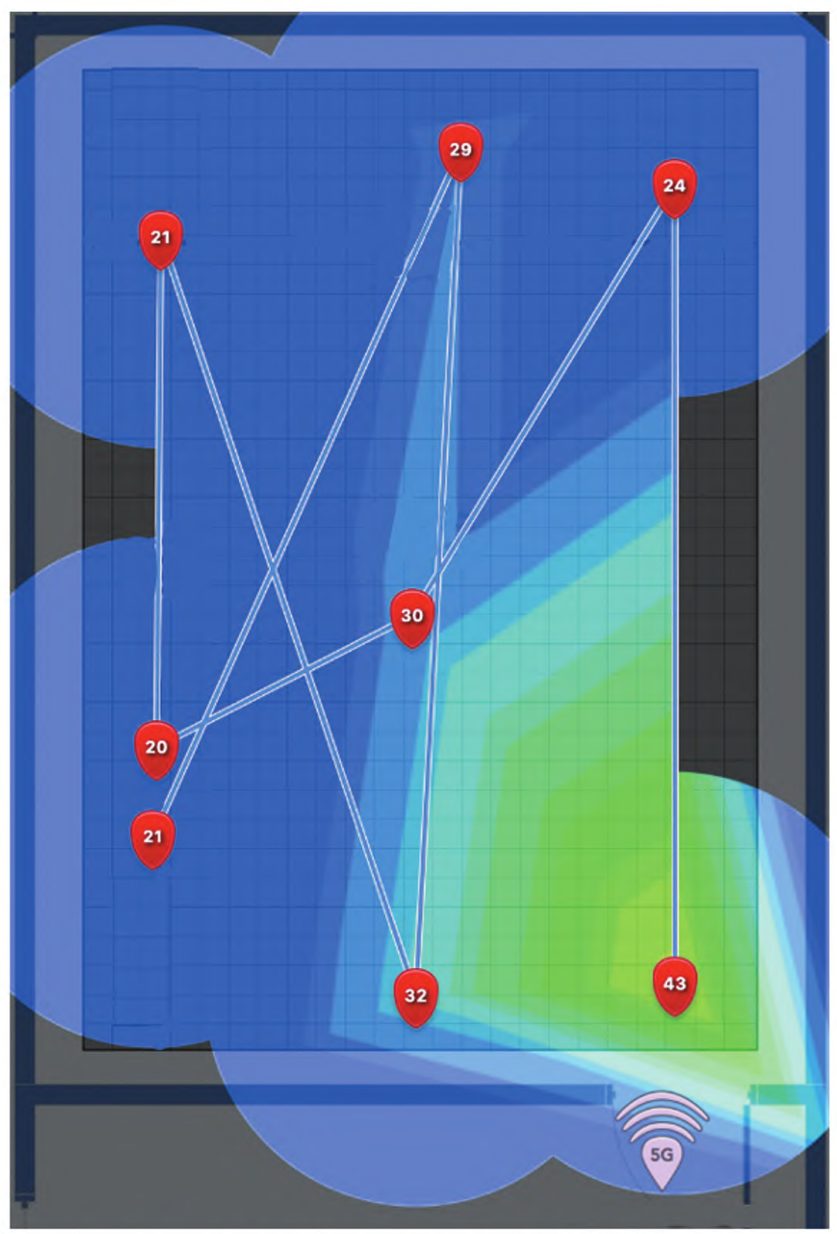}}
\subfloat[RIS without phase optimization.]{
\includegraphics[width=.3015\linewidth]{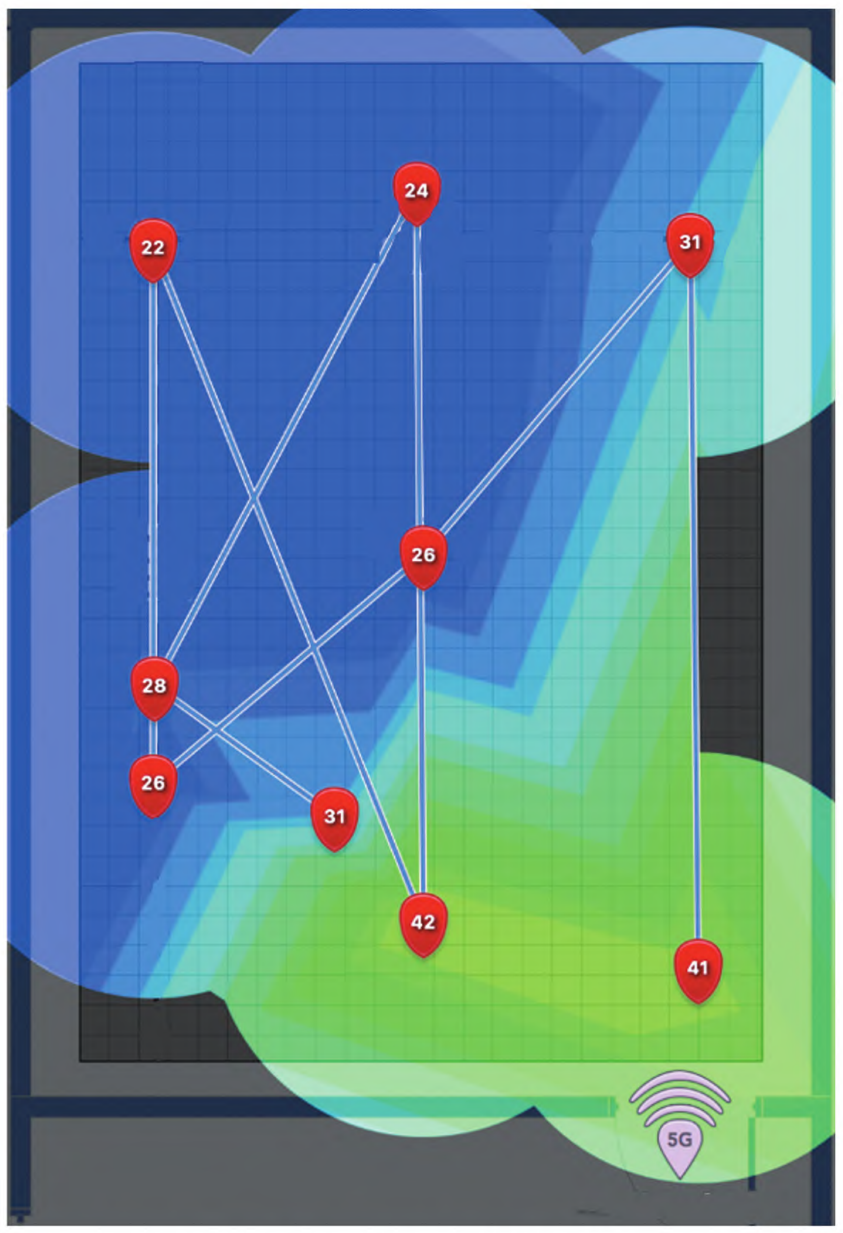}}
\subfloat[ RIS with phase optimization.]{
\includegraphics[width=.2965\linewidth]{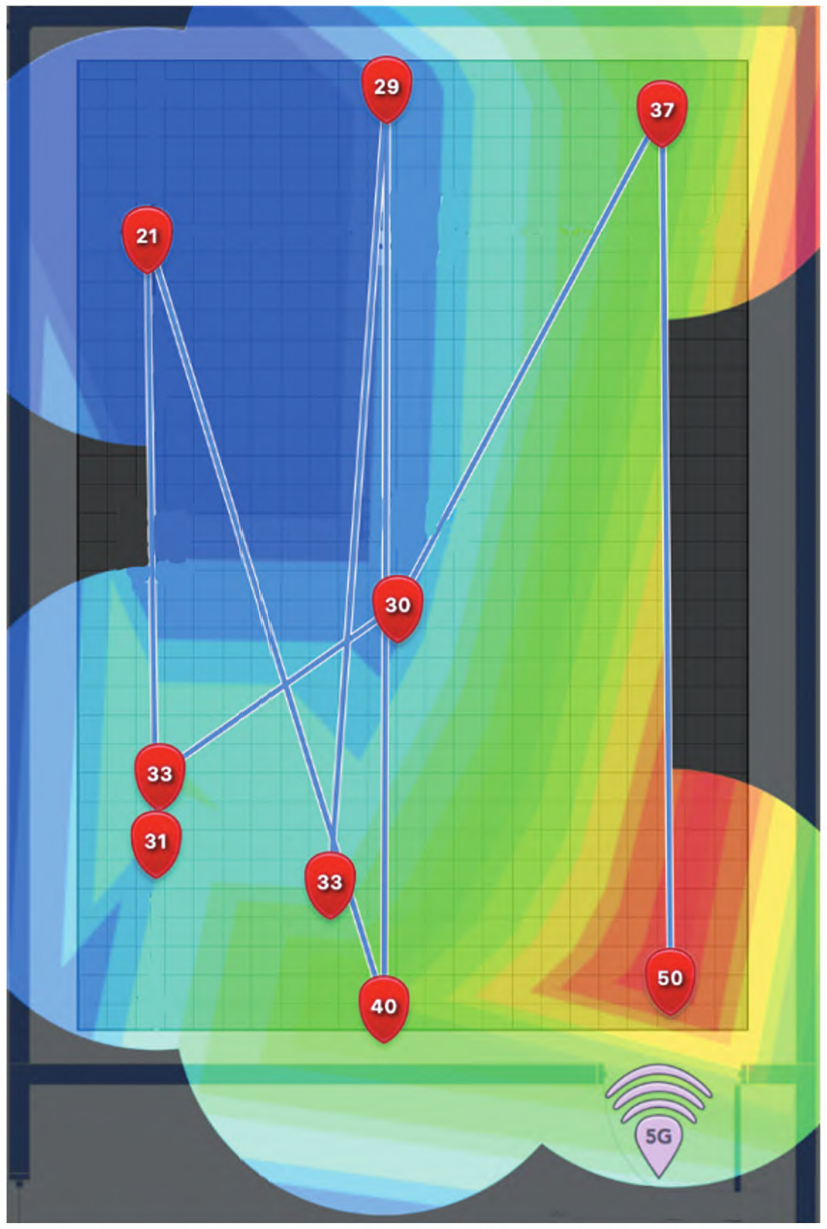}}
\caption{Heat map of signal coverage in the chamber.}
\label{F2-22}
\end{figure*}

The indoor Wi-Fi signal coverage enhancement test scenario is shown in Fig.\ref{F2-21}, and the experiment is completed in a microwave chamber of 5$\times$7$\times$6 ${\rm m^3}$. In this experiment, a Huawei Honor XD3 router supporting 3 000 Mbit/s Wi-Fi 6 is used as the transmitter of the wireless signal, and a RIS supporting 5.8 GHz band with 10$\times$16 array units is placed at the entrance of the chamber to complete the enhancement of the signal coverage. Firstly, we test the signal strength in the environment without the RIS. We run the NetSpots software on the PC, record the Wi-Fi signal strength at 8 sampling points in real-time, and use the signal strength to draw the heat map of the whole room, as shown in Fig.\ref{F2-22}. The signal strength of Fig.\ref{F2-22}(a) is less than 70 dB due to the lack of a direct link from the router to the area in the red circle. In order to solve this problem, the DaS algorithm \cite{xiong2022optimal} is used to enhance the signal by shaping the RIS beams to the area with a distance of 350 cm and a pinch angle of $45^{\circ}$.

After beamforming, the signal strength of all sampling points is measured again using NetSpots, and the results are shown in Fig.\ref{F2-22}(b). Compared with the area before the RIS is placed, the SNR of the area covered by the RIS beamforming is improved by more than 10 dB on average; at the same time, the SNR of the area covered by the RIS beam forming is improved by 5 dB on average compared with the area covered by only placing the RIS in the off state (the diodes of all the RIS units are not on).

\begin{figure}
\centering
\subfloat[Plan view.]{
\includegraphics[width=.52\columnwidth]{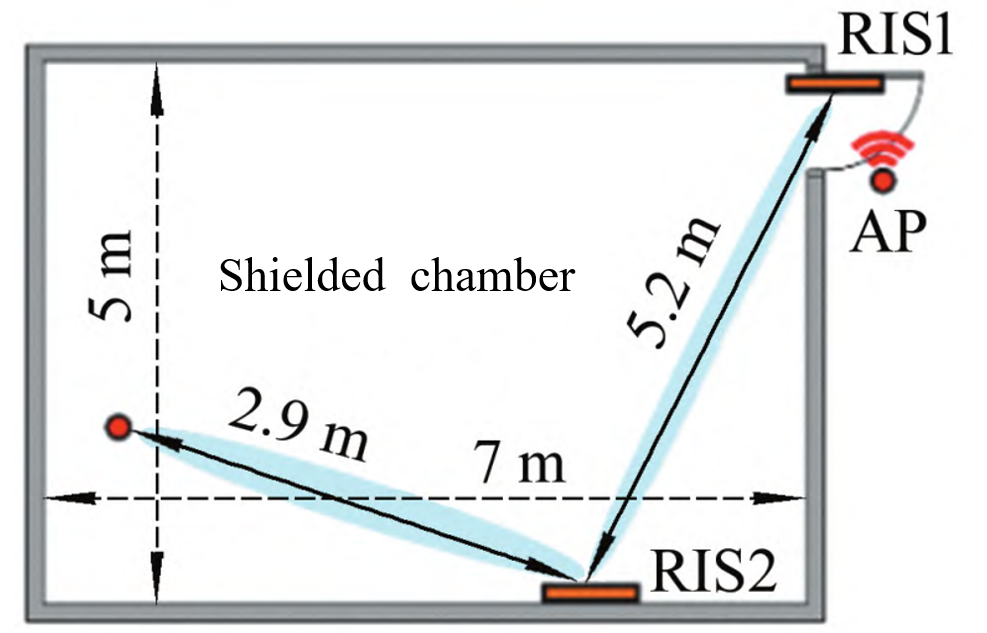}}
\subfloat[Actual scene.]{
\includegraphics[width=.42\columnwidth]{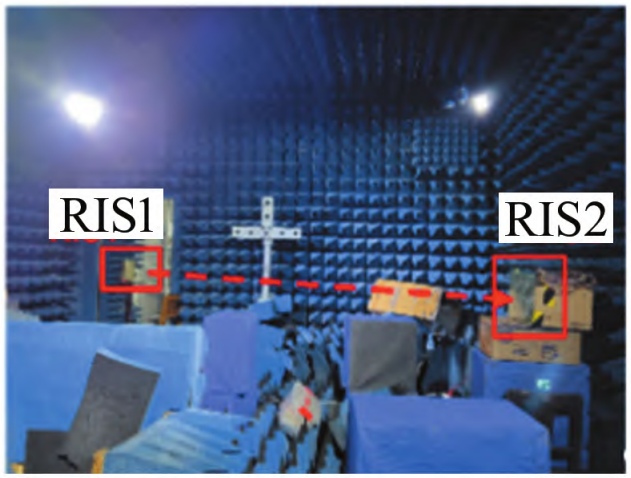}}
\caption{Dual-hop RIS-assisted signal coverage enhancement in the chamber.}
\label{F2-23}
\end{figure}

\begin{figure}[htbp]
\centerline{\includegraphics[width=1.0\linewidth]{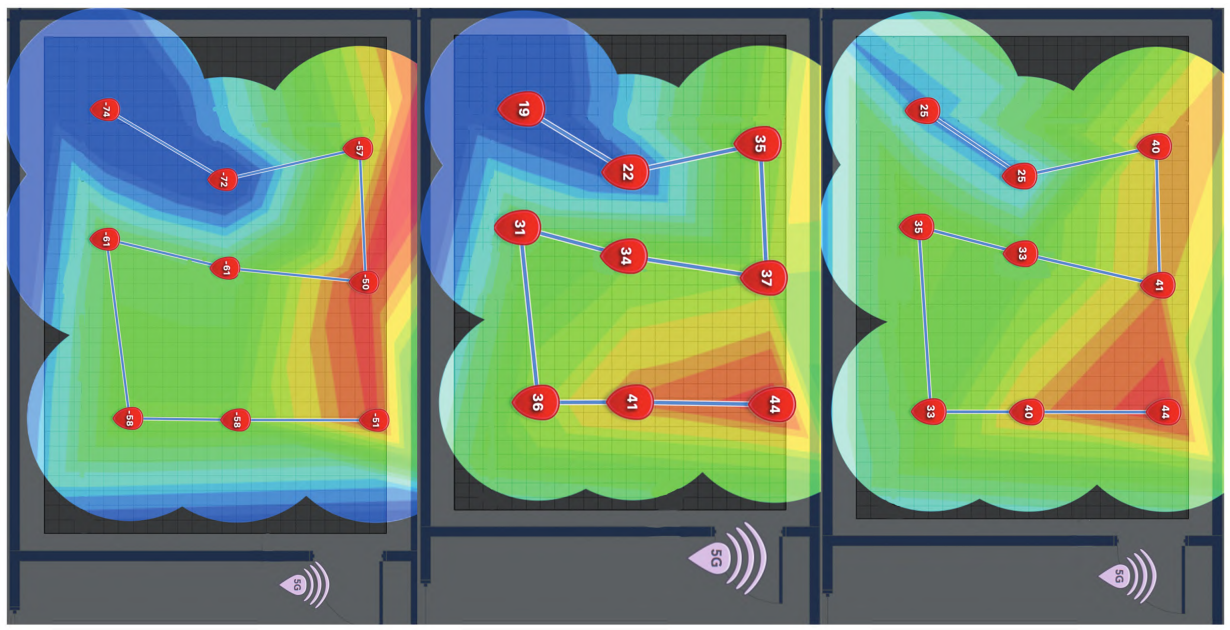}}
\caption{Heat map of signal strength distribution in dual-hop RIS coverage experiment.}
\label{F2-24}
\end{figure}

The central console of the chamber results in the presence of non-directed links in some areas. In order to improve the signal coverage and enhance the signal strength in the blind spots, a double-hop signal relay algorithm is proposed, and the experimental scenario is shown in Fig.\ref{F2-23}. The DaS-based beam optimization algorithm is applied to RIS\#1 to RIS\#2 focusing and RIS\#2 to blind spot signal coverage. RIS\#2 is blinded to the area at 2.9 meters. The results of the thermal map are shown in Fig.\ref{F2-23}. The heat map results are shown in Fig.\ref{F2-23}. The results of the heat map are shown in Fig.\ref{F2-24}. Compared with the single RIS coverage, the signal strength of the blind spot area can be improved by more than 8 dB after the double-hop RIS coverage.

\section{Conclusion and Future directions}\label{Section4}

This paper primarily delineates and summarizes the developmental trajectory of existing RIS units and the design of prototypes, with a detailed analysis of the major outdoor field tests. It highlights the shortcomings present in current RIS systems. Furthermore, addressing these limitations provides a prospective outlook on potential future research directions.
\subsection{Unit Design}

(i) Tailored unit design for both far-field and near-field scenarios. In the past, component unit designs mostly focused on far-field scenarios. However, in indoor environments, near-field applications such as smart homes are prevalent. Researching the electromagnetic distribution theory of near-field and far-field and analyzing the factors influencing the design of these components can effectively enhance the efficiency of RIS systems in indoor scenarios.

(ii) Multifunctional RIS unit design with diverse operational modes, encompassing reflection, refraction, and absorption. Presently, RIS units tend to be relatively straightforward, primarily functioning in the domain of wave reflection. However, imbuing components with versatile properties tailored to specific requirements, and incorporating multifunctional structural design within a single unit, can significantly enhance the adaptability of RIS across various scenarios.

(iii) Unit design for high frequency. As frequency increases, the robustness of electromagnetic components tends to weaken. Especially in millimeter-wave, terahertz, or even higher frequency bands, there are elevated demands for the electromagnetic properties, structural stability, and precision of components like diodes, which play a pivotal role in the design.

(iv) Design for broadband frequency. Presently, RIS component units are typically designed for narrow frequency bands, often only accommodating a few tens of megahertz or even a single frequency point. In real-world applications, it becomes imperative to devise distinct RIS unit architectures tailored to the demands of various frequency bands. Additionally, as communication frequencies ascend towards higher spectrums, necessitating wider communication frequency bands, there remains a dearth of unit designs capable of delivering consistent performance for ultra-wide bandwidth requirements.

(v) Randomness and Ultra-Density Units. The majority of RIS unit designs adhere to regular patterns and often abide by the $\lambda/2$ size principle. Akin to high-frequency sampling in the time domain, high-frequency spatial sampling implies densely packed RIS units. Concentrating the units within a restricted spatial area can increase the sampling rate and enhance the signal-to-noise ratio. Furthermore, the statistical characteristics of random arrays can be utilized for rapid responses in particular scenarios, such as wide-area signal coverage.

(vi) Advancements in materials: A pivotal research direction involves the development of novel materials tailored for RIS unit design on specialized media, including materials like liquid crystal and plexiglass integrated into substrates like windows.

\subsection{RIS Prototype Systems}

(i) Theoretical derivation of system-level performance. Existing RIS systems predominantly employ a single or a limited number of RIS units for elementary functions like signal enhancement. Scaling these applications to complex functions, such as multi-RIS cooperation or large-scale deployment, remains uncharted territory.

Theoretical boundaries of performance in single RIS-assisted communication setups, and the interplay between diverse performance metrics and the physical attributes of the RIS, are subjects of ongoing exploration. Furthermore, large-scale RIS network systems necessitate refined theoretical models and analyses for crucial parameters like channel capacity and throughput. In scenarios with randomly distributed large-scale RIS, discerning system capacity and, from an information theory standpoint, employing apt coding schemes to align RIS system performance with Shannon's classical limits pose unresolved challenges. In the future, amalgamating theoretical derivation with real-world implementation in expansive RIS network environments will bolster the advancement of RIS prototype systems, bridging the gap toward theoretical limits to a considerable extent.

Moreover, in multi-user scenarios, noise in the channel has a limited impact on capacity under high signal-to-noise ratios. At this time, the mutual interference of signals between users is the primary factor affecting capacity. Theoretically, combining phase modulation of RIS with interference management strategies like zero-forcing and interference alignment can greatly enhance system degrees of freedom~\cite{fu2021reconfigurable,bafghi2021degrees,jiang2022interference,xu2023enhanced}. In passive RIS with a sufficiently large number of units, it's possible to completely eliminate interference between users~\cite{bafghi2021degrees,jiang2022interference}. Integrating RIS prototype design with interference management strategies to mitigate interference in actual multi-user communication environments is a pressing challenge for the future, and it plays a crucial role in enhancing the quality of wireless communication networks under the scenario of connecting a massive number of users.

(ii) The performance characteristics and application research of RIS prototypes with special structures. Existing experimental studies indicate that certain RIS prototypes exhibit distinctive features. For example, a 1-bit RIS forms symmetrical beams in two directions. These phenomena need in-depth exploration from both theoretical and experimental perspectives.

(iii) RIS prototype design with adjustable phase shift resolution. Due to hardware constraints, the phase shift caused by individual units to incident electromagnetic waves is often discrete. However, in the RIS prototype, it is possible to create a `super unit' with continuous phase variation through the coordinated operation of different units and controllable coupling. This enables the construction of an RIS prototype with arbitrarily adjustable phase shift resolution to meet the requirements of high-precision scenes.

(iv) Collaborative design of the BS and RIS prototype. Current theoretical and experimental results reveal that stronger performance gains can be obtained when the RIS is located closer to the base station or user. However, our experimental results have shown that proximity does not necessarily translate into higher gains, as it is influenced by the non-uniform aperture field distribution on the RIS surface. Additionally, user positions are often not fixed, making it necessary to explore the co-design of BS-RIS prototypes. However, determining how to coordinate and optimize the parameters, such as the distance, angle, and area ratio between the RIS and BS, still lacks accurate models and theoretical analysis.

(v) Design of Beamforming Algorithms. In the context of RIS-assisted communication, when a single RIS is employed, the aim of beamforming is typically to maximize the reception of the desired signal power (or signal-to-noise ratio), or alternatively, to optimize the transmission rate. However, in scenarios involving the collaborative operation of multiple RIS units, the principal objective shifts towards enhancing the system's overall capacity. It's important to note that beamforming in such contexts is a sophisticated orchestration of the individual contributions of each RIS, far from a mere superposition. The coordinated interplay of multiple RIS units often yields notably superior performance. Furthermore, in multi-hop scenarios, the focus is frequently on maximizing the efficiency of each reflective node. Hence, research into coordinated beamforming algorithms tailored for multiple RIS represents a crucial research avenue.

(vi) Optimization of large-scale RIS networks. The future landscape of RIS networks is destined to be an expansive domain, characterized by the integration of numerous RISs into the broader Internet infrastructure. Investigating the avenues through which multi-RIS collaboration can be achieved across the realms of the physical layer, network architecture, and protocols is an imperative and noteworthy research endeavor.

(vii) Integration with other systems. In the diversified landscape of the information society, RIS prototypes, as fundamental infrastructures, have the potential to seamlessly integrate their functionalities with established systems. This convergence holds the promise of enhancing user experiences, facilitating aspects like integrated sensing, pervasive intelligence, and seamless communication between virtual and physical realms.


\bibliographystyle{IEEEtran}
\bibliography{Reference}

\end{document}